\documentclass[preprint,12pt]{elsarticle}

\PassOptionsToPackage{margin=0.81in}{geometry}

\usepackage{amssymb, bm}
\usepackage{amsmath}
\usepackage{subfig}
\usepackage{caption}
\usepackage{subfig}
\usepackage{graphicx}
\usepackage{epstopdf}
\usepackage{fancyhdr}
\usepackage{geometry}
\usepackage{setspace}
\usepackage{booktabs}
\usepackage{nomencl}
\usepackage{tikz}
\usepackage{multirow}
\makenomenclature
\RequirePackage{ifthen}
\renewcommand{\nomgroup}[1]{\ifthenelse{\equal{#1}{R}}{\item[\textbf{Roman letters}]}{\ifthenelse{\equal{#1}{S}}{\item[\textbf{Subscripts}]}{\ifthenelse{\equal{#1}{G}}{\item[\textbf{Greek letters}]}{\ifthenelse{\equal{#1}{A}}{\item[\textbf{Acronyms}]}}}}}

\journal{International Journal of Heat and Mass Transfer}

\begin{document}

\begin{frontmatter}

%% Title, authors and addresses

%% use the tnoteref command within \title for footnotes;
%% use the tnotetext command for theassociated footnote;
%% use the fnref command within \author or \address for footnotes;
%% use the fntext command for theassociated footnote;
%% use the corref command within \author for corresponding author footnotes;
%% use the cortext command for theassociated footnote;
%% use the ead command for the email address,
%% and the form \ead[url] for the home page:
%% \title{Title\tnoteref{label1}}
%% \tnotetext[label1]{}
%% \author{Name\corref{cor1}\fnref{label2}}
%% \ead{email address}
%% \ead[url]{home page}
%% \fntext[label2]{}
%% \cortext[cor1]{}
%% \address{Address\fnref{label3}}
%% \fntext[label3]{}

\title{DropletSMOKE++: a comprehensive multiphase CFD framework for the evaporation of multidimensional fuel droplets}

%% use optional labels to link authors explicitly to addresses:
%% \author[label1,label2]{}
%% \address[label1]{}
%% \address[label2]{}

\author[a]{A.E. Saufi}
\author[a]{A. Frassoldati}
\author[a]{T. Faravelli}
\author[a]{A. Cuoci}

\address[a]{Department of Chemistry, Materials, and Chemical Engineering “G. Natta”, P.zza Leonardo da Vinci 32, Milano, Italy}

\begin{abstract}
	
	This paper aims at presenting the DropletSMOKE++ solver, a  comprehensive multidimensional computational framework for the evaporation of fuel droplets, under the influence of a gravity field and an external fluid flow. The Volume Of Fluid (VOF) methodology is adopted to dynamically track the interface, coupled with the solution of energy and species equations. The evaporation rate is directly evaluated based on the vapor concentration gradient at the phase boundary, with no need of semi-empirical evaporation sub-models.  \\
	The strong surface tension forces often prevent to model small droplets evaporation, because of the presence of parasitic currents. In this work we by-pass the problem, eliminating surface tension and introducing a  centripetal force  toward the center of the droplet. This expedient represents a major novelty of this work, which allows to numerically hang a droplet on a fiber in normal gravity conditions without modeling surface tension. Parasitic currents are completely suppressed, allowing to accurately model the evaporation process whatever the droplet size.\\
	DropletSMOKE++ shows an excellent agreement with the experimental data in a wide range of operating conditions, for various fuels and initial droplet diameters, both in natural and forced convection. The comparison with the same cases modeled in microgravity conditions highlights the impact of an external fluid flow on the evaporation mechanism, especially at high pressures. Non-ideal  thermodynamics for phase-equilibrium is included to correctly capture evaporation rates at high pressures, otherwise not well predicted by an ideal gas assumption. Finally, the presence of flow circulation in the liquid phase is discussed, as well as its influence on the internal temperature field.\\
	DropletSMOKE++ will be released as an open-source code, open to contributions from the scientific community.

\end{abstract}


\end{frontmatter}

%% \linenumbers
%	\printnomenclature	

\section{Introduction}
The high energy density of liquid fuels is nowadays exploited in many engineering devices such as diesel engines and industrial burners. Spray injection systems are widely used in order to disperse the liquid fuel in an oxidizing environment and the control of this process is currently an active area of research. In particular, the study of a single isolated droplet allows to neglect complex interaction phenomena (coalescence, breakup, etc.) among droplets and to focus on the evaporation and combustion mechanisms. 
In the last decades the numerical modeling of droplets evaporation and combustion has considerably improved. Starting from the works of Spalding \cite{spalding1953combustion} and Godsave \cite{godsave1953studies} concerning the well known "$d^2$ law", derived under the hypothesis of a perfectly spherical, motionless and constant liquid temperature droplet, Abramzon  and Sirignano \cite{abramzon1989droplet, sirignano1999fluid} started to account for a convective flow, providing numerous correlations for the heat and mass transfer outside and within the droplet in presence of a relative gas motion. In this context, Dwyer et al. analyzed the droplet dynamics in high $T$ fields \cite{dwyer1989calculations, dwyer1985detailed}. In particular, the transient surface heating has been investigated by Law \cite{law1976unsteady}, assuming a constant spatial liquid temperature, while Kotake and Okazaki  \cite{kotake1969evaporation} started to analyze the influence of the liquid internal circulation on the vaporization rate. More recently,  Sazhin et al. analyzed the heating and evaporation process \cite{sazhin2014modelling, sazhin2006advanced}, as well as the effect of thermal radiation \cite{abramzon2005droplet}.\\
With the growing available computational power, the direct numerical solution of the equations governing the evolution of isolated fuel droplets became affordable, allowing the detailed 1D modeling of spherical fuel droplets in microgravity conditions \cite{marchese1999numerical, cuoci2005autoignition, farouk2014isolated}. Microgravity conditions can be reproduced conducting the experiments in free-falling chambers \cite{okajima1975further, kumagai1971combustion, chauveau2000effects}, both for suspended and free droplets. Recently, some  experiments have been conducted aboard the International Space Station (ISS) for the study of evaporation and multi-stage combustion phenomena. The absence of gravity allows to adopt a simple 1D mathematical description and to focus on physical aspects such as differential species diffusion, non-ideal thermodynamics and combustion kinetics. In particular, the implementation of detailed mechanisms for gas-phase chemistry in 1D models has paved the way for a better understanding of low-T chemistry \cite{farouk2017combustion, farouk2017isolated,  dietrich2014droplet} , ignition and extinction phenomena \cite{cuoci2017flame, stagni2018numerical}.\\
However, most of experiments on droplets are conducted under the influence of a gravity field, since experimental devices are much cheaper and the operating conditions are closer to the real ones where droplets are commonly adopted. 1D models are intrinsically not able to predict physical phenomena such as buoyancy, droplet deformation, relative gas motion and internal circulation, which are always present in real conditions. In particular, the external flow field (both induced and forced) around the droplet strongly modifies the evaporation rate at the interface,  while the liquid circulation governs the internal heat transfer.\\ 
From the modeling point of view, the description of these phenomena  requires at least a 2D multiphase model, able to predict the anisotropic deformation of the droplet and to provide both liquid and gas velocity fields. This must be coupled with a reliable evaporation model to describe the droplet shrinking, the  diffusion of vapor towards the gas-phase and its further transport, both by the external convection and by the induced Stefan flow. The energy transfer between the two phases is also necessary, as well as proper correlations for the fluid properties.\\
Among the several methods available for multiphase flows, the Volume Of Fluid (VOF) methodology \cite{hirt1981volume} is widely known for its simplicity, robustness and especially for the excellent mass conservation properties. Its application for evaporation and condensation phenomena has been investigated in recent years \cite{nabil2016interthermalphasechangefoam, georgoulas2017enhanced, nieves2015openfoam, guo2014review}. However, most of these works rely on phase-change models based on experimental data, kinetic theories and semi-empirical correlations. A more general evaporation model based on the interface concentration gradient has been implemented by Shlottke \cite{schlottke2008direct}, neglecting however the detailed characterization of thermodynamic equilibrium at the interface and without a comparison of the results with experimental data. More recently, Gatha et al. \cite{ghata2014computational, ghata2015computational} used a VOF technique to model droplet evaporation and combustion, limiting however the investigation to microgravity conditions.\\
A major numerical issue of VOF method concerns surface tension. Small droplets dynamics involve strong surface tension forces, which are extremely difficult to model due to the presence of the so-called parasitic currents \cite{brackbill1992continuum}. These currents tend to numerically deform and eventually destroy the droplet, due to inaccuracies on the evaluation of the surface curvature. Some techniques \cite{cummins2005estimating, popinet2018numerical} have been developed to mitigate parasitic currents for particular cases (e.g. rising bubbles, inviscid static droplets, capillary oscillations), but small droplets evaporation still suffers from this issue. The rapid size reduction of the droplet makes surface tension to be more and more dominant, further amplifying the problem. 
Even if some works concerning droplets vaporization with VOF methodology are available in literature \cite{strotos2016predicting, banerjee2013numerical, george2017detailed}, the problem of parasitic currents is often not even mentioned. These works are based on commercial CFD codes (mainly $\texttt{Ansys FLUENT}^{\textregistered}$ and $\texttt{COMSOL}^{\textregistered}$), making very difficult to reproduce the results with open-source codes since the adopted numerical algorithms are not provided in detail. Concerning the numerical framework,  promising results have been reached by  codes such as $\texttt{FS3D}$ \cite{eisenschmidt2016direct} and $\texttt{Gerris}$ \cite{popinet2009accurate}, whereas the $\texttt{OpenFOAM}^{\textregistered}$ environment is still lacking in comprehensive solvers for the detailed analysis of droplets evaporation.\\ 
This work aims at presenting an open-source general computational tool (called \texttt{DropletSMOKE++}) for the modeling of isolated droplets evaporation under convective conditions in the presence of a gravity field, overcoming the aforementioned limits of the currently available solvers. In particular it includes:

\begin{itemize}
	\item An accurate interface tracking method, based on a geometrical advection, coupled with the Navier-Stokes equations,  numerically solved for both liquid and gas phase.
	\item A reliable evaporation model based on the surface mass flux, coupled with a species equation, with no need of semi-analytical approaches or correlations based on mass-heat transfer dimensionless numbers.  
	\item The detailed description of the vapor-liquid equilibrium thermodynamics, including fugacity coefficients and Poynting correction calculation using a cubic Equation of State. 
	\item A numerical technique which suppresses parasitic currents and by-passes the problem of curvature evaluation, introducing an external centripetal force to mimic surface tension effects.
\end{itemize} 

The \texttt{DropletSMOKE++} code is fully implemented within the open-source $\texttt{OpenFOAM}^{\textregistered}$ framework to manage the computational mesh and the discretization of governing equations. The numerical code will be freely available and open for contributions from the scientific community. The code version used for this work is provided in the supplementary material.\\
The paper organization includes a thorough description of the mathematical model, as well as the numerical methodology adopted. The centripetal force substituting surface tension is discussed in detail, including its numerical implementation, its effect on the droplet shape and how it impacts evaporation. \\
A quantitative assessment of the code has been  performed, comparing the \texttt{DropletSMOKE++} results with a reference  finite-differences based solver for droplets evaporation in microgravity conditions \cite{cuoci2005autoignition}. Afterwards, the gravity field has been introduced, allowing the analysis of the natural and forced convection regime.\\
An extensive validation with available experimental data has been carried on for n-heptane, n-decane and n-hexadecane droplets in a very wide range of operating conditions, both for natural and forced convection. Experimental data from various authors include the normalized equivalent diameter decay and some data on the surface temperature behavior, which have been compared with the  \texttt{DropletSMOKE++} numerical results. Equivalent numerical cases (with the same initial conditions) have been simulated also in microgravity conditions in order to highlight the main differences between the models and the intrinsic inability of microgravity solvers to correctly predict the experimental data.
Finally, the presence of internal circulation in the liquid phase is briefly analyzed.

\section{Mathematical model}
\subsection{Interface advection}
The VOF methodology is often referred to as a "one-fluid" approach, where the two phases are treated as a single fluid whose properties vary abruptly at the phase boundary. A scalar marker function $\alpha$ represents the liquid volumetric fraction, varying from value 0 in the gas-phase to value 1 in the liquid phase. The $\alpha$  advection equation in the most general form is: 

\begin{equation}
\frac{\partial\left(\rho \alpha\right)}{\partial t}+\nabla\cdot\left(\rho \alpha \textbf{v} \right)=-\dot{m}
\end{equation}

The source term $\dot{m}$  represents the evaporation/condensation contribution to the liquid. Rewriting the equation in function of $\alpha$:

\begin{equation}
\frac{\partial \alpha}{\partial t} + \nabla\cdot\left(\textbf{v}\alpha\right)=-\frac{\dot{m}}{\rho}-\frac{\alpha}{\rho}\frac{D\rho}{Dt}
\end{equation}

This equation is solved in two steps, treating the advection and the source terms in a segregated approach, similarly to what happens in an operator-splitting technique. The interface tracking is solved first, using the \texttt{isoAdvector} library developed by Roenby and Jasak \cite{roenby2016computational} :

\begin{equation}
\frac{\partial \alpha}{\partial t} + \nabla\cdot\left(\textbf{v}\alpha\right)=0
\label{advection}
\end{equation}

\begin{figure}
	\centering
	\subfloat[]
	{\includegraphics[width=.32\textwidth,height=0.25\textheight]{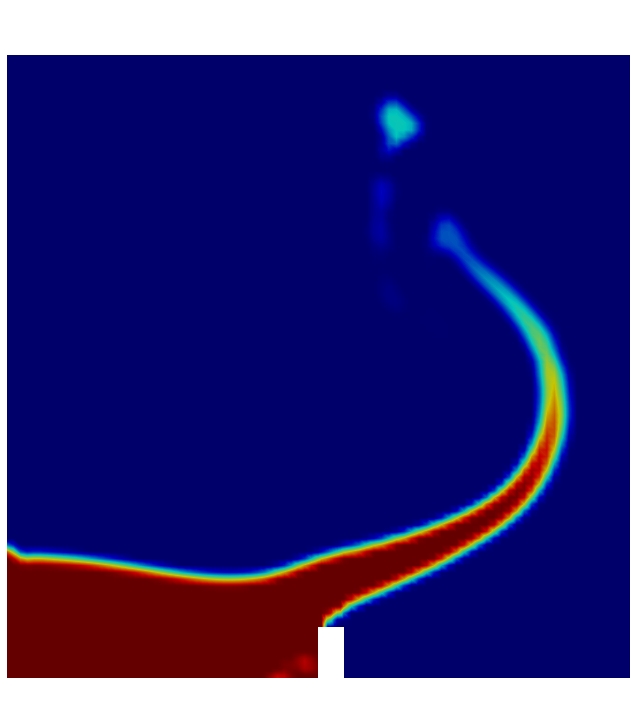}}~~~
	\subfloat[]
	{\includegraphics[width=.32\textwidth,height=0.25\textheight]{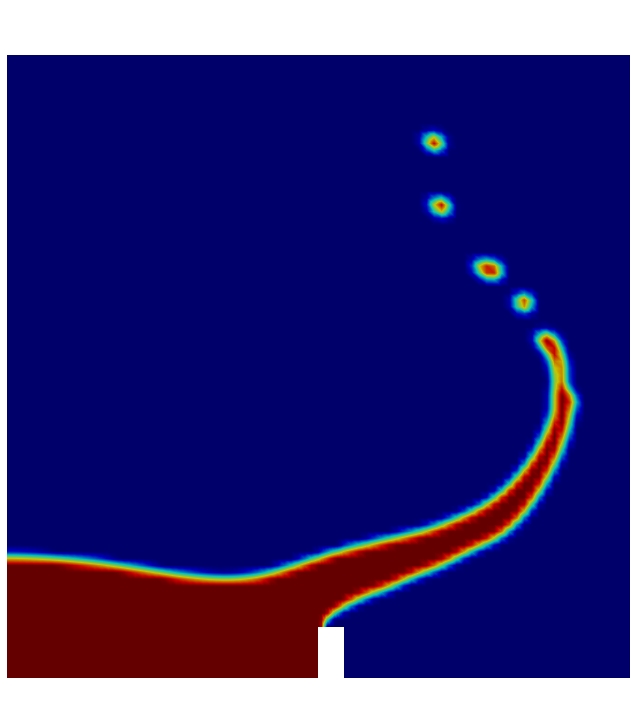}}

	\caption{Comparison of the interface resolution between the MULES compressive scheme (Figure a) and \texttt{isoAdvector} (Figure b) for the well known \texttt{damBreak} case \cite{greenshields2015openfoam}. }
	\label{isoAdvector}
\end{figure}

\texttt{isoAdvector} performs a geometric advection of the interface, whose quality is superior (Figure \ref{isoAdvector}) to the MULES (Multidimensional Universal Limiter with Explicit Solution) compressive scheme by Weller \cite{damian2013extended} usually used in $\texttt{OpenFOAM}^{\textregistered}$ multiphase solvers. The source terms are included in a second step:

\begin{equation}
\frac{\partial \alpha}{\partial t} = -\frac{\dot{m}}{\rho}-\frac{\alpha}{\rho}\frac{D\rho}{Dt}
\label{alphasource}
\end{equation}

where the first term accounts for the evaporation, while the second term describes the liquid density variation, and eventually the droplet expansion due to external heating.
\subsection{Pressure and velocity fields}
A single Navier-Stokes equation is solved for both phases:

\begin{equation}
\frac{\partial\left(\rho\textbf{v}\right)}{\partial t} + \nabla\cdot\left(\rho\textbf{v}\otimes\textbf{v}\right)= \nabla\cdot\mu\left(\nabla\textbf{v}+\nabla\textbf{v}^T\right)-\nabla p + \rho\textbf{g}
\label{navierstokes}
\end{equation}

It is convenient to separate the hydrostatic pressure $\rho\textbf{g}\cdot\textbf{r}$ from the total pressure $p$, in order to simplify the definition of boundary conditions for pressure:

\begin{equation}
\frac{\partial\left(\rho\textbf{v}\right)}{\partial t} + \nabla\cdot\left(\rho\textbf{v}\otimes\textbf{v}\right)= \nabla\cdot\mu\left(\nabla\textbf{v}+\nabla\textbf{v}^T\right)-\nabla p_{rgh} -\textbf{g}\cdot\textbf{r}\nabla\rho
\end{equation}

where $p_{rgh}=p-\rho\textbf{g}\cdot\textbf{r}$ is the dynamic pressure and $\textbf{r}$ is the position vector.\\
The continuity equation (needed for the pressure-velocity coupling) accounts for the density variations and for the abrupt change of density at the interface occurring during the evaporation, generating the  Stefan flow:

\begin{equation}
\frac{\alpha}{\rho_L}\frac{D\rho_L}{Dt}+ \frac{1-\alpha}{\rho_G}\frac{D\rho_G}{Dt}+\dot{m}\left(\frac{1}{\rho_L}-\frac{1}{\rho_G}\right) + \nabla\cdot\textbf{v}=0
\label{pequation}
\end{equation}

In case of constant density and no evaporation Equation \ref{pequation} reduces to the continuity equation for incompressible flows $\nabla\cdot\textbf{v}=0$.

\subsection{Temperature and species fields}
Additionally, energy and species equations are included:

\begin{equation}
\rho C_p\left(\frac{\partial T }{\partial t}+\textbf{v}\nabla T\right) = \nabla\cdot\left(k\nabla T\right) + \beta\frac{Dp}{Dt}-\sum_{i=0}^{Ns}\textbf{j}_{d,i}C_{p,i}\nabla T - \sum_{i=0}^{Ns_{L}}\dot{m}_i\Delta h_{ev,i}
\label{Teqn}	
\end{equation}

\begin{equation}
\rho \left(\frac{\partial \omega_i }{\partial t}+\textbf{v}\nabla \omega_i\right) = -\nabla\cdot\textbf{j}_{d,i}
\label{speciesequation} 
\end{equation}

The last term  in Equation \ref{Teqn} accounts for the interface cooling during evaporation. The diffusion fluxes $\textbf{j}_{d,i}$ are discussed in the next section.

\subsection{Evaporation model}
In this work, the mass evaporation flux is calculated based on the interface concentration gradient of the vapor. Its evaluation  is not trivial, since it strongly depends on how the diffusion coefficients are computed. The species diffusion coefficients $\mathcal{D}_i$ in the mixture are evaluated based on the binary diffusion coefficients and the mole fractions:

\begin{equation}
\mathcal{D}_i=\frac{\sum_{j\neq i}^{}y_j M_{w,j}}{M_w \sum_{j \neq i}^{}\frac{y_i}{\mathcal{D}_{i,j}}}
\end{equation}

which means that the diffusion velocities $\textbf{v}_{d,i}$ must be evaluated based on the mole fraction gradient of the species \cite{bird2002transport}:

\begin{equation}
\textbf{v}_{d,i}=-\mathcal{D}_i\frac{\nabla y_i}{y_i}
\end{equation}

The diffusion fluxes $\textbf{j}_{d,i}$ are defined as:

\begin{equation}
\textbf{j}_{d,i}=\rho\omega_i\textbf{v}_{d,i}=-\rho\omega_i\mathcal{D}_i\frac{\nabla y_i}{y_i}=-\rho\mathcal{D}_i\frac{M_{w,i}}{M_w}\nabla y_i
\end{equation}

This expression of the diffusion flux is also used in Equation \ref{speciesequation}.\\\\
A convective flux $\textbf{j}_{c,i}$ is generated by the interfacial density change during the evaporation process, as stated in Equation \ref{pequation}:

\begin{equation}
\textbf{j}_{c,i}= \rho\omega_i \textbf{v}
\label{convectiveFlux}
\end{equation}

The total evaporating flux $\textbf{j}_{i}$ is the sum of diffusive and convective fluxes:

\begin{equation}
\textbf{j}_{i}= \textbf{j}_{d,i}+\textbf{j}_{c,i}
\label{globalFlux}
\end{equation}

\begin{figure}
	\centering
	\subfloat[]
	{\includegraphics[width=.45\textwidth,height=0.24\textheight]{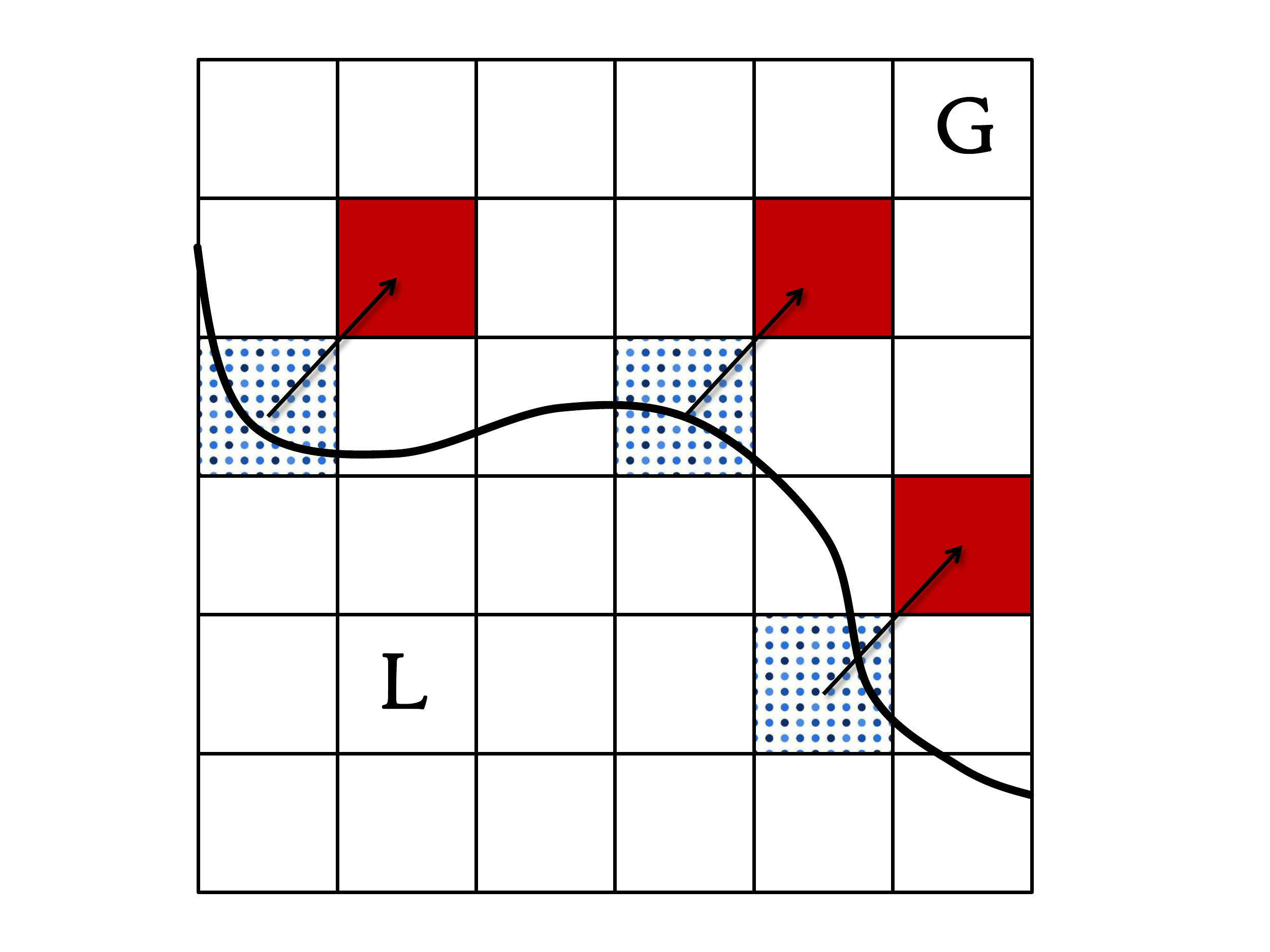}}~~
	\subfloat[]
	{\includegraphics[width=.45\textwidth,height=0.25\textheight]{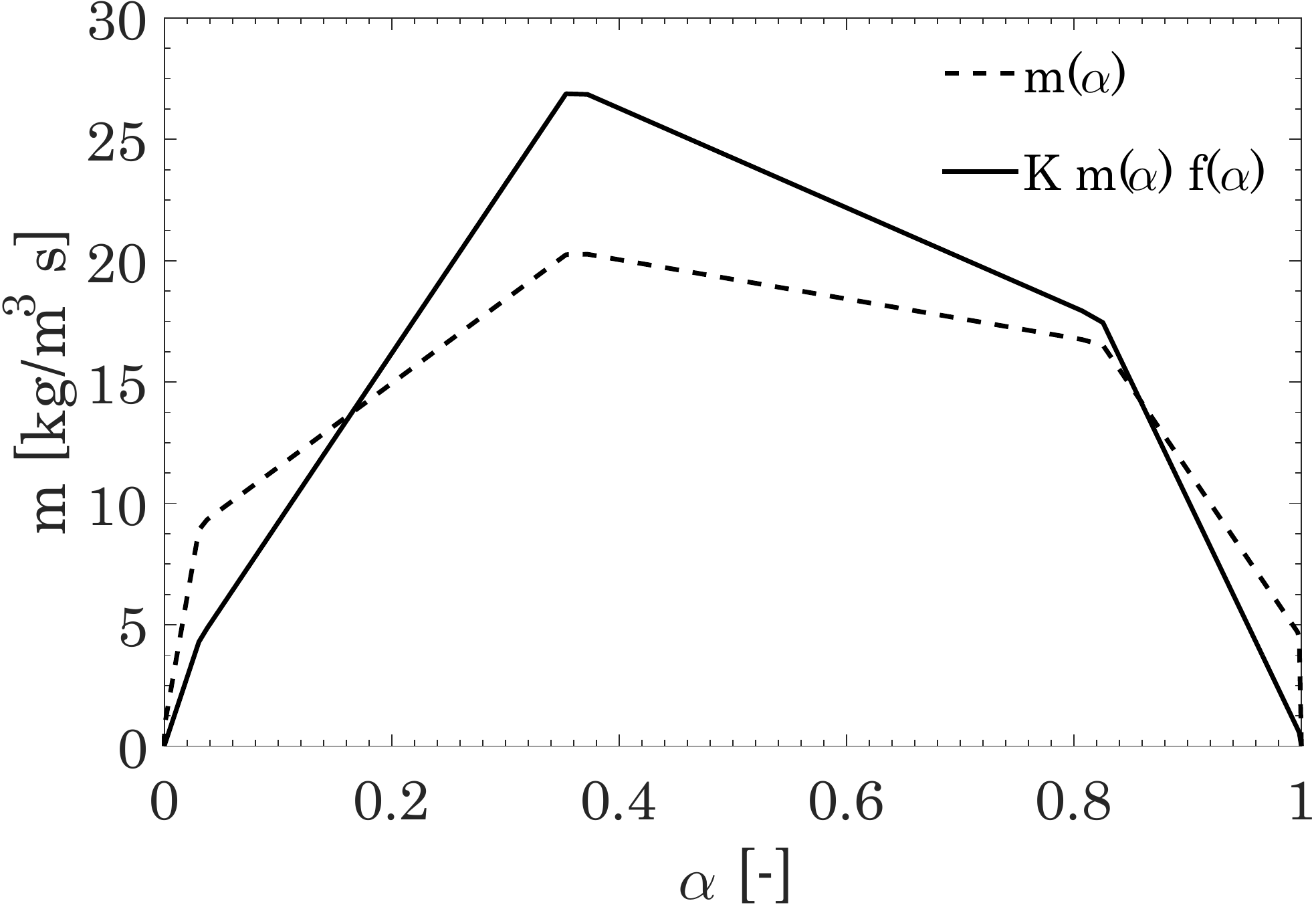}}		
	
	\caption{Calculation of $\dot{m}_i=\textbf{j}_i|\nabla\alpha|$ in Figure (a): interface cells (in dotted blue) associated to the adjacent gas-phase cells (in red). In Figure (b): evaporation flux $\dot{m}$ distribution over the interface for standard implementation (dashed line, Equation \ref{massflux}) and corrected implementation (solid line, Equation \ref{alphafunction}).}
	\label{massfluxdistribution}
\end{figure}

The local surface area per unit volume is usually computed from the $\alpha$ function as $|\nabla\alpha|$, since its volume integral is the surface area by definition:
\begin{equation}
S=\int_{V}^{}|\nabla\alpha| dV
\end{equation}

The evaporation rate for species $i$  is then:
\begin{equation}
\dot{m}_i = \textbf{j}_i\nabla\alpha
\label{massflux}
\end{equation}

It is important to point out that $|\nabla\alpha|$ is calculated at the interface cells (where  $|\nabla\alpha|\neq 0 $), while $\textbf{j}_i$ must be computed at the gas cells adjacent to the interface cells. Referring to Figure \ref{massfluxdistribution} (a), a specific algorithm has been developed to associate every cell at the interface (represented in dotted blue) to the closest cell in the gas-phase (represented in red) along the interface normal (black arrow). Indicating with subscript $int$ a generic interface cell and with $adj$ the closest adjacent gas-phase cell with respect to $int$, we have:

\begin{equation}
\dot{m}_i\vert_{int}=\textbf{j}_i\vert_{adj}|\nabla \alpha|_{int}
\end{equation}

which has to be evaluated for every cell at the interface.\\

The total evaporation rate $\dot{m}$ is the sum over the liquid (evaporating) species:

\begin{equation}
\dot{m}=\sum_{i=0}^{Ns_{L}}\dot{m}_i
\label{sumoverm}
\end{equation}

When solving Equation \ref{alphasource} it is important to ensure a sharp interface. However, it is possible for $\alpha$ to reach negative values in certain cells, especially if $\dot{m}$ is high enough. In order to avoid this, $\dot{m}$ is multiplied by a proper function of $\alpha$, whose aim is to force $\dot{m}$ to be zero when $\alpha$ approaches zero. Differently from previous works \cite{guo2014review}, the evaporating flux is evaluated as follows:
\begin{equation}
\dot{m} = K f(\alpha)\sum_{i=0}^{Ns_{L}}\dot{m}_i
\label{alphafunction}
\end{equation}
where the constant $K$ is introduced to ensure the conservation of the integral mass flux $\int_{V}^{}\dot{m}dV$. It is simply recovered equaling Equations \ref{sumoverm} and \ref{alphafunction} after a volume integration:

\begin{equation}
K = \frac{\int_{V}^{}\dot{m}dV}{\int_{V}^{}f\left(\alpha\right)\dot{m}dV}
\end{equation}

We have chosen the function $f(\alpha)$ as follows:
\begin{equation}
f(\alpha) = \sqrt{\alpha\left(1-\alpha\right)}
\end{equation}

which is a redistribution of $\dot{m}$ over the interface, as shown in Figure \ref{massfluxdistribution}. This function goes to $0$ and to $1$ much faster, which helps to maintain a sharp interface after Equation \ref{alphasource} solution.

\subsection{Thermodynamic equilibrium}
The characterization of equilibrium thermodynamics at the phase boundary is crucial for a correct prediction of experimental data, especially for non-ideal mixtures and high pressure cases. In this work a detailed thermodynamics is included, based on Peng and Robinson equation of state \cite{peng1976new} for the evaluation of the compressibility factor $Z$ and the fugacity coefficients $\phi_i$.\\
The vapor-liquid equilibrium relation for a two-phase systems is \cite{smithintroduction}:

\begin{equation}
p_i^0(T)x_i\phi_i\left(T,p_i^0\right)e^{\int_{p_i^0}^{p}\frac{v_{L,i}}{RT}dp}= py_i\hat{\phi}_i(T,p,y_i)
\label{equilibrium}
\end{equation}

where $p_i^0(T)$ is the vapor pressure of species $i$, $\phi_i$ is the gas-phase fugacity coefficient for the pure species and $\hat{\phi}_i$ is the gas-phase mixture fugacity coefficient. The exponential term represents the  Poynting correction, while $x_i$ and $y_i$ are the liquid and gas mole fractions of species $i$.  The gaseous mole fraction $y_i$ is evaluated in a segregated approach, from equation \ref{equilibrium}:
\begin{equation}
y_i=\frac{p_i^0(T)\phi_i\left(T,p_i^0\right)e^{\int_{p_i^0}^{p}\frac{v_{L,i}}{RT}dp}}{p\hat{\phi}_i(T,p,y_i)}x_i
\label{equilibirumyi}
\end{equation}

This saturation mole fraction is assigned to the whole liquid phase and then advected in the gas phase through Equation \ref{speciesequation}, following the approach described in \cite{banerjee2004algorithm}.\\
The gas density is calculated by:
\begin{equation}
\rho_G=\frac{p M_w}{Z(T,p,y_i)RT}
\label{densitygas}
\end{equation}

where $Z$ is provided by the Peng-Robinson cubic equation of state.\\
The molar volume $v_L$ in the Poynting correction is simply evaluated as:
\begin{equation}
v_{L,i} = \frac{M_{w,i}}{\rho_L}
\end{equation}

\subsection{The problem of surface tension: introduction of the centripetal force field}
The droplet-gas interaction is characterized by high density ratios between the two phases and strong surface tension forces, particularly enhanced in small droplet sizes. Even though VOF methods are widely recognized for the great efficiencies in describing topologically complex interfaces, the accurate representation of surface tension is still a major problem in multiphase flows \cite{popinet2018numerical}. \\
The continuous representation of surface tension force $\textbf{f}_s$ by Brackbill \cite{brackbill1992continuum} requires an accurate evaluation of the interface curvature $\kappa$:
\begin{equation}
\textbf{f}_s=\sigma\kappa\nabla\alpha
\label{surfacetension}
\end{equation}
where curvature $\kappa$ is evaluated as:

\begin{equation}
\kappa = \nabla\cdot\left(\frac{\nabla \alpha}{|\nabla\alpha|}\right)
\end{equation}

The $\alpha$ gradient evaluation is numerically challenging, because of the discontinuous nature of the $\alpha$ function. A direct implementation of Equation \ref{surfacetension} creates a numerical imbalance between the surface tension force $\textbf{f}_s$ and the pressure gradient $\nabla p$ in the Navier-Stokes equation. As a consequence, unphysical flows and pressure spikes are generated around the interface, whose magnitude is proportional to $\sigma$ and inversely proportional to the droplet size. For the typical droplet diameters of the experiments (usually no more than 1-1.5 mm) these flows can strongly deform the interface and eventually break the droplet apart. Several methods have been proposed in this context, such as balanced force algorithms \cite{francois2006balanced} to reduce the numerical imbalance in surface tension force evaluation, height functions \cite{francois2006balanced} to compute the interface curvature directly from the $\alpha$ field and artificial viscosity methods \cite{denner2017artificial}. A sophisticated combination of quad/octree spatial discretization, height functions and balanced algorithms has been implemented by Popinet \cite{popinet2009accurate}  in Gerris solver.\\
Concerning the $\texttt{OpenFOAM}^{\textregistered}$ environment, Albadawi \cite{albadawi2013influence} proposed a coupled VOF-Level Set to mitigate the parasitic currents, while Raeini \cite{raeini2012modelling} introduced smoothing factors for $\kappa$ field combined with the introduction of a capillary pressure equation and a curtailing of the $\alpha$ marker. Even if these latter  methods significantly improve the modeling of surface tension forces, they have been validated on bubbles growth, stationary droplets and capillary flows, whereas no test case can be found on small droplets dynamics. Their application on the cases of our interest did not provide satisfactory results, mainly because the droplet shrinking due to the evaporation exponentially amplified the problem: the lower was the droplet size, the stronger were the parasitic currents around the interface, even if the initial stationary droplet was stabilized enough. 

In this work, in order to avoid these problems, parasitic currents have been suppressed directly from their source imposing zero surface tension. In presence of a gravity field surface tension is essential to hang the droplet on a fiber or a wire. Therefore, its elimination required the introduction of an additional force to compensate the droplet weight.\\
A small spherical fiber is initially introduced inside the liquid droplet. A centripetal force directed toward the center of the fiber have been imposed, defined as:

\begin{equation}
\textbf{f}_m=\rho\alpha\nabla\xi
\label{magneticfield}
\end{equation}

where $\xi$ is a scalar potential field defined as:

\begin{equation}
\xi=\xi_0\frac{D_f}{2r}
\label{magneticpotential}
\end{equation}

where $D_f$ is the fiber radius ($D_f=0.05$ mm in this work), $r$ is the distance from the fiber center and $\xi_0$ the intensity of the potential field. The $\alpha$ term in Equation \ref{magneticfield} forces $\textbf{f}_m$ to be applied to the sole  liquid phase and makes $\textbf{f}_m$ to be proportional to the droplet volume, as it is for gravity forces.\\
The result is a force field directed towards the droplet center and only applied to the liquid phase, similarly to what happens in a magnetic attraction of a ferrofluid. The Navier-Stokes equation becomes:

\begin{equation}
\frac{\partial\left(\rho\textbf{v}\right)}{\partial t} + \nabla\cdot\left(\rho\textbf{v}\otimes\textbf{v}\right)= \nabla\cdot\mu\left(\nabla\textbf{v}+\nabla\textbf{v}^T\right)-\nabla p_{rgh} -\textbf{g}\cdot\textbf{r}\nabla\rho+\textbf{f}_m
\label{NavierStokesModified}
\end{equation}

It is important to notice that the combination of a geometrical advection (i.e. \texttt{isoAdvector}) and the absence of surface tension $\sigma$ allows to track the interface maintaining a very high resolution, without any additional correction (such as filtering kernels, smoothing factors etc.) usually needed in the VOF methodology because of the strong gradients involved. The detailed numerical description and the implementation of the centripetal force $\textbf{f}_m$ are presented in the next section, as well as a sensitivity analysis on the value $\xi_0$ to be used in the simulations. The effect of the initial droplet shape on the evaporation will be also discussed.

\subsection{Fluid properties}
The evaluation of transport and thermodynamics properties is based on the OpenSMOKE++ library \cite{cuoci2015opensmoke++}. Gas properties (diffusion coefficients, thermal conductivity, heat capacities and viscosity) are based on the kinetic theory of gases, while liquid properties (vapor pressure, density, conductivity, heat capacity, viscosity and vaporization heat) are evaluated based on the correlations available in the Yaws database \cite{yaws2015yaws}.\\
Mixture properties to be used in the governing equations can be then computed. For a generic property $\chi$:
\begin{equation}
\chi=\chi_L\alpha + \chi_G\left(1-\alpha\right)
\end{equation}

with $\chi = \rho, \mu, C_{p}, k, \mathcal{D}_i$.

\section{Numerical methodology}

\begin{figure}
	\centering
	{\includegraphics[width=.99\textwidth,height=0.55\textheight]{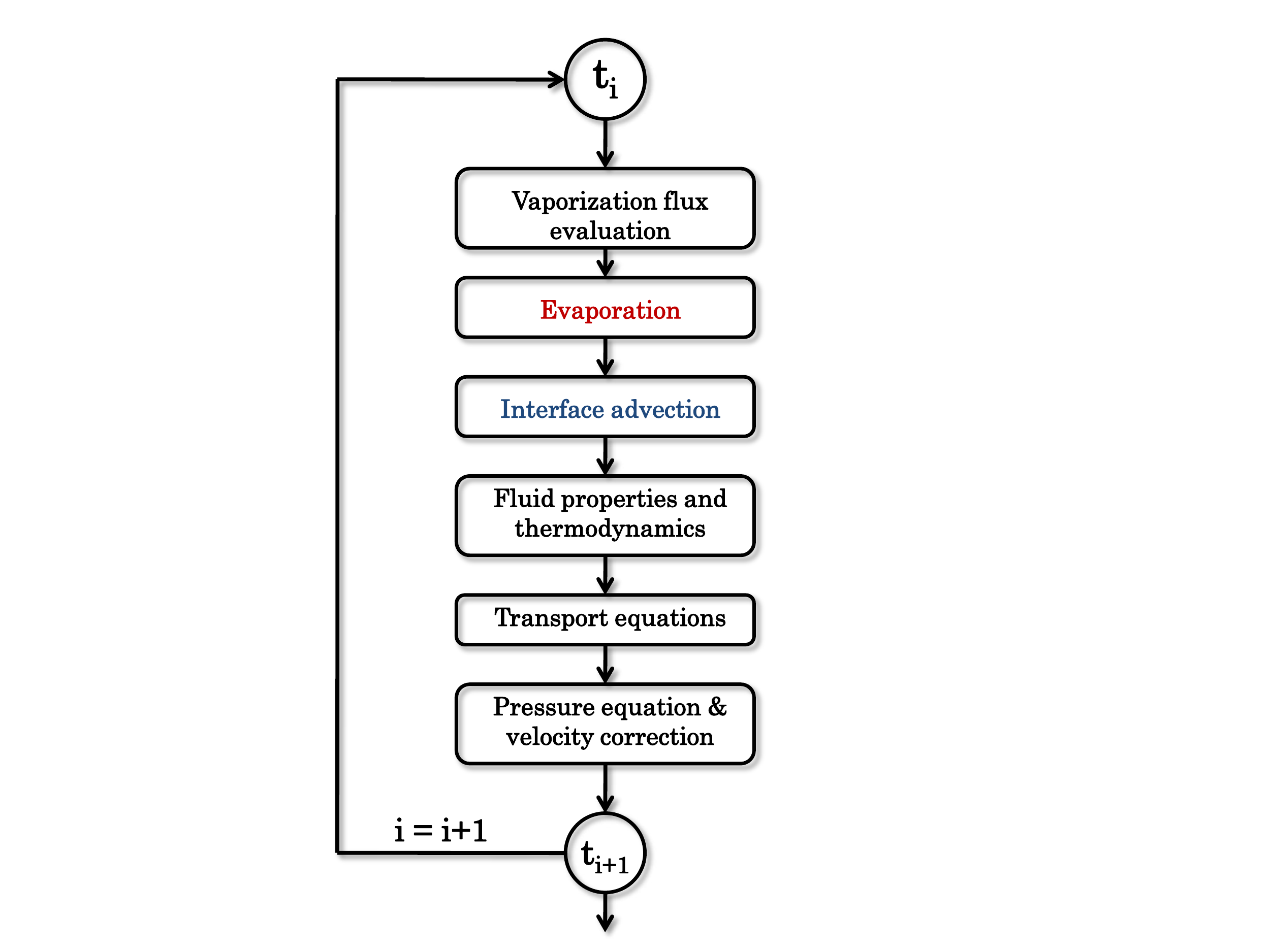}}
	
	\caption{Steps of the numerical algorithm used in  \texttt{DropletSMOKE++}}
	\label{Scheme}
\end{figure}

\subsection{Description of the DropletSMOKE++ solver}
The \texttt{DropletSMOKE++} code is embedded within the $\texttt{OpenFOAM}^{\textregistered}$   framework which allows to manage the spatial discretization of the governing equations on arbitrary geometries. The PIMPLE algorithm \cite{greenshields2015openfoam}, a combination between SIMPLE  (Semi-Implicit Method for Pressure-Linked Equations) and PISO  (Pressure Implicit Splitting of Operators), is used to manage the pressure-velocity coupling, computing a velocity field which satisfies both momentum and continuity equation through an iterative procedure.\\
For every time step, \texttt{DropletSMOKE++} encompasses the following steps in a segregated approach (Figure \ref{Scheme}):

\begin{enumerate}
	\item Evaluation of the evaporation flux (Equation \ref{massflux})
	\item Liquid evaporation or expansion (Equation \ref{alphasource})
	\item Interface advection  (Equation \ref{advection})
	\item Update of gas-liquid properties and thermodynamics
	\item Momentum, temperature and species equation (Equations \ref{NavierStokesModified}, \ref{Teqn}, \ref{speciesequation})
	\item Iterative calculation of the pressure field (Equation \ref{pequation}) and velocity correction
\end{enumerate}

The time discretization follows an implicit Euler method, where the time step is automatically computed based on a threshold Courant number. As reported by Brackbill et al. \cite{brackbill1992continuum}, the VOF time step constraint states that $ \Delta t <\left[\frac{\rho \left(\Delta x\right)^3}{2\pi\sigma}\right]^{\frac{1}{2}}$, which depends on surface tension. It is worth noticing that the suppression of surface tension introduced in this work alleviates this constraint, allowing larger time steps to be used.\\
Finally,   Gauss linear upwind scheme is used for spatial discretization of convective terms, while an orthogonal correction is adopted for Laplacian terms.

\begin{table}
	\centering
	
	\begin{tabular}{lllll}
		\toprule
		
		Boundary & $\alpha$ & $\textbf{v}$ & T, $\omega_i$ & $p_{rgh}$ \\
		\midrule
		inlet (natural convection)   & $\alpha=0$ &  inletOutlet & inletOutlet & $p_{rgh}=p_{ext}$  \\
		inlet (forced convection)   & $\alpha=0$ &   $\textbf{v}=\textbf{v}_{ext}$ & $T,\omega_i=T_{ext},\omega_{i, ext}$ & $\nabla p_{rgh} =0$  \\
		outerWall & $\alpha=0$ & noSlip & $\nabla T,\nabla\omega =0$ & $\nabla p_{rgh} =0$  \\
		outlet   & $\alpha=0$ &  inletOutlet & inletOutlet & $p_{rgh}=p_{ext}$  \\
		sphere  & $\nabla \alpha =0$ & noSlip & $\nabla T,\omega =0$ & $\nabla p_{rgh} =0$  \\
		\bottomrule
		
	\end{tabular}
	\caption{Boundary conditions for $\alpha$, velocity, temperature, species $i$ and pressure field, to be compared with Figure \ref{mesh}.}
	\label{tableboundary}
\end{table}

\subsubsection{Computational mesh}

The computational mesh used for all the numerical cases proposed in this work has been built with the commercial CFD code $\texttt{Ansys FLUENT}^{\textregistered}$  v17.2. The geometry is 2D and axisymmetric, representing a slice of a cylinder having a radius  $W=5$ mm and a height $H=30$ mm. At 10 mm from the base a spherical fiber (diameter $D=0.1$ mm) is introduced, which is needed as a source for the  force field $\textbf{f}_m$ previously described. The liquid droplet will be place around this small spherical fiber.
\begin{figure}
	\centering
	{\includegraphics[width=.98\textwidth,height=0.41\textheight]{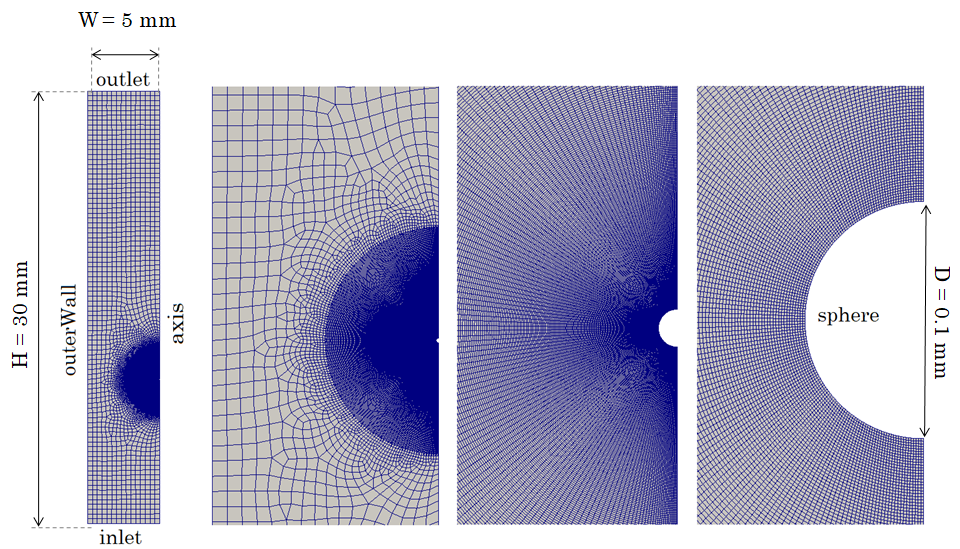}}
	
	\caption{Computational mesh represented in increasing levels of detail, from left to right. The boundary names are also shown, to be compared with Table \ref{tableboundary}.}
	\label{mesh}
\end{figure}
The geometry is meshed in a non-structured way (Figure \ref{mesh}), with a particular attention to the gas-liquid boundary: the mesh is structured and orthogonal on the cylinder boundaries. The region including the liquid droplet and the gas in its proximity is meshed with a concentric pattern, getting finer while approaching the inner fiber. Therefore, the presence of a spherical boundary is also necessary for the mesh construction.\\
In this work, the mesh is composed by 70,000 cells, with a maximum mesh non-orthogonality equal to 56.6 and a maximum Skewness of 1.3. It has been verified that the numerical solution  does not change doubling the number of cells, proving a complete mesh independence.

\begin{figure}
	\centering
	\subfloat[]
	{\includegraphics[width=.32\textwidth,height=0.23\textheight]{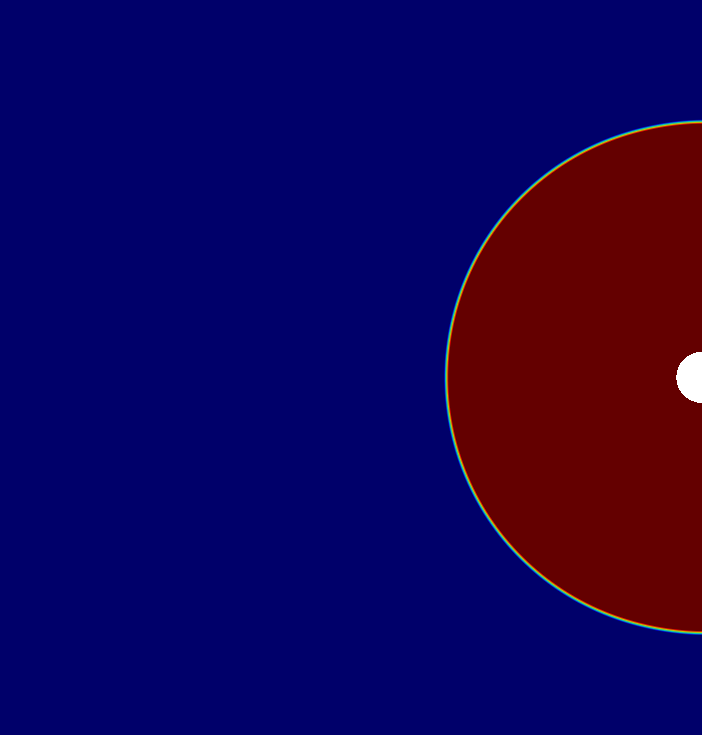}}~
	\subfloat[]
	{\includegraphics[width=.31\textwidth,height=0.28\textheight]{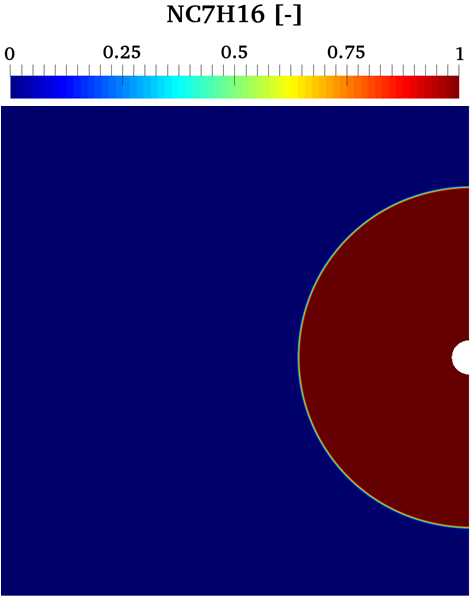}}~
	\subfloat[]
	{\includegraphics[width=.31\textwidth,height=0.28\textheight]{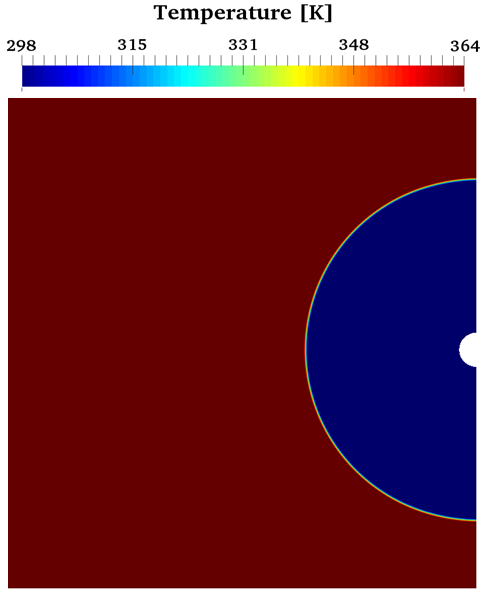}}~	
	
	\caption{Example of initial conditions in droplet evaporation simulations. Liquid volume fraction ($\alpha$ field) in Figure (a), n-heptane mass fraction in Figure (b) and temperature field in Figure (c).}
	\label{initialconditions}
\end{figure}

\subsection{Boundary conditions}
The definition of proper boundary conditions is fundamental to reach reliable results. The geometry has four boundaries, shown in Figure \ref{mesh}: 
\begin{enumerate}
	\item outerWall: external lateral surface of the cylinder
	\item inlet: the bottom base surface of the cylinder
	\item outlet: the upper base of the cylinder
	\item sphere: the spherical fiber surface
\end{enumerate}

The boundary conditions are summarized in Table \ref{tableboundary}. The $inlet$ boundary conditions change depending on the case (natural or forced convection). The $inletOutlet$ condition is basically a zero gradient condition, that switches to a fixed value condition if the velocity vector next to the boundary points inside the domain (backward flow). The $noSlip$ condition imposes a zero relative velocity between the fluid and the boundary. It is worth noticing that the separation of the hydrostatic pressure, allows a simple definition of boundary conditions for $p_{rgh}$.\\

\subsection{Initial conditions}
Referring to the mesh presented in Figure \ref{mesh}, the liquid droplet must be positioned around the spherical fiber. In Figure \ref{initialconditions} the initial conditions for the evaporation of a cold n-heptane droplet in a hot environment are presented as an example.

\subsection{Code parallelization}

\begin{table}
	\centering
	
	\begin{tabular}{lllllll}
		\toprule
		
		Case ~~& Species~~ & $D_0$ ~~& $T_L$ ~~& $T_{G}$ ~~& p ~~&  Results \\
		~~&  ~~& [mm] ~~& [K]~~ & [K] ~~& [atm] ~~&  [Figure \ref{1Dcomparison}]  \\
		\midrule
		1 & n-Decane  & 0.5 & 435 & 435 & 1 & (a, b, c)   \\
		2 & n-Heptane & 0.5 & 360 & 360 & 1 & (d, e, f) \\ 
		3 & n-Heptane & 1.03   & 300 & 364 & 20& (g, h, i)  \\
		4 & Water   & 0.7 & 360 & 360 & 1 & (j, k, l)  \\
		\bottomrule
		
	\end{tabular}
	\caption{Numerical cases for the comparison with 1D model \cite{cuoci2005autoignition}. The related plots are reported in Figure \ref{1Dcomparison}.}
	\label{table1Dcases}
\end{table}

The numerical cases presented in this work are run on a multi-processor (Intel Xeon X5675, 3.07 GHz) machine. The Domain Decomposition Method is used to split the mesh into sub-domains and allocate them to separate processors. The \texttt{DropletSMOKE++} code can then run in parallel mode, with communication between processors with MPI communication protocol, allowing a significant reduction of the computational time. The optimal number of processors is found to be around 12, which corresponds to an average of 6000 cells for each processor and a speed-up performance $\sim$ 6 times higher with respect to the serial mode.\\
The simulations required from 2 to 20 hours, depending on the operating conditions.

%		\begin{figure}
%			\centering
%			%\subfloat[]
%			{\includegraphics[width=.35\textwidth,height=0.2\textheight]{Immagini/processors}}%\quad
%		%	\subfloat[]
%		%	{\includegraphics[width=.4\textwidth,height=0.2\textheight]{Immagini/radialV}}~			
%			
%			\caption{Speed up of the code in function of the number of processor used. The speed up is here defined as the ratio between the execution time in serial and the execution time in parallel.}
%			\label{processors}
%		\end{figure}

\section{Validation of microgravity cases}
In this section \texttt{DropletSMOKE++} is validated in microgravity conditions (imposing $\textbf{g}=\textbf{0}$), against the numerical results of the solver developed by Cuoci et al. \cite{cuoci2005autoignition}. This latter code has been validated in a wide range of operating conditions over the last 10 years \cite{stagni2018numerical, cuoci2017flame} and represents a reliable reference for comparison. This validation is done in order to assess the validity of the equations (in terms of diameter decay, temperature profiles and vaporization velocity) before the activation of the gravity field. Obviously, the centripetal force $\textbf{f}_m = 0$ in these cases.

\begin{figure}
	\centering
	\subfloat[]
	{\includegraphics[width=.34\textwidth,height=0.19\textheight]{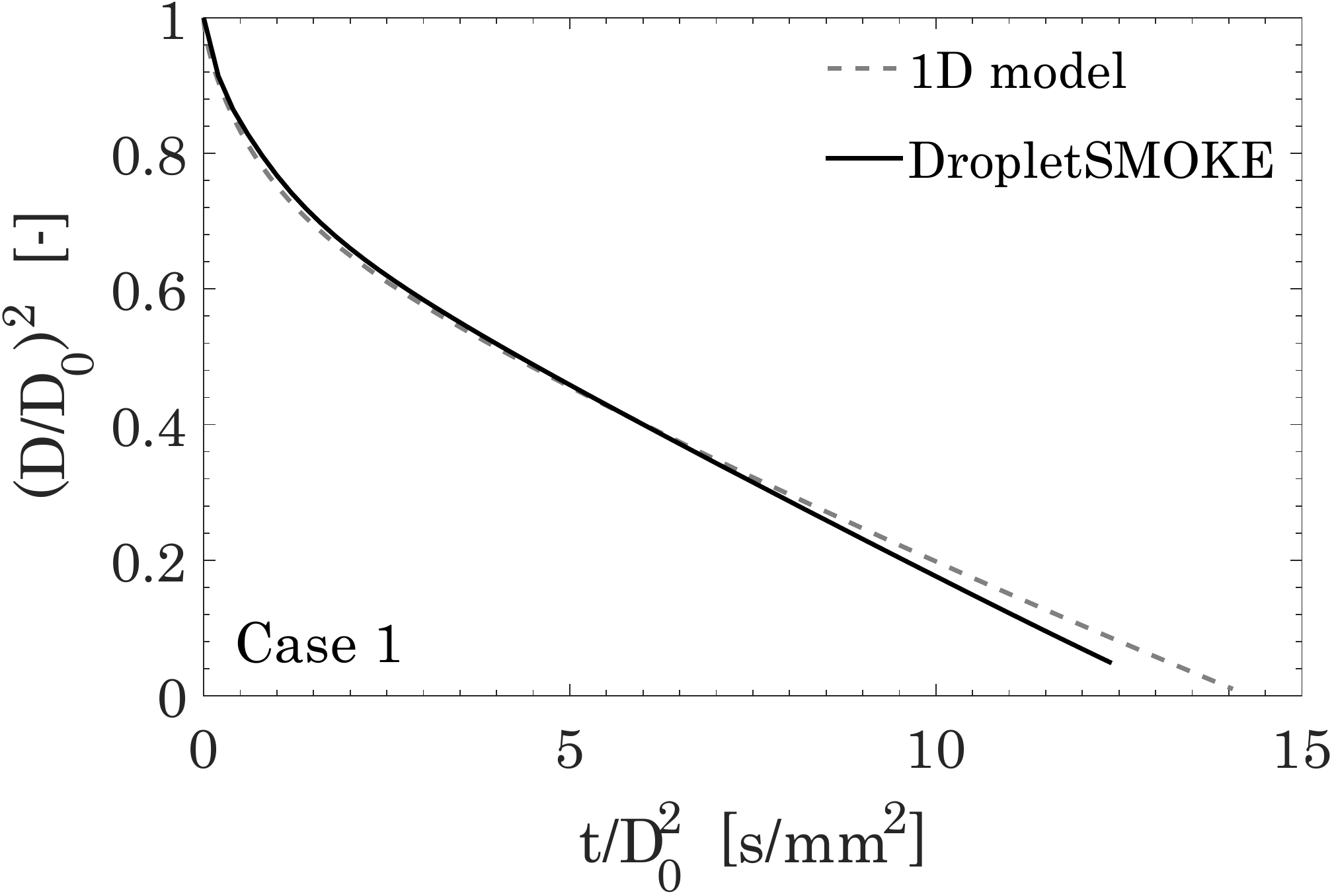}}~
	\subfloat[]
	{\includegraphics[width=.34\textwidth,height=0.19\textheight]{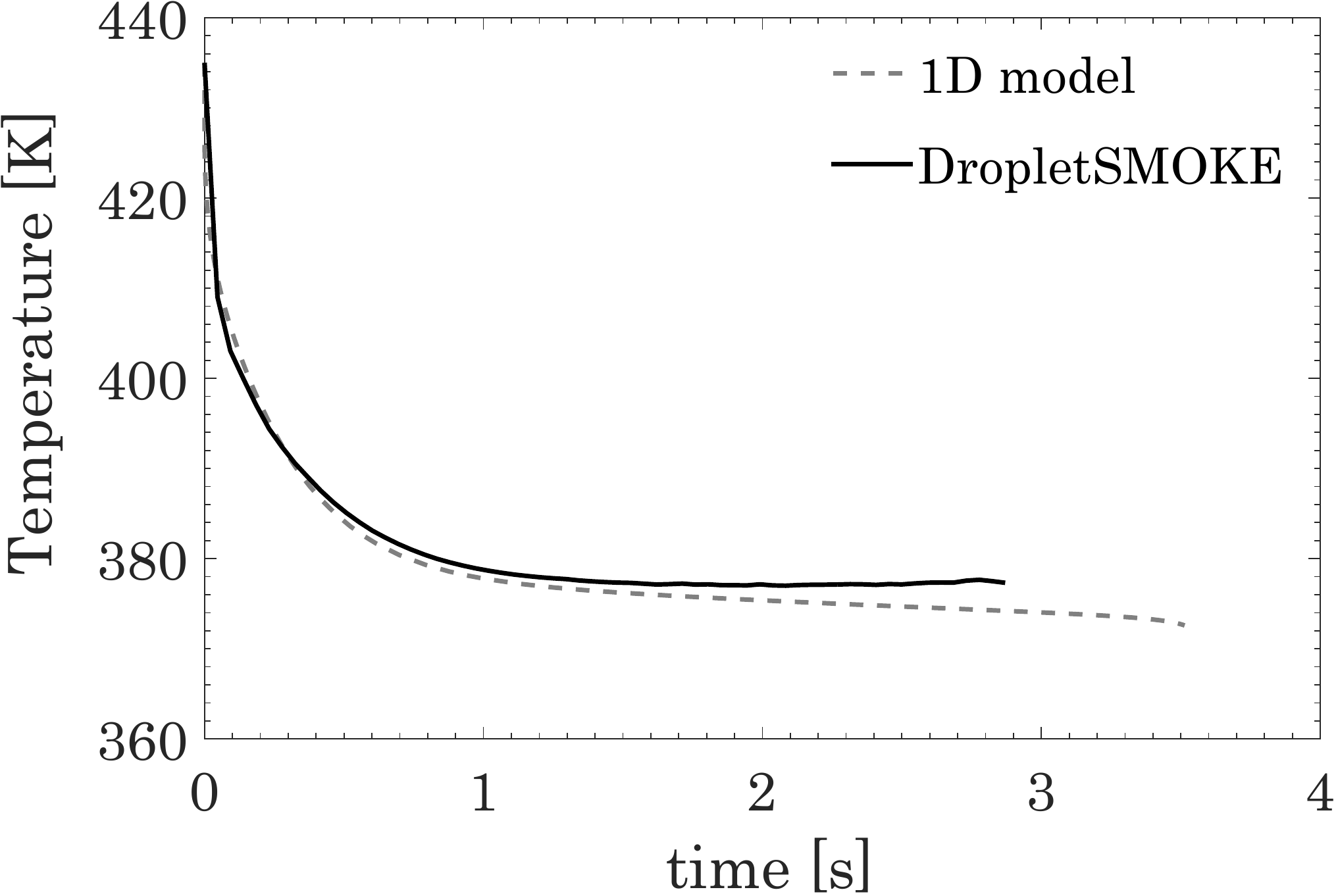}}~
	\subfloat[]
	{\includegraphics[width=.34\textwidth,height=0.19\textheight]{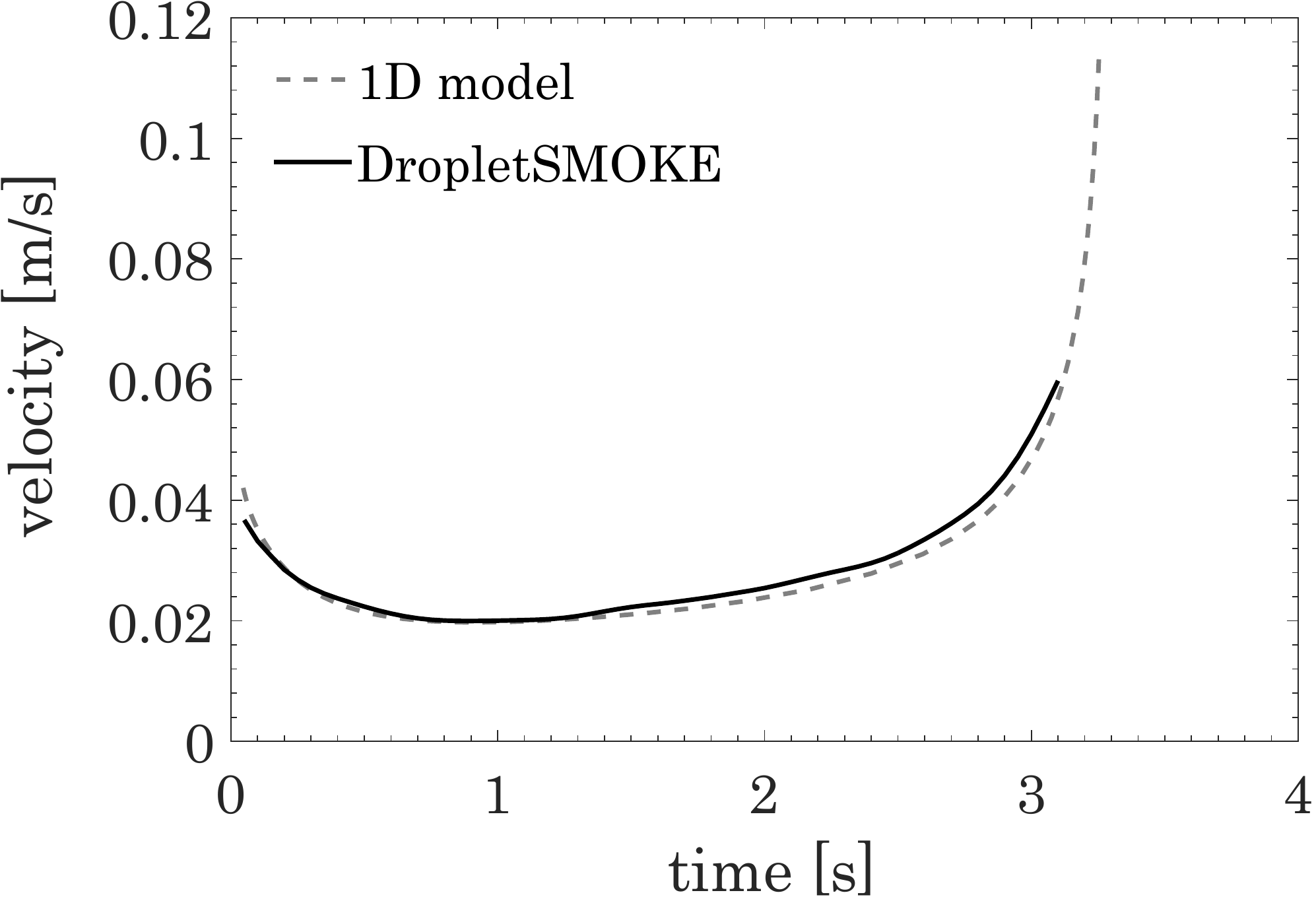}}\\
	\subfloat[]
	{\includegraphics[width=.34\textwidth,height=0.19\textheight]{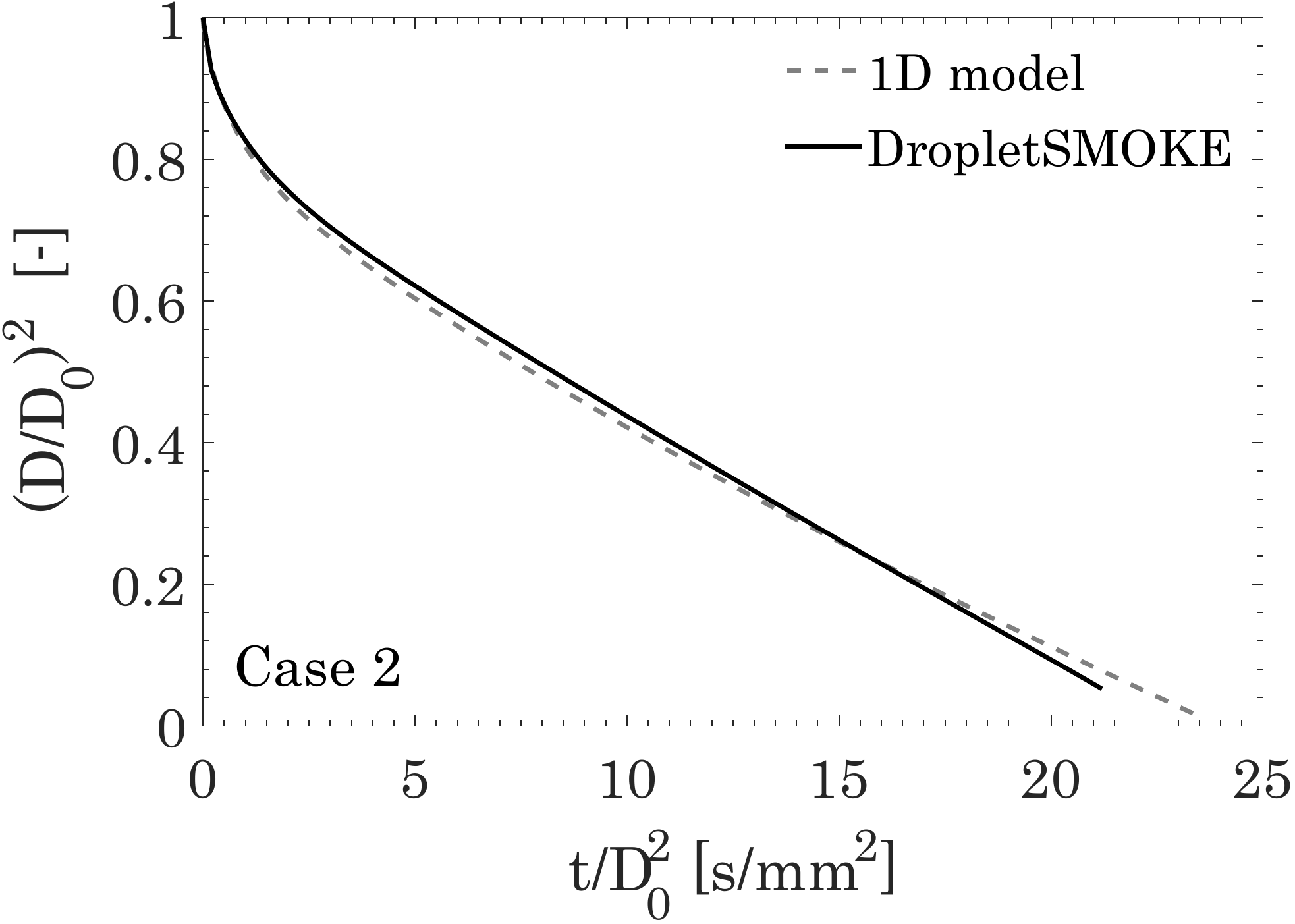}}~
	\subfloat[]
	{\includegraphics[width=.34\textwidth,height=0.19\textheight]{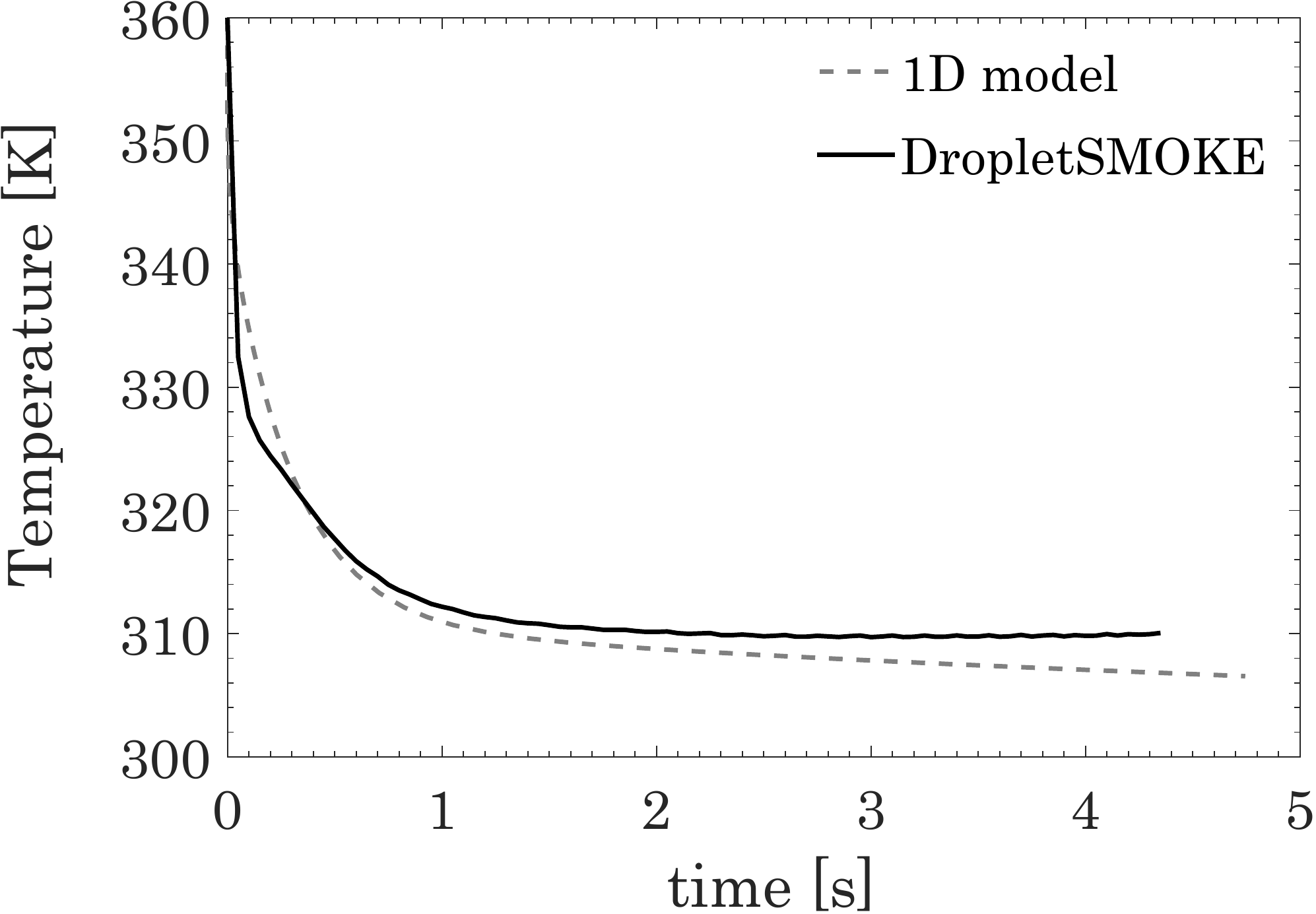}}~
	\subfloat[]
	{\includegraphics[width=.34\textwidth,height=0.19\textheight]{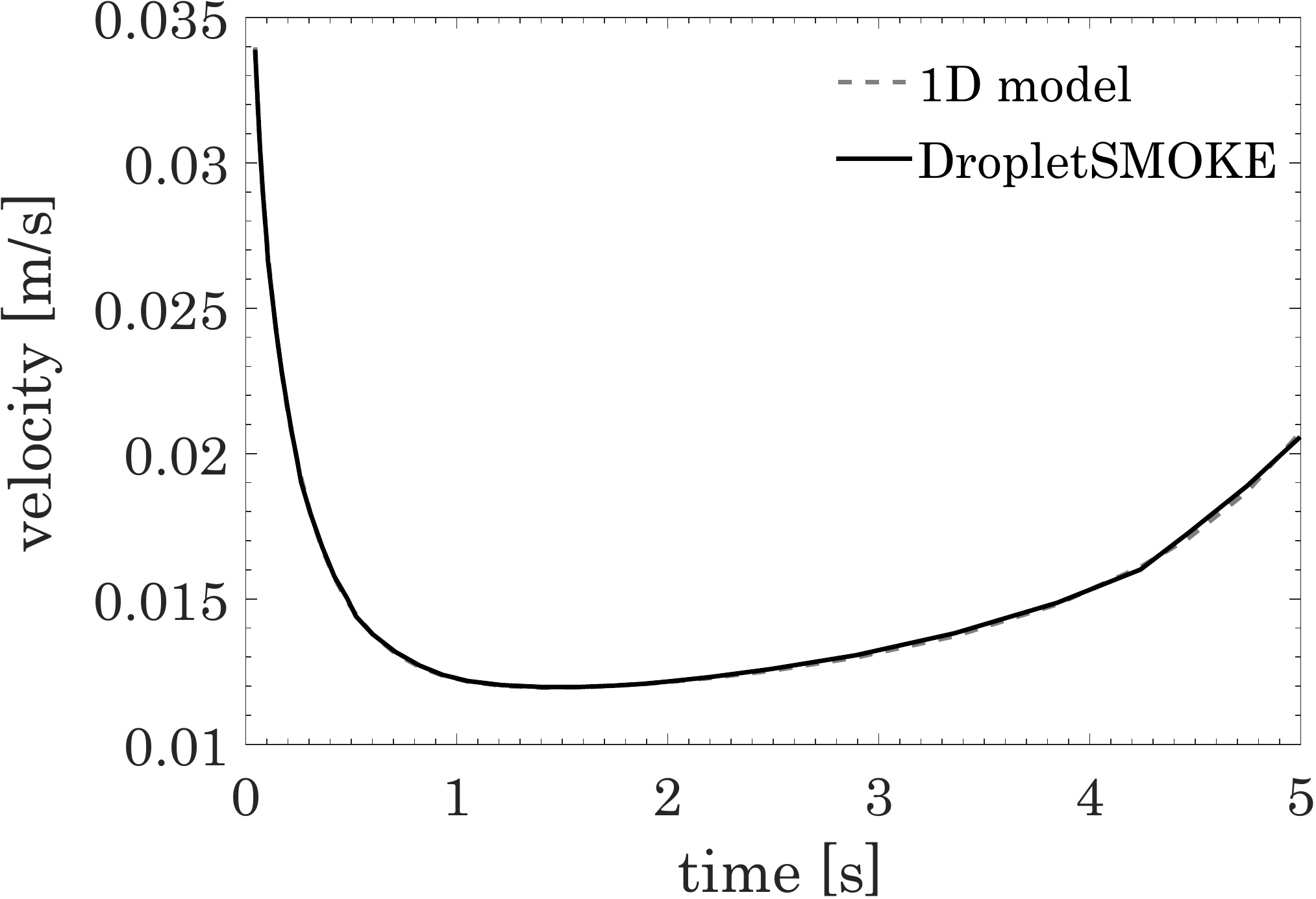}}\\
	\subfloat[]
	{\includegraphics[width=.34\textwidth,height=0.19\textheight]{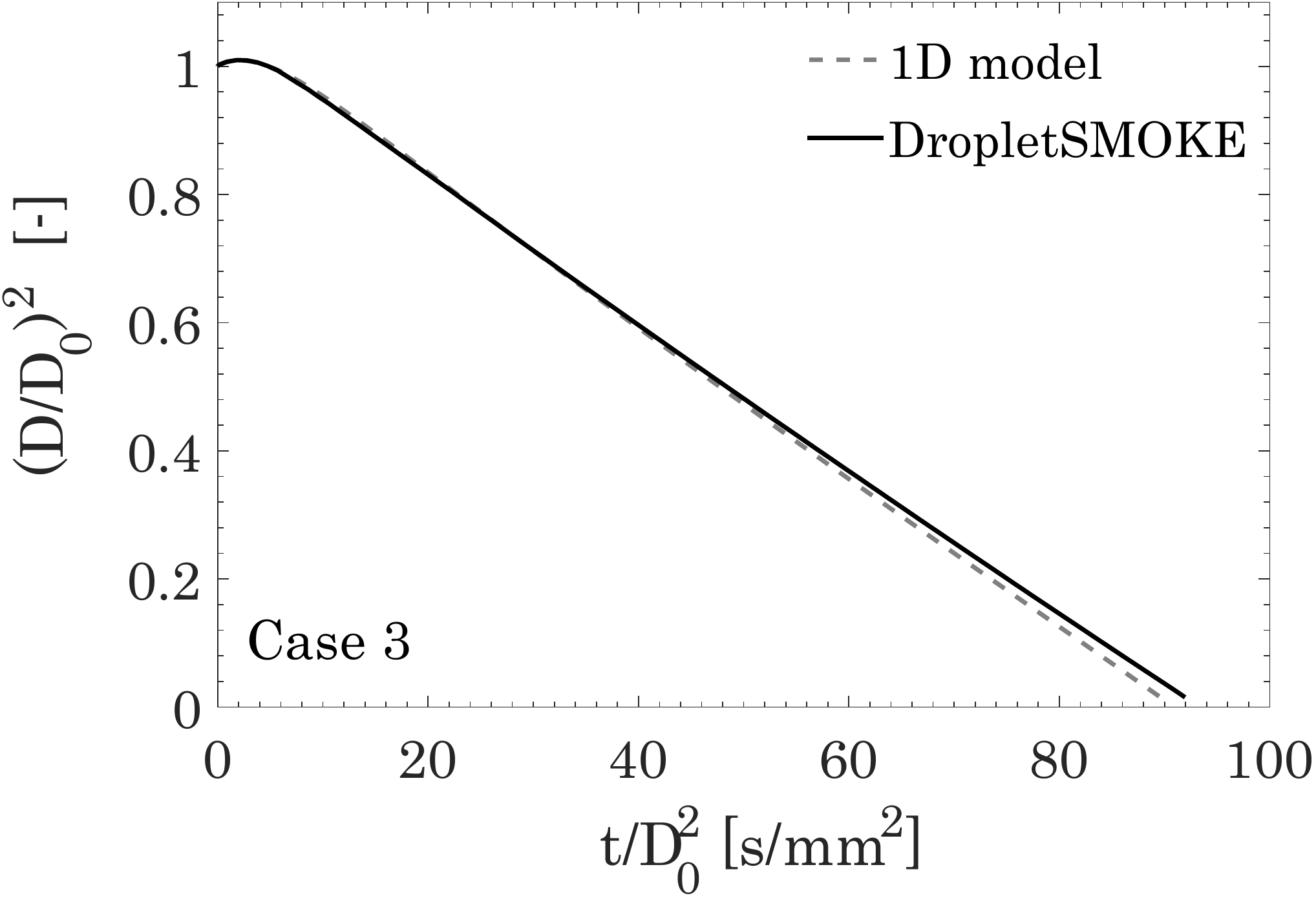}}~
	\subfloat[]
	{\includegraphics[width=.34\textwidth,height=0.19\textheight]{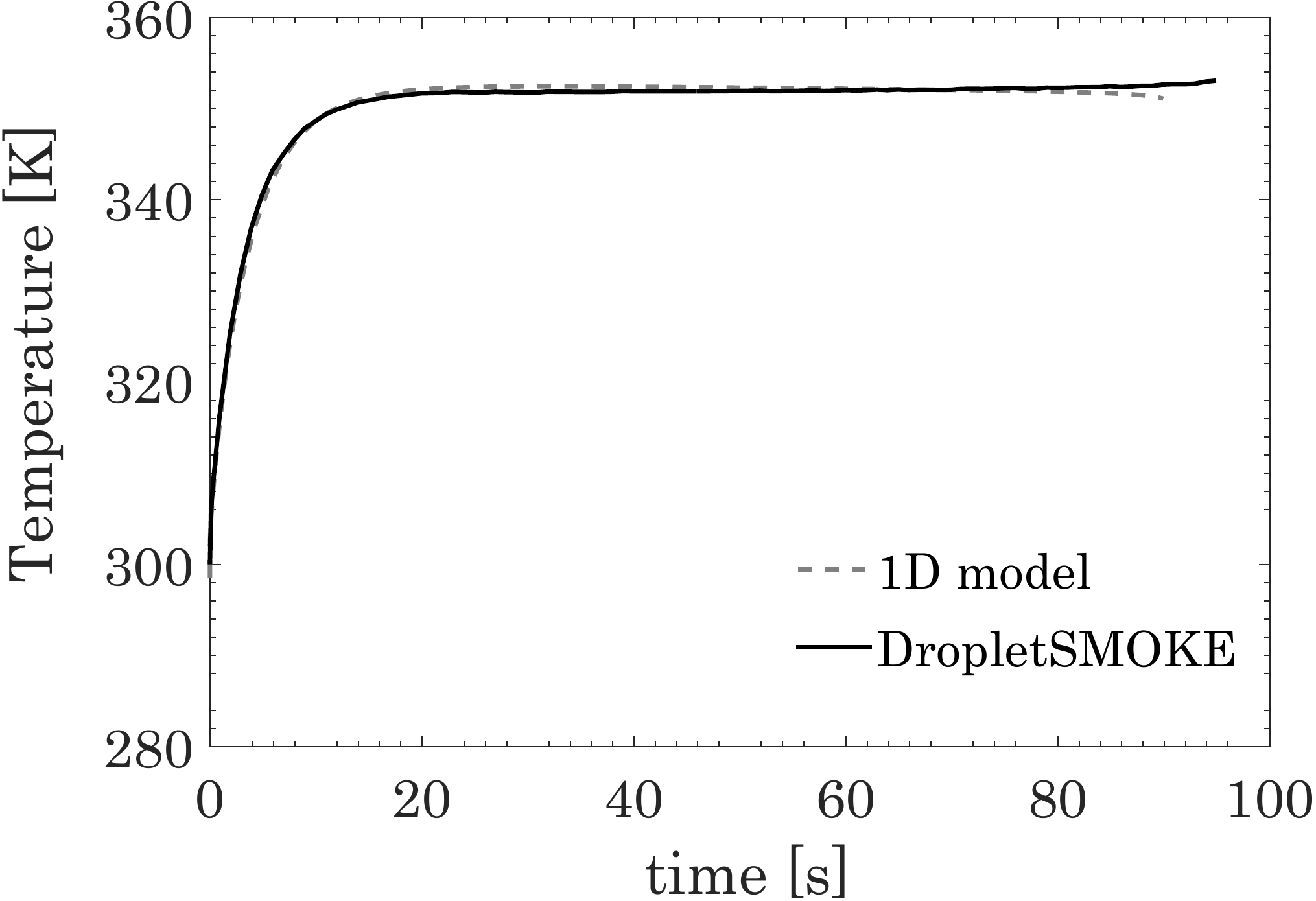}}~
	\subfloat[]
	{\includegraphics[width=.34\textwidth,height=0.2\textheight]{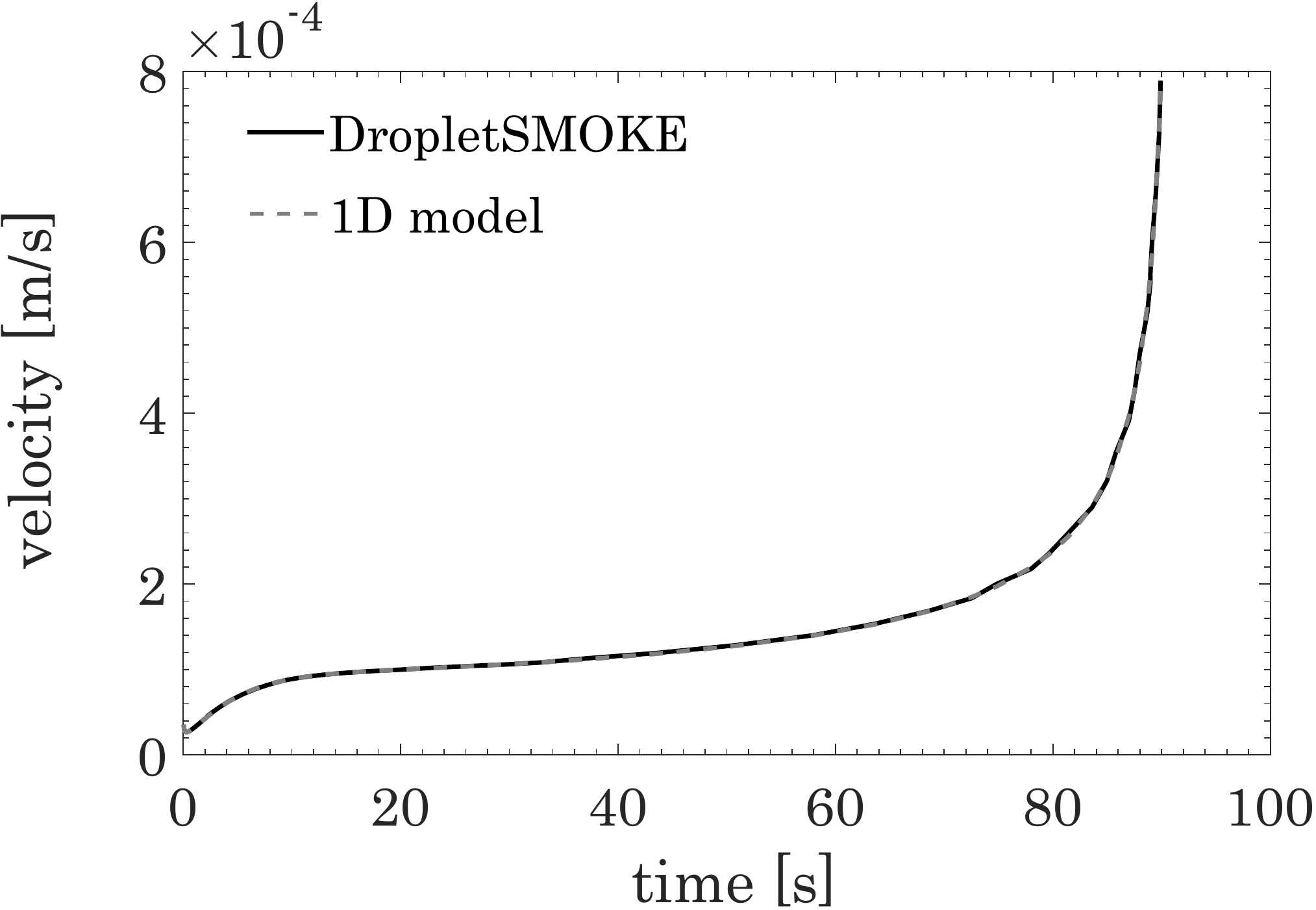}}\\	
	\subfloat[]
	{\includegraphics[width=.34\textwidth,height=0.19\textheight]{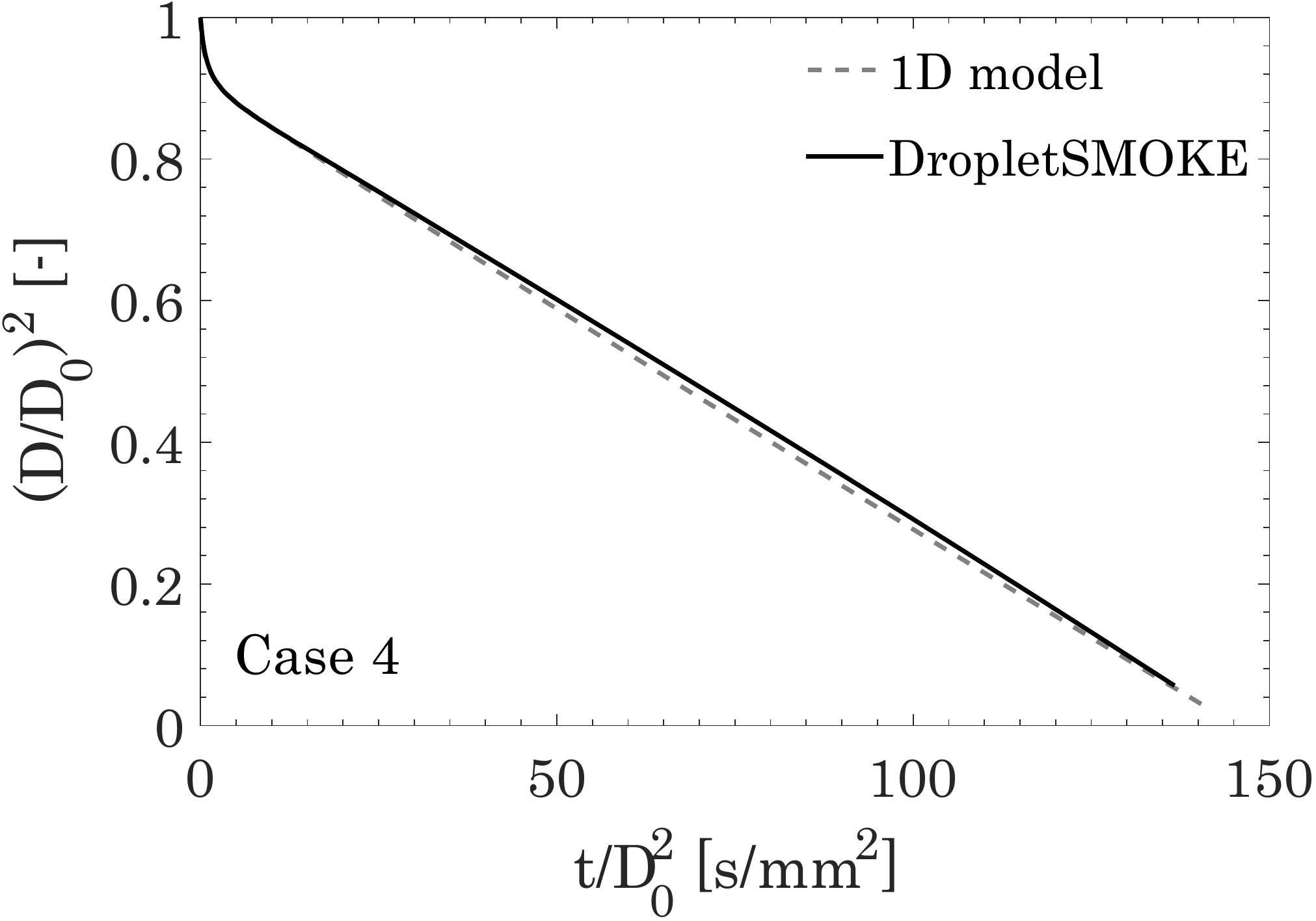}}~
	\subfloat[]
	{\includegraphics[width=.34\textwidth,height=0.19\textheight]{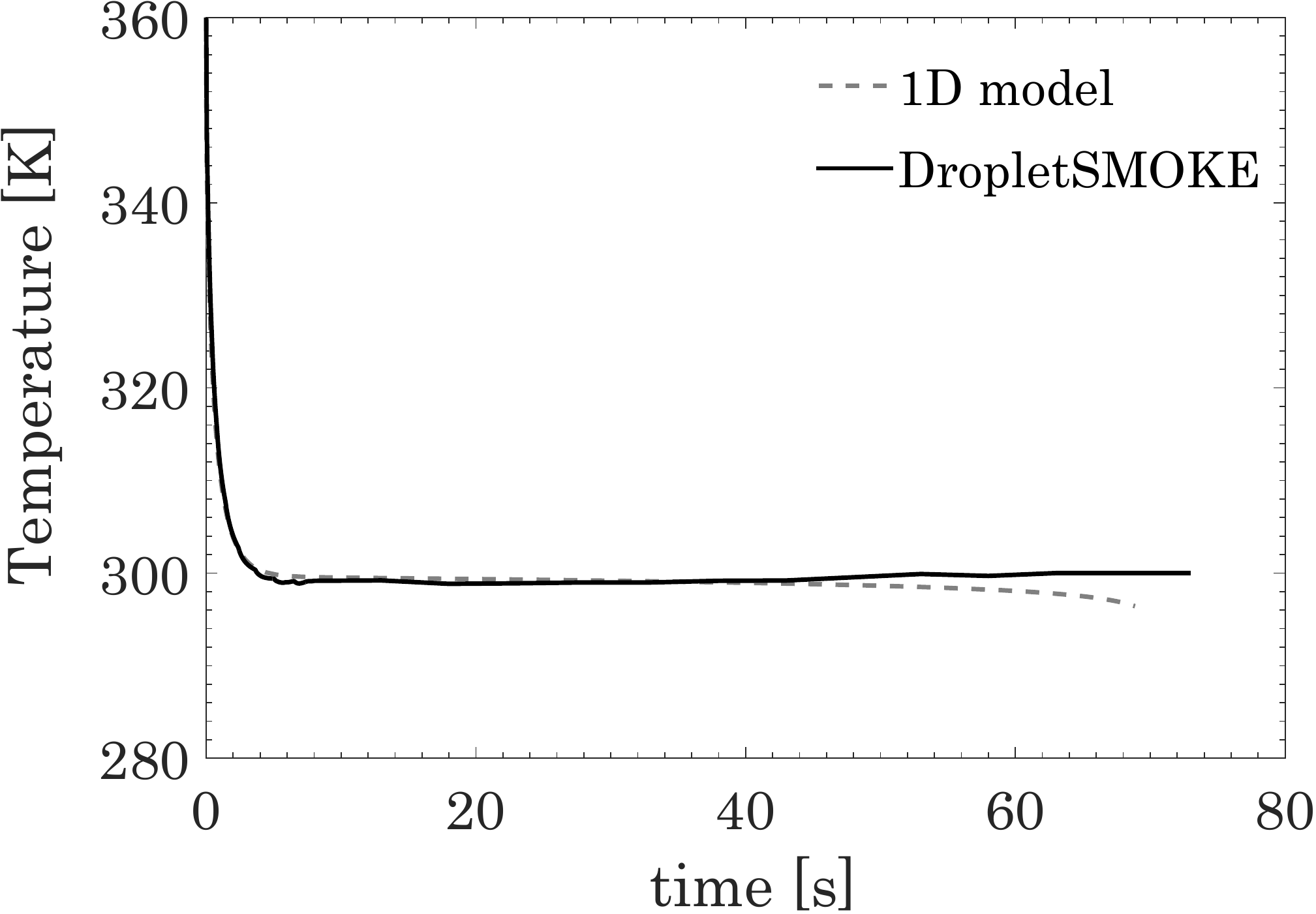}}~
	\subfloat[]
	{\includegraphics[width=.34\textwidth,height=0.19\textheight]{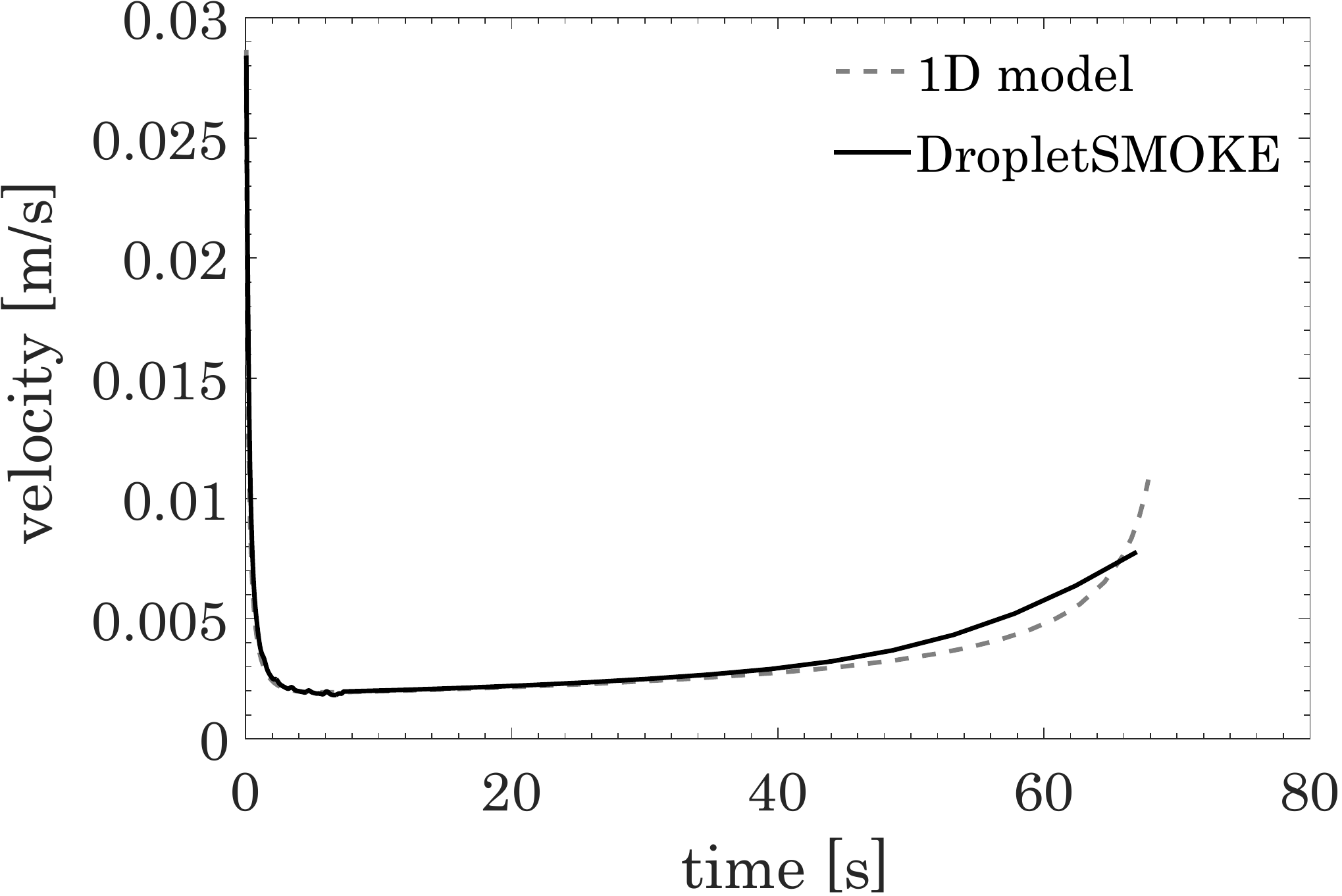}}							
	\caption{Numerical comparison of the 1D model of Cuoci et al. \cite{cuoci2005autoignition} and \texttt{DropletSMOKE++} (Table \ref{table1Dcases}).}
	\label{1Dcomparison}
\end{figure}

The 4 numerical cases analyzed are listed in Table \ref{table1Dcases}. Three liquids have been chosen (n-decane, n-heptane and water) at different temperature and pressure conditions. Numerical results are summarized in Figure \ref{1Dcomparison}.\\
For each case (lines) three plots are reported (columns): the temporal evolution of the droplet size, the liquid surface temperature and the vaporization velocity. More precisely, this latter is the convective velocity of the gas generated at the phase boundary.

\subsection{Results and discussion}
For all cases the squared diameter, after the initial transient period, reduces linearly with time, as the $d^2$ law states. A wet-bulb temperature is reached by the interface and maintained along the whole evaporation process, due to the  equilibrium between the incoming heat flux and the outcoming vaporization enthalpy. The transient period depends on the initial conditions. In cases 1-2-4 (initially isothermal) the interface is immediately cooled, diminishing the vaporization flux.  In case 3 the liquid is firstly heated (reducing $\rho_L$ and increasing the droplet size) providing an increasing vaporization flux.\\
The agreement between the models can be considered satisfactory, even if small differences exist. In particular the surface temperature tends to be slightly over-predicted, especially in the final steps of the evaporation. This is probably due to the spherical boundary of the fiber in the \texttt{DropletSMOKE++} simulations (Figure \ref{mesh}), which is absent in the 1D model. The presence of a small sphere inside the liquid provides a lower droplet mass if compared to a 1D droplet with the same initial diameter. As can be seen in Figure \ref{1Dcomparison} its effect is completely negligible for most of the simulation (the "missing" mass is less than $0.01\%$ of the initial total mass). While the droplet surface approaches the fiber, this effect becomes important: the incoming heating flux is now distributed over a lower liquid mass, providing a slightly higher equilibrium temperature. This will not be a problem when modeling evaporation under convection, since fibers are always present in experiments  to hang liquid droplets.

\section{Implementation of the centripetal force $\textbf{f}_m$}

\begin{figure}
	\centering
	%	\subfloat[]
	%	{\includegraphics[width=.32\textwidth,height=0.33\textheight]{Immagini/magnetic.png}}~
	%	\subfloat[]
	%	{\includegraphics[width=.32\textwidth,height=0.264\textheight]{Immagini/force2.png}}~
	%	\subfloat[]
	%	{\includegraphics[width=.32\textwidth,height=0.33\textheight]{Immagini/pressuredifference.png}}~	
	{\includegraphics[width=.96\textwidth,height=0.31\textheight]{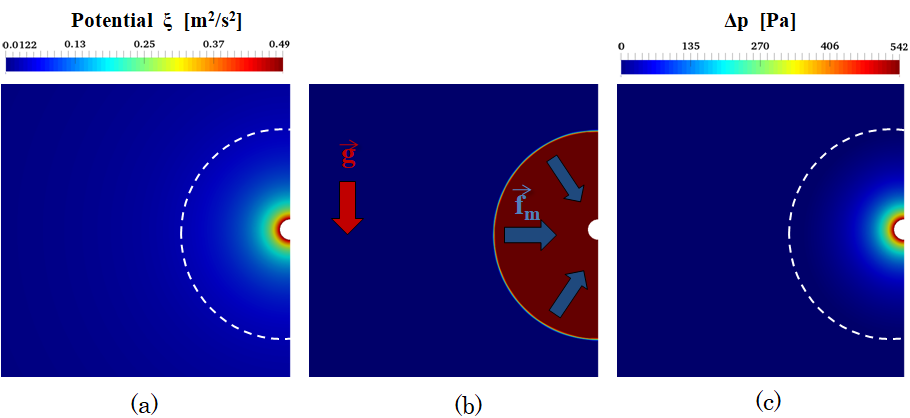}}~
	
	\caption{Potential field $\xi$ (Figure a) around fiber, evaluated with $\xi_0=1$ $m^2/s^2$. In Figure (b) the droplet is placed around the fiber and sustained against gravity $\textbf{g}$ by the force $\textbf{f}_m=\rho\alpha\nabla\xi$. The $\Delta p = p-p_{ext}$ field is presented in Figure (c). The white dashed lines in (a,c) represent the droplet surface.}
	\label{magneticforce}
\end{figure}

In order to sustain the droplet against the gravity field, a centripetal force $\textbf{f}_m$ (Equation \ref{magneticfield}) is introduced, based on a potential $\xi$ defined by Equation \ref{magneticpotential}, represented in Figure \ref{magneticforce} (a) and previously described in the mathematical model. The liquid droplet is then fixed around the fiber (Figure \ref{magneticforce} b). Inside the droplet a small pressure gradient is established (from $\sim$100 to $\sim$500 Pa, depending on the value $\xi_0$), similarly to what happens for a liquid column subjected to a gravity field (Figure \ref{magneticforce} c). This is actually consistent with surface tension forces, which provide an internal pressure higher than the external one (capillary pressure). \\

\subsection{Control of the droplet shape}
A droplet suspended under the effect of a gravity field reaches a steady-state shape imposed by the equilibrium between the liquid weight, the pressure gradient and the surface tension forces: 
\begin{equation}
-\nabla p + \sigma\kappa\nabla\alpha + \rho\textbf{g} = \textbf{0}
\end{equation}

In this work surface tension in not considered, and the centripetal force $\textbf{f}_m$ governs the equilibrium shape of the droplet:
\begin{equation}
-\nabla p + \textbf{f}_m + \rho\textbf{g} = \textbf{0}
\end{equation}

\begin{figure}
	\centering
	\subfloat[]
	{\includegraphics[width=.19\textwidth,height=0.145\textheight]{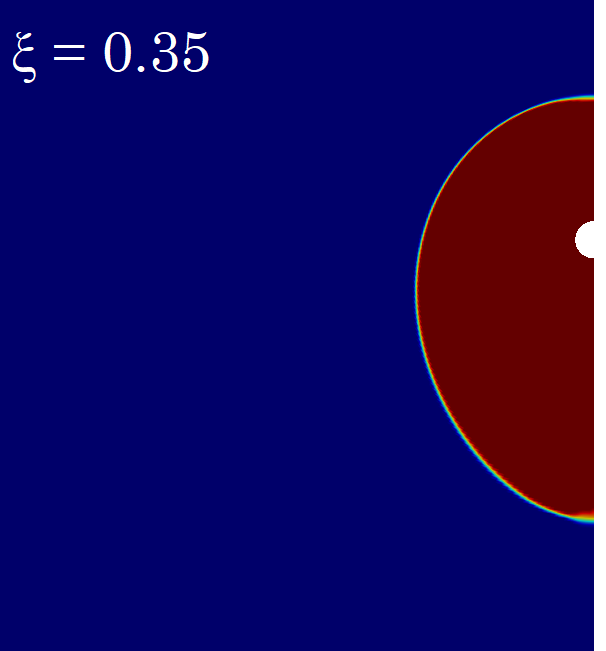}}~
	\subfloat[]
	{\includegraphics[width=.19\textwidth,height=0.145\textheight]{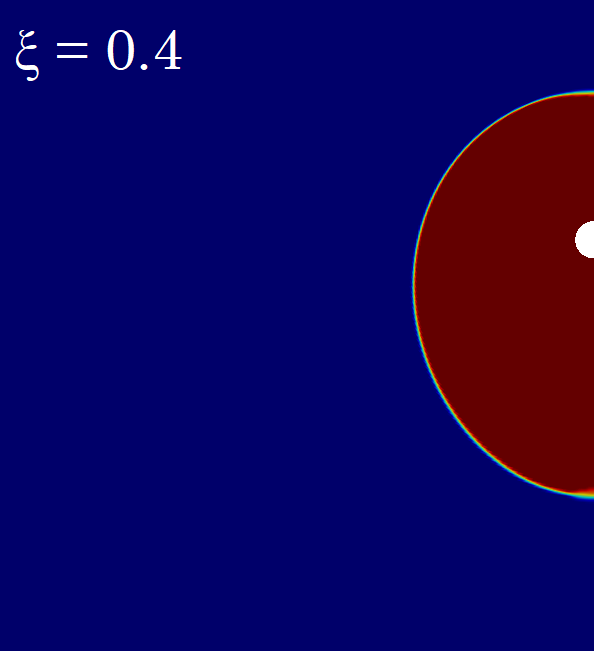}}~
	\subfloat[]
	{\includegraphics[width=.19\textwidth,height=0.145\textheight]{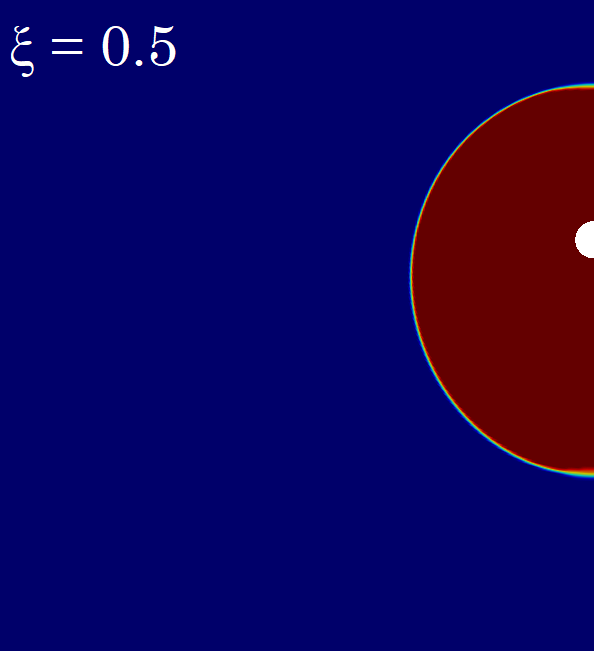}}~	
	\subfloat[]
	{\includegraphics[width=.19\textwidth,height=0.145\textheight]{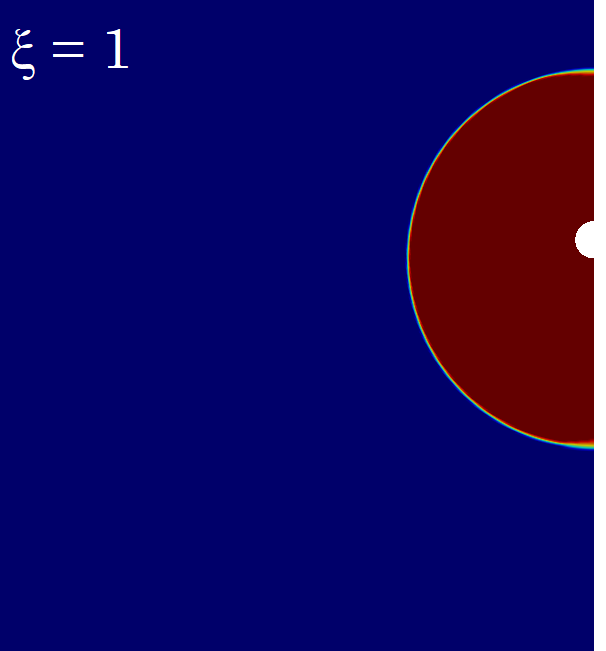}}~
	\subfloat[]
	{\includegraphics[width=.19\textwidth,height=0.145\textheight]{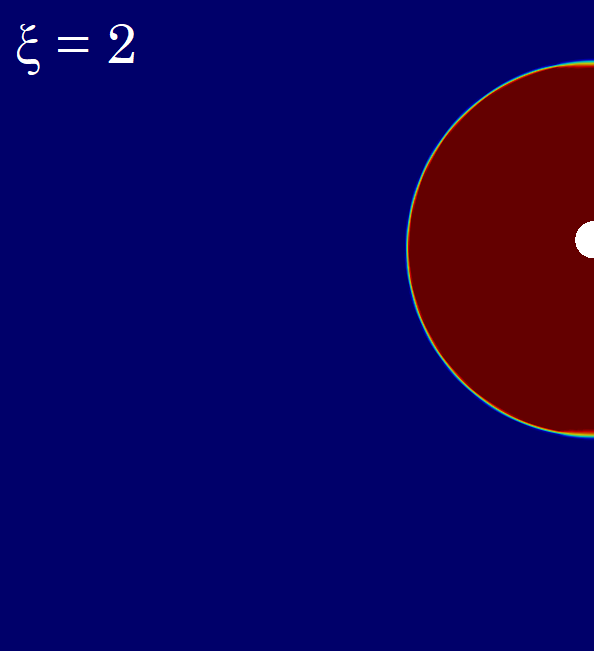}}~		
	\caption{Equilibrium shape of a n-heptane droplet ($D_0=1$ mm, ambient $T$) under the effect of gravity $\textbf{g}$ and different intensities of the potential field $\xi_0 \left[\frac{m^2}{s^2}\right]$.}
	\label{DropletShape}
\end{figure}

The intensity of $\textbf{f}_m$ can be controlled by the value of the constant $\xi_0$, which defines the potential field $\xi$ (Equation \ref{magneticpotential}). Figure \ref{DropletShape} shows the equilibrium shape assumed by a n-heptane droplet (1 mm diameter, ambient $T$) for different values of $\xi_0$, from 0.35 to 2 $\frac{m^2}{s^2}$. For values $\xi_0<0.35$ $\frac{m^2}{s^2}$  the droplet falls down, while for $\xi_0>2$ $\frac{m^2}{s^2}$ the droplet shape remains spherical. These threshold values have been found to be nearly the same for different liquids. \\
The effect is very similar to surface tension forces: for low values of $\xi_0$ (or surface tension $\sigma$) the droplet tends to assume an elongated shape, because of the dominant gravity forces. Increasing $\xi_0$ (or surface tension $\sigma$), the droplet approaches a spherical geometry. In this work $\xi_0$ represents a degree of freedom of the problem, which we can saturate imposing a $\xi_0$ value that better describes the droplet shape of the experimental cases. 

\subsection{Effect of the droplet shape on evaporation}
\begin{figure}
	\centering
	\subfloat[]
	{\includegraphics[width=.32\textwidth,height=0.18\textheight]{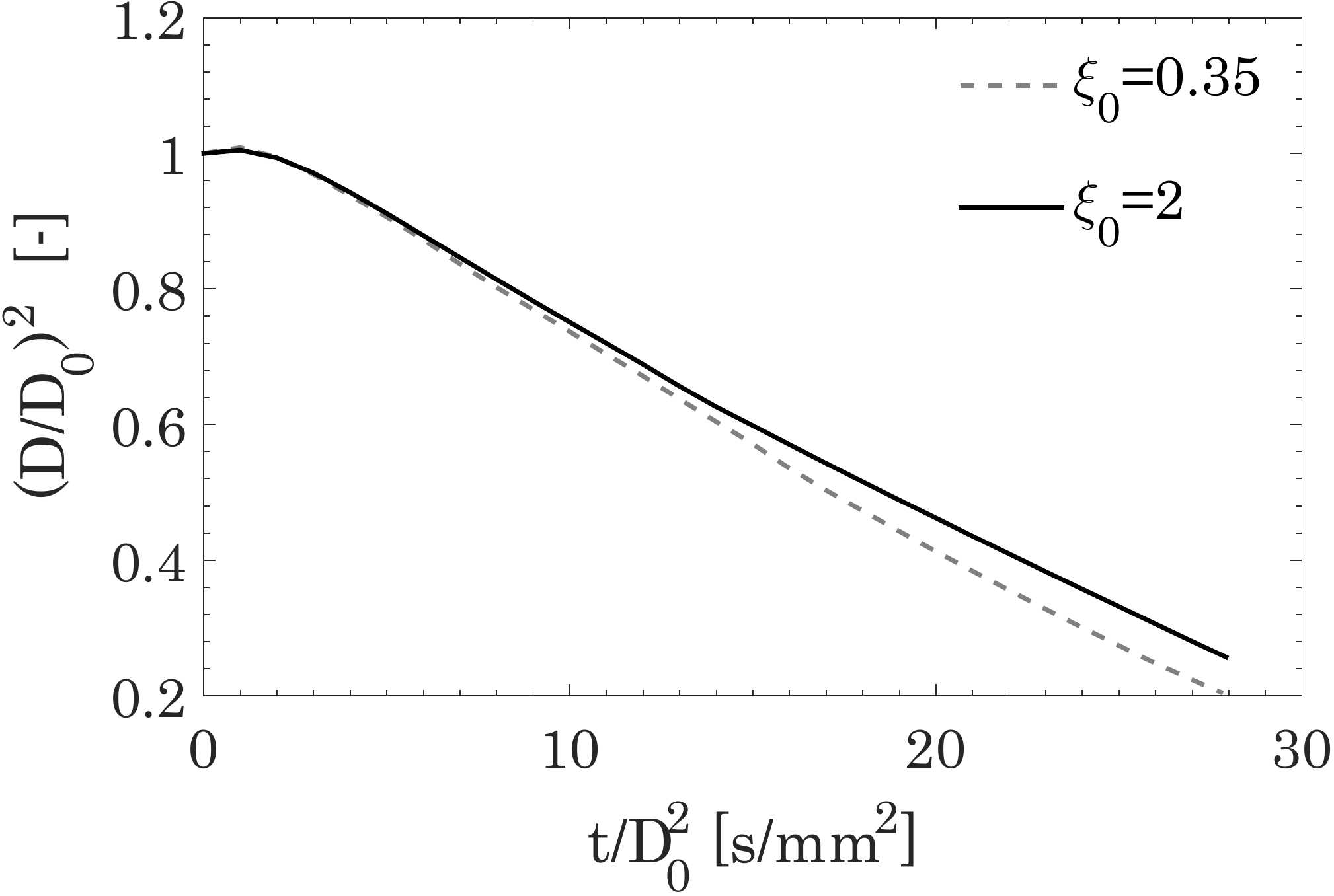}}~
	\subfloat[]
	{\includegraphics[width=.32\textwidth,height=0.18\textheight]{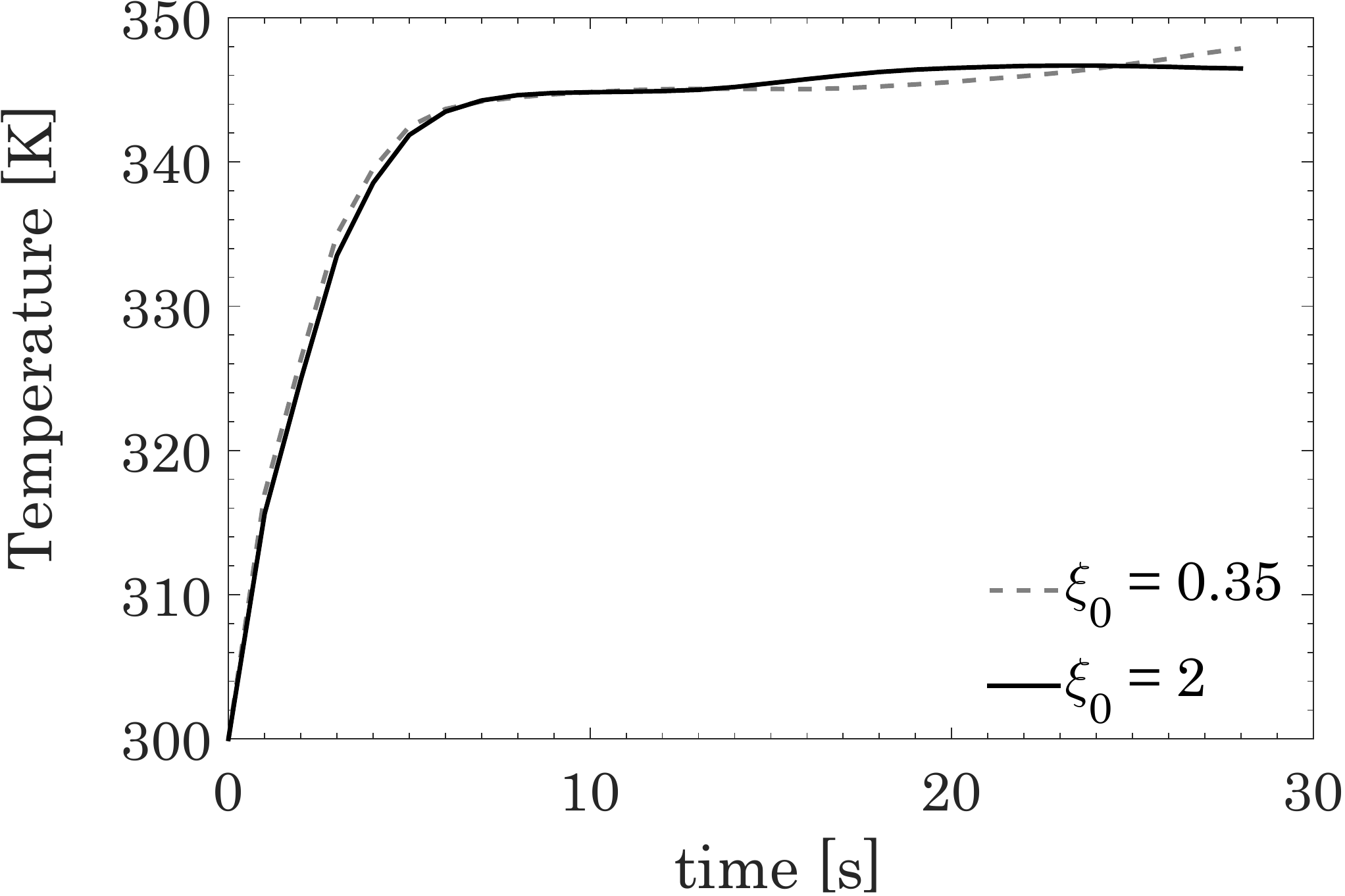}}~
	\subfloat[]
	{\includegraphics[width=.32\textwidth,height=0.19\textheight]{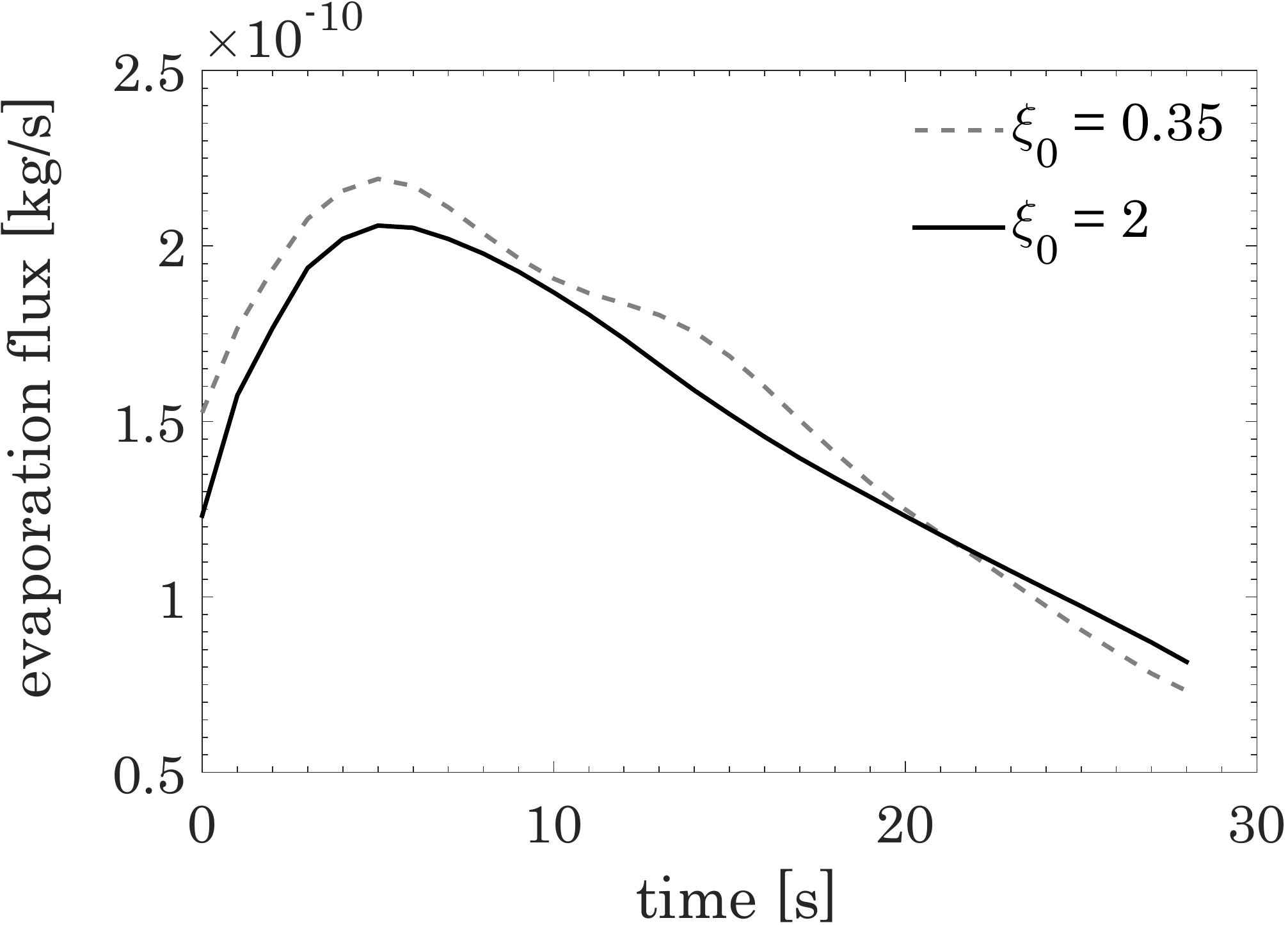}}		
	\caption{Effect of the droplet shape on the evaporation process. The dashed grey line corresponds to Figure \ref{DropletShape} (a), while the black line refers to Figure \ref{DropletShape} (e).}
	\label{ShapeAndEvaporation}
\end{figure}

\begin{figure}
	\centering
	\subfloat[]
	{\includegraphics[width=.45\textwidth,height=0.23\textheight]{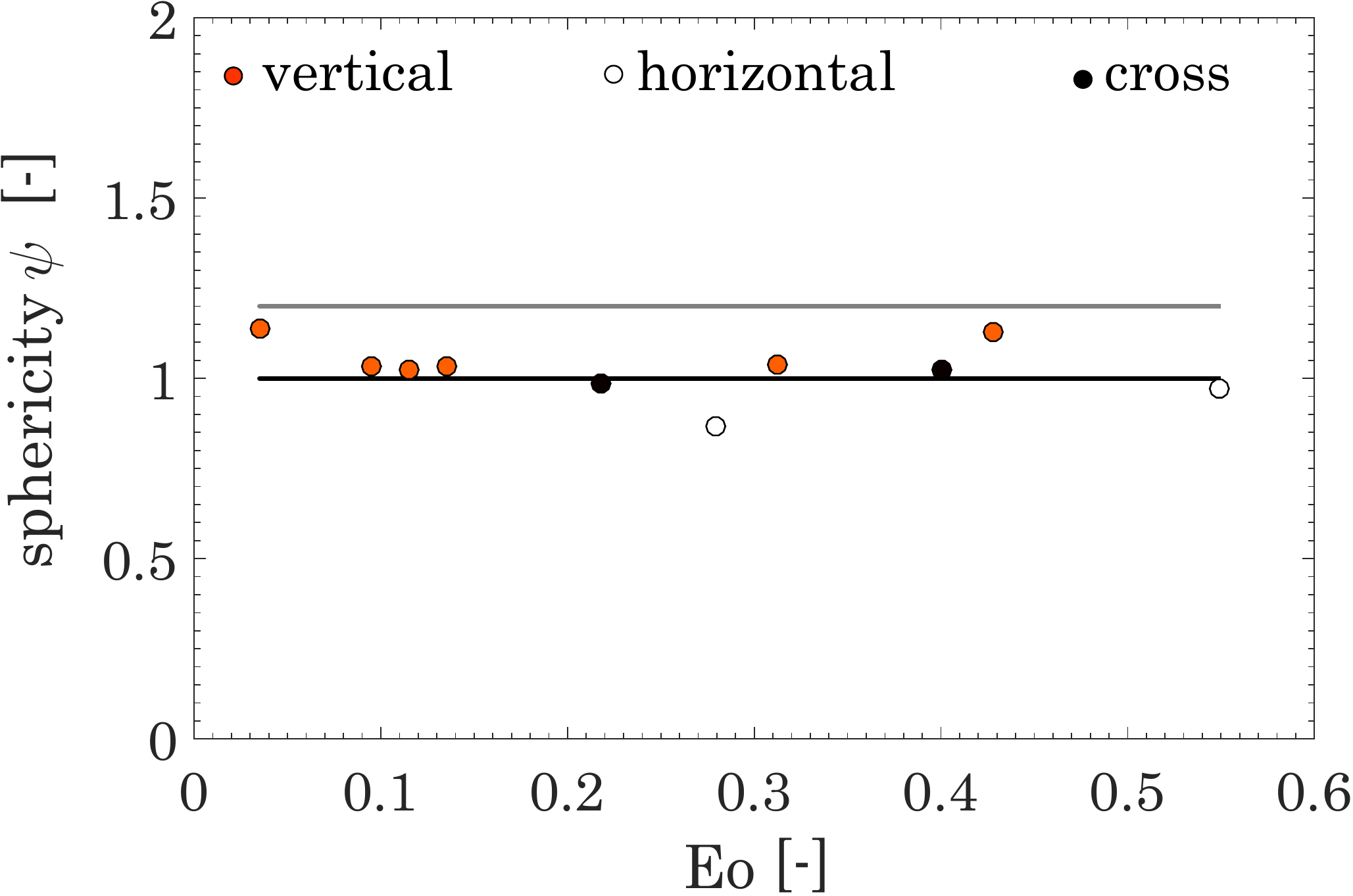}}~~~~	
	\centering
	\subfloat[]
	{\includegraphics[width=.32\textwidth,height=0.23\textheight]{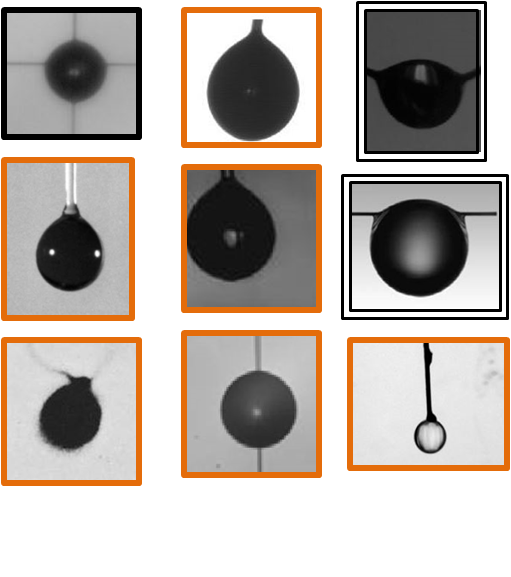}}	
	\caption{Scatterplot of experimental cases of suspended droplets (Figure a), based on their E\"{o}tv\"{o}s number $Eo=\frac{\rho_L g D^2}{\sigma}$ and sphericity $\psi=\frac{D_y}{D_x}$. Colors represent the supporting fiber: vertical (orange), horizontal (white) and cross fiber (black). The black line represents a perfect sphere ($\psi=1$), the gray line represents the limit case in Figure \ref{DropletShape} a ($\psi=1.2$). Figure (b) reports the 9 of the 10 droplet pictures, signed with the same colors for comparison.}
	\label{sphericity}
\end{figure}

In order to set a proper initial shape of the droplet, a sensitivity analysis on the evaporation process has been carried out. Two numerical simulations of an evaporating n-heptane droplet have been run for the threshold values  $\xi_0 = 0.35$ $\frac{m^2}{s^2}$ and $\xi_0 = 2$ $\frac{m^2}{s^2}$, (Figures \ref{DropletShape} a, e). The initial droplet diameter $D_0=1$ mm, the liquid temperature is $T_L=300$ K, the ambient temperature is $T_{G}=364$ K and the pressure is $p=20$ bar.\\
From Figure \ref{ShapeAndEvaporation} we can see that the droplet shape does not have a major impact on the evaporation process. The surface area increases for lower values of $\xi_0$, explaining the slightly larger evaporation rates (Figure \ref{ShapeAndEvaporation} c), while the surface temperature (Figure \ref{ShapeAndEvaporation} b) is almost insensitive to $\xi_0$.\\
We expect the experimental cases of our interest to behave somewhere in-between these two cases, closer to the upper limit described by the black lines in Figure \ref{ShapeAndEvaporation}\\
In order to verify this hypothesis, a simple analysis  has been conducted on the droplet shapes usually involved in experiments. 10 pictures of suspended droplets available in literature from experimental works \cite{ghassemi2006experimental, matlosz1972investigation, morin2004vaporization, han2015evaporation, walton2004evaporation, chauveau2008experimental, suzuki2009development} have been collected (Figure \ref{sphericity} b) and organized based on two dimensionless numbers, the sphericity $\psi$ and the E\"{o}tv\"{o}s number $Eo$, defined as: 

\begin{equation}
\psi=\frac{D_y}{D_x}
\end{equation}

\begin{equation}
Eo=\frac{\left(\rho_L-\rho_G\right) g D^2}{\sigma}\sim\frac{\rho_L g D^2}{\sigma}
\end{equation}

where $D_y$ and $D_x$ are the maximum lengths of the droplet, in vertical and horizontal directions. The sphericity $\psi$ is a measure of how the droplet shape approaches that of a perfect sphere (which has $\psi=1$). The E\"{o}tv\"{o}s number $Eo$ is the ratio between gravitational and surface tension forces. A further classification is introduced based on the tipe of supporting fiber (vertical, horizontal or cross fiber). The 10 experimental cases are reported in Figure \ref{sphericity} (a), together with the black line $\psi=1$, representing a perfect sphere (Figure \ref{DropletShape} e), and the gray line $\psi=1.2$ representing the limit case in Figure \ref{DropletShape} (a).

The E\"{o}tv\"{o}s number is below 1 for all cases, indicating that surface tension forces always dominate over gravity. The sphericity $0.86<\psi<1.12$ indicates nearly spherical shapes for all the droplets, especially when cross fibers are used (black points). As expected, the horizontal fibers tend to slightly deform the droplet along $D_x$ direction, providing $\psi<1$ (white points) while the vertical ones along $D_y$, providing $\psi>1$ (orange points). Figure \ref{sphericity} clearly shows that most of experimental cases lay  between the two lines $\psi=1$ (Figure \ref{DropletShape} e) and $\psi=1.2$ (Figure \ref{DropletShape} a).
A rigorous modeling of the surface topology would require  different computational meshes to describe the large variety of fibers used to support the droplets (vertical, horizontal, thermocouples, cross fibers...), as well as precise data regarding the dynamic contact angle between the solid and the different liquids. In order to simplify the problem and based on Figures \ref{ShapeAndEvaporation} and \ref{sphericity}, we assume the evaporation regime not to be significantly influenced by small deviations of the droplet shape from the spherical one, which probably better describes on average the different experimental cases analyzed in this work.  For these reasons we decided to adopt an intermediate  value $\xi_0=1$ $\frac{m^2}{s^2}$  (Figure \ref{DropletShape} d).\\

\section{Cases in natural convection}
\subsection{Description of the experimental cases}
The numerical model is now validated against 11 sets of experimental data in natural convection regime, taken from 4 papers in literature from 1997 to 2018. All the experimental cases are summarized in Table \ref{tableGravityCases}.\\
Ghassemi \cite{ghassemi2006experimental} used a fine quartz fiber of 0.125 mm of diameter (similarly to the spherical fiber used in our simulations) to hang and evaporate small n-heptane droplets in a hot nitrogen environment. Nomura et al. \cite{Nomura1997} examined the evaporation of cold n-heptane droplets in a hot air environment from 300 K to 450 K at various pressures (from 5 atm to 50 atm). The experimental data are taken from the paper of Gogos et al. \cite{gogos2003effects}, who proposed a very complex axysimmetric model to model them. An aluminum cross-fiber support frame has been utilized by Verwey et al. \cite{verwey2018experimental}, in order to maintain a spherical shape of the droplet. N-decane and n-heptane evaporation has been investigated. Finally, n-hexadecane  droplets evaporation was examined through a horizontal fiber by Han et al. \cite{han2015evaporation}. Where needed, electric furnaces have been used to heat up the environment and nitrogen to pressurize the vessel. High-speed video cameras have been adopted to follow the droplet lifetime. For all cases, the experimental data provided concern the droplet "equivalent" diameter decay. Han et al. \cite{han2015evaporation} also provided some data on the mean droplet temperature by using a thermocouple as a fiber.\\
To our knowledge, the cases from Verwey et al.  \cite{verwey2018experimental} and Han et al. \cite{han2015evaporation} have never been modeled. Experimental data from Ghassemi et al. \cite{ghassemi2006experimental} and Nomura et al. \cite{Nomura1997} have only been approached so far either with simplified  semi-analytical methods, or specific correlations based on mass-heat transfer dimensionless numbers for the evaporation sub-model \cite{azimi2017effect, fang2017new}.\\  
In principle, the \texttt{DropletSMOKE++} code can handle the thermal perturbation of the fiber on the droplet. This could be done either imposing a heat flux of the sphere boundary (Figure 5) which mimics its presence, or directly meshing the fiber geometry and solve the multi-region heat transfer between the fluid and the solid. Because of the small size of the fibers used in the experiments and the relatively low temperatures involved (since we are not modeling combustion processes), we found the evaporation process to be marginally affected by the thermal perturbation of the fiber, deciding to neglect it and postponing the problem to future works.

\begin{table}
	\centering
	
	\begin{tabular}{llllllll}
		\toprule
		
		Case ~~& Species~~ & $D_0$ ~~& $T_L$ ~~& $T_{G}$ ~~& p ~~& Refs. ~~& Results  \\
		~~&  ~~& [mm] ~~& [K]~~ & [K] ~~& [atm] ~~& ~~&   \\
		\midrule
		1 & n-Heptane     & 1     & 300 & 773 & 10 & \cite{ghassemi2006experimental}& Figure \ref{GhassemiEtAl} (a) \\
		2 & n-Heptane     & 1     & 300 & 673 & 10 & \cite{ghassemi2006experimental}& Figure \ref{GhassemiEtAl} (b)  \\
		3 & n-Heptane     & 0.698 & 300 & 398 & 50 & \cite{Nomura1997} & Figure \ref{NomuraEtAl} (a)  \\
		4 & n-Heptane     & 0.607 & 300 & 474 & 10 & \cite{Nomura1997}  & Figure \ref{NomuraEtAl} (b)\\ 
		5 & n-Heptane     & 0.604 & 300 & 450 & 5  & \cite{Nomura1997}   & Figure \ref{NomuraEtAl} (c)\\
		6 & n-Decane      & 0.546 & 300 & 373 & 10 & \cite{verwey2018experimental} & Figure \ref{VerweyEtAl} (a)\\  
		7 & n-Heptane     & 0.539 & 300 & 300 & 5  & \cite{verwey2018experimental} & Figure \ref{VerweyEtAl} (b)\\ 
		8 & n-Decane      & 0.5   & 300 & 300 & 1 & \cite{verwey2018experimental} & Figure \ref{VerweyEtAl} (c)\\ 	
		9 & n-Decane      & 0.245 & 300 & 300 & 10 & \cite{verwey2018experimental}  & Figure \ref{VerweyEtAl} (d)\\ 
		10 & n-Hexadecane & 1     & 300 & 773 & 1 & \cite{han2015evaporation} & Figure \ref{HanEtAl} (a, b, c)\\ 
		11 & n-Hexadecane & 1     & 300 & 673 & 1& \cite{han2015evaporation}  & Figure \ref{HanEtAl} (d, e, f)\\ 		
		\bottomrule
		
	\end{tabular}	
	
	\caption{Experimental cases in natural convection regime examined in this work.}
	\label{tableGravityCases}
\end{table}

% \begin{figure}
% 	\centering
%
% 	{\includegraphics[width=.95\textwidth,height=0.12\textheight]{Immagini/HanEtAl773K}}
% 	
% 	\caption{Sequential photographs of the n-hexadecane droplet vaporization at 773 K by Han et al. \cite{han2015evaporation}, case 10 in Table \ref{tableGravityCases}.}
% 	\label{ExperimentalData}
% \end{figure}

\subsection{Numerical simulations: results and discussion}

\begin{figure}
	\centering	
	%	\quad\quad\quad\quad\quad\quad\quad\quad \quad \quad
	%	{\includegraphics[width=.23\textwidth,height=0.03\textheight]{Immagini/ScreenshotsArticolo/nc7massfraction}}~
	%	{\includegraphics[width=.23\textwidth,height=0.03\textheight]{Immagini/ScreenshotsArticolo/temperature}}~
	%	{\includegraphics[width=.23\textwidth,height=0.03\textheight]{Immagini/ScreenshotsArticolo/velocity}} \\			
	{\includegraphics[width=.25\textwidth,height=0.19\textheight]{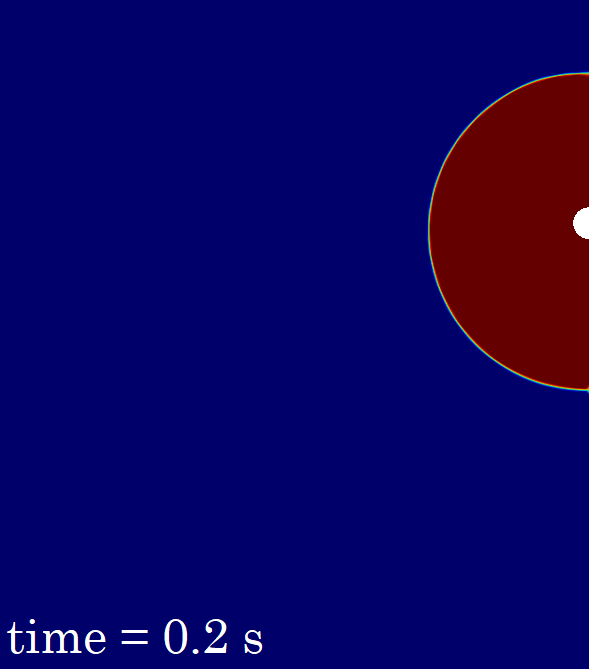}}~
	{\includegraphics[width=.245\textwidth,height=0.23\textheight]{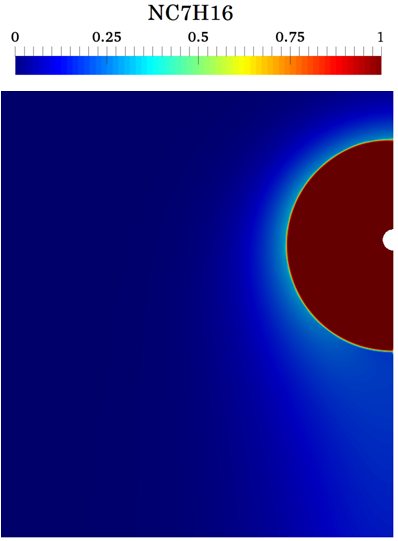}}~
	{\includegraphics[width=.25\textwidth,height=0.23\textheight]{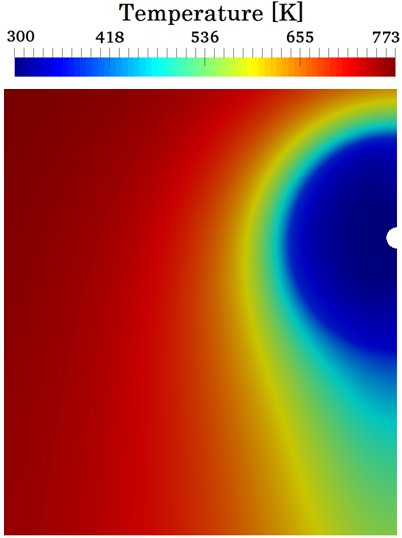}}~
	{\includegraphics[width=.245\textwidth,height=0.23\textheight]{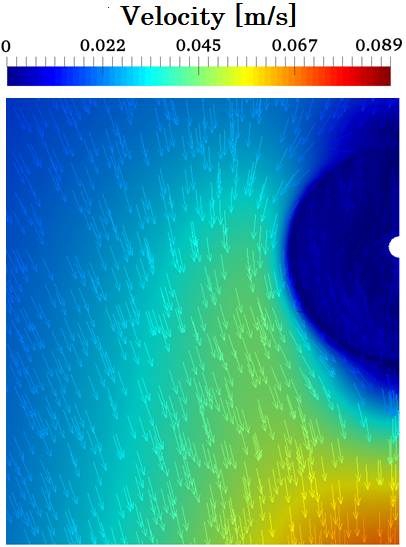}}\\ 
	{\includegraphics[width=.25\textwidth,height=0.19\textheight]{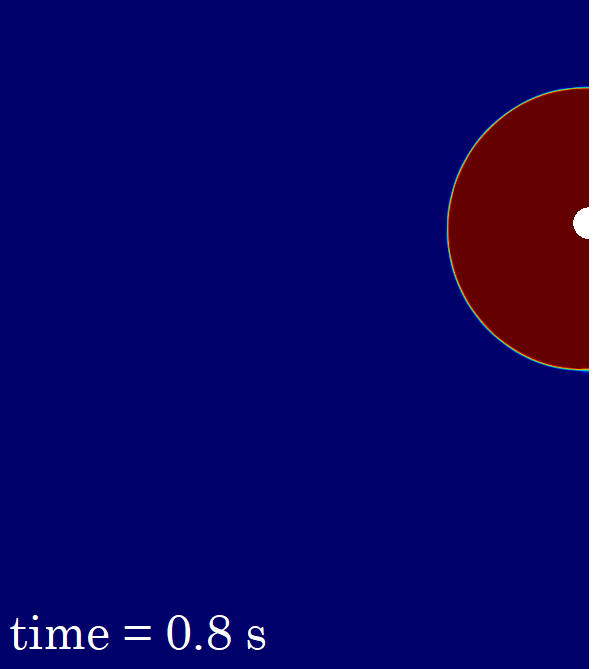}}~
	{\includegraphics[width=.25\textwidth,height=0.19\textheight]{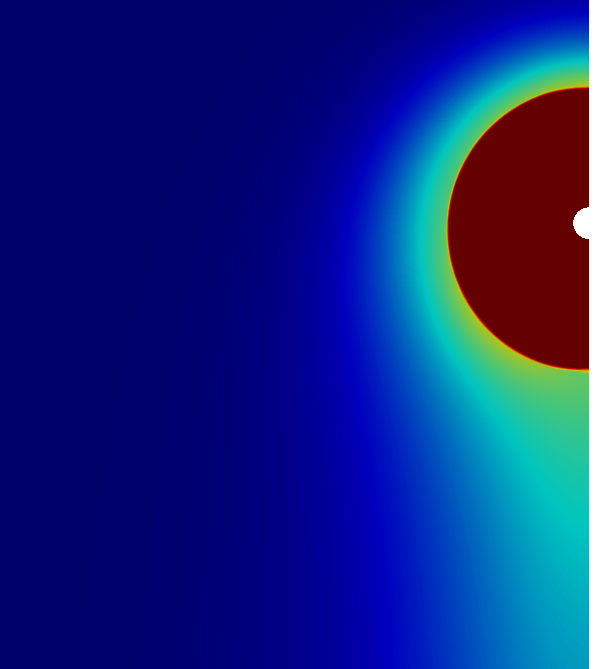}}~
	{\includegraphics[width=.25\textwidth,height=0.19\textheight]{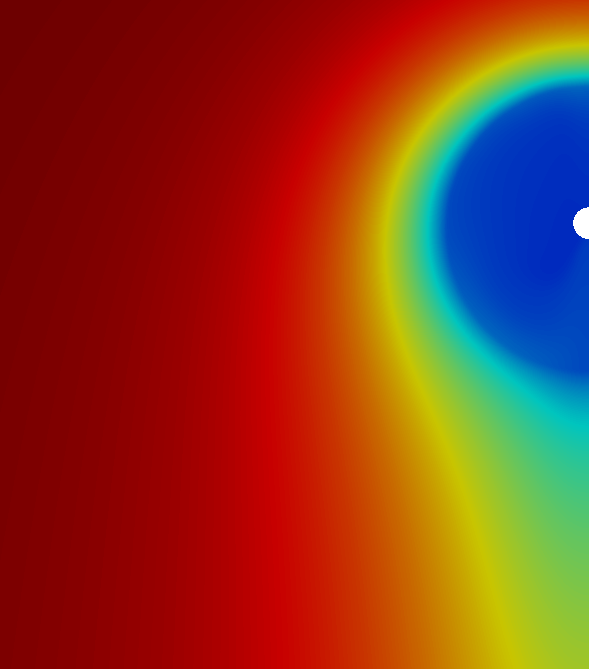}}~
	{\includegraphics[width=.25\textwidth,height=0.19\textheight]{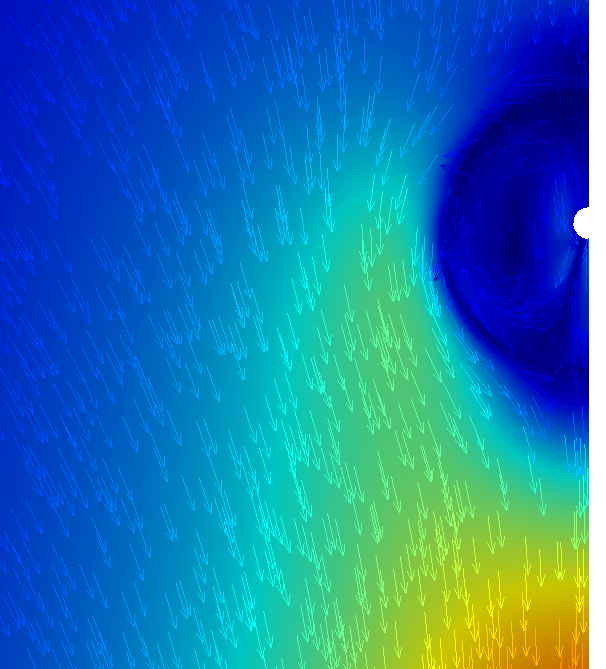}}\\	
	{\includegraphics[width=.25\textwidth,height=0.19\textheight]{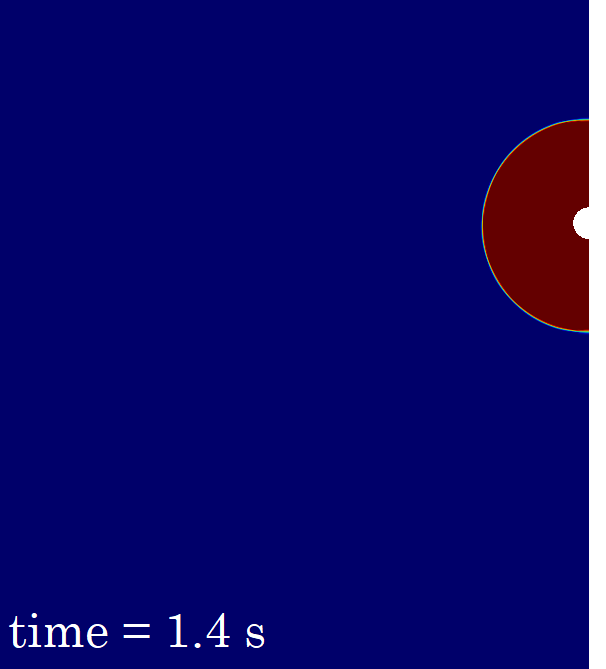}}~
	{\includegraphics[width=.25\textwidth,height=0.19\textheight]{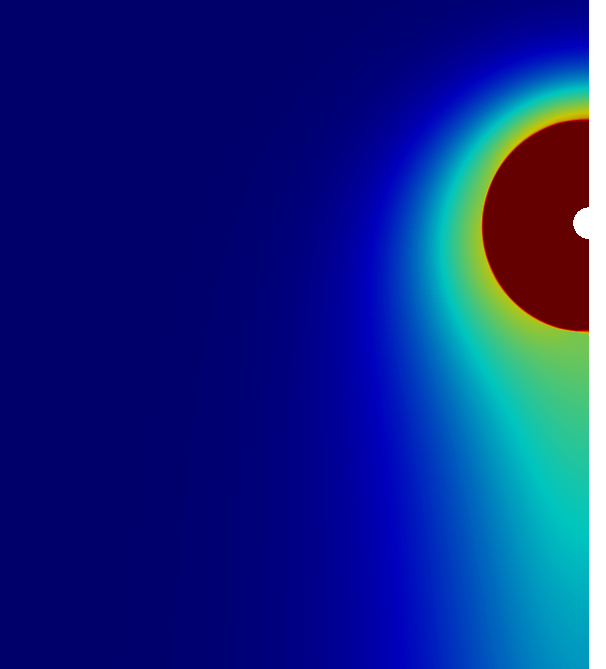}}~
	{\includegraphics[width=.25\textwidth,height=0.19\textheight]{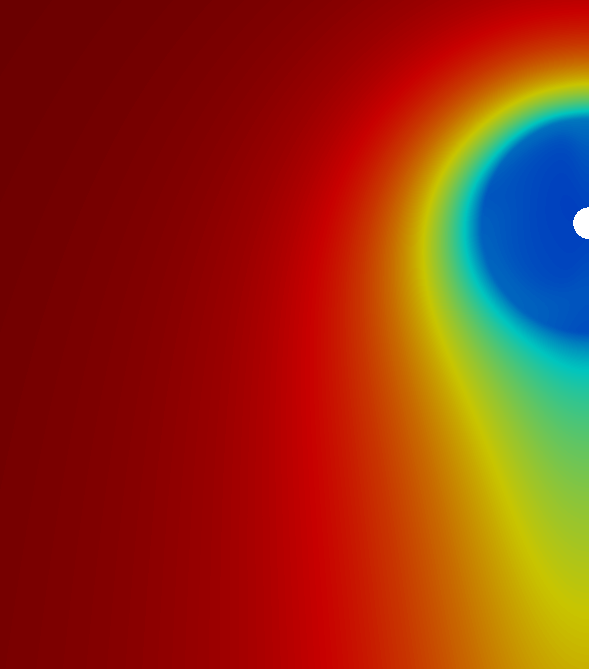}}~
	{\includegraphics[width=.25\textwidth,height=0.19\textheight]{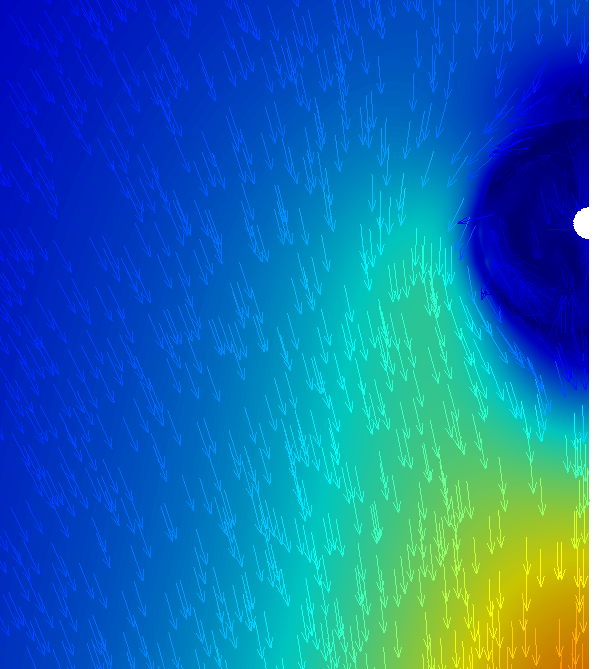}}
	\subfloat[]
	{\includegraphics[width=.25\textwidth,height=0.19\textheight]{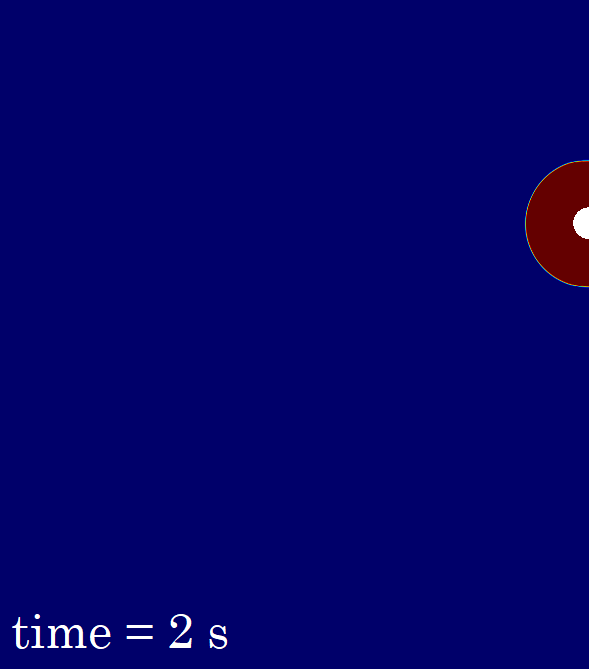}}~
	\subfloat[]
	{\includegraphics[width=.25\textwidth,height=0.19\textheight]{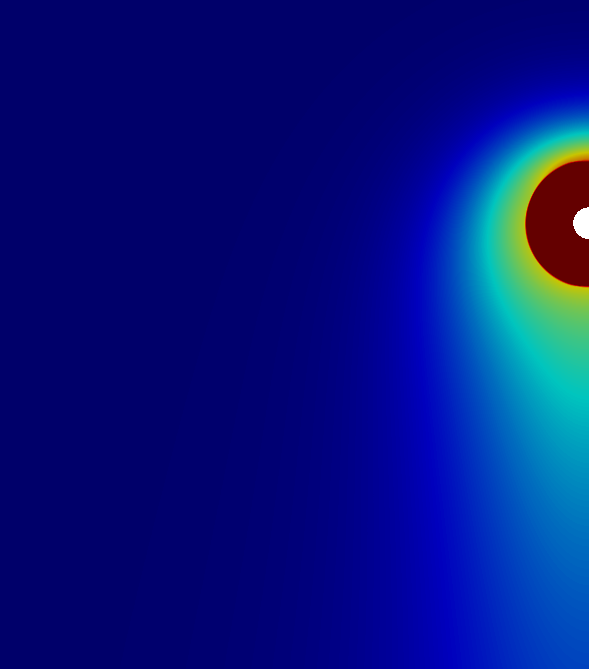}}~
	\subfloat[]
	{\includegraphics[width=.25\textwidth,height=0.19\textheight]{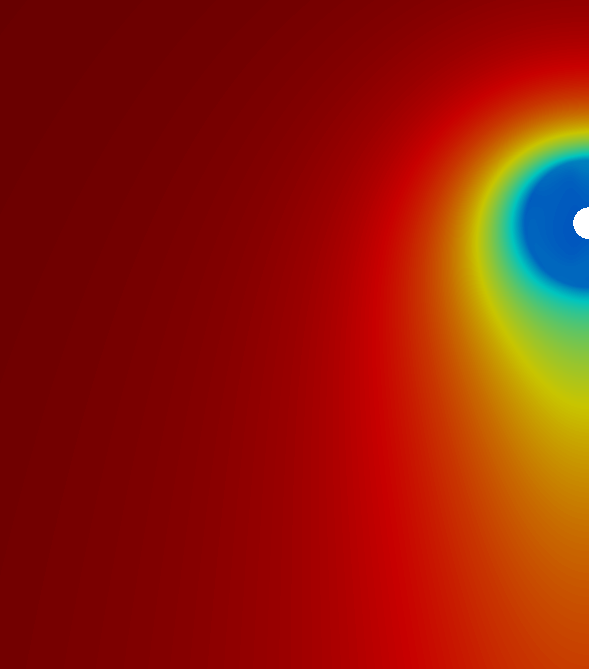}}~
	\subfloat[]
	{\includegraphics[width=.25\textwidth,height=0.19\textheight]{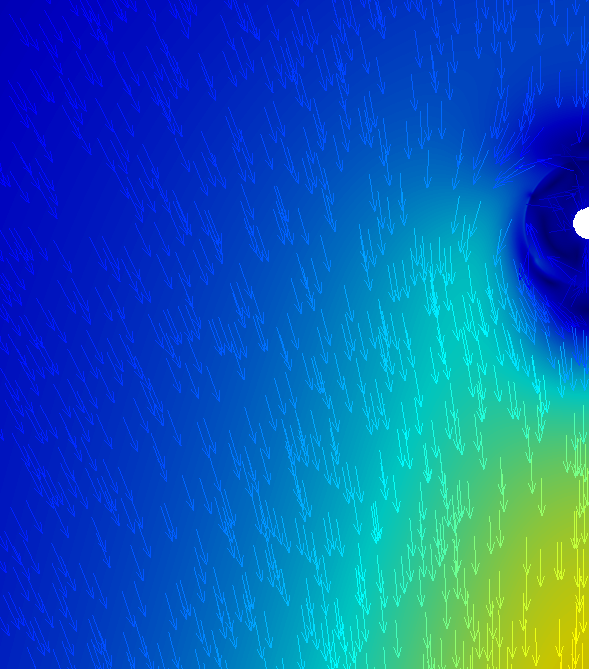}}										
	
	\caption{$\alpha$ (a), n-heptane mass fraction (b), temperature (c) and velocity vector fields (d) evolution. Numerical simulation of case 1 from Ghassemi et al. \cite{ghassemi2006experimental}.}
	\label{Screenshots}
\end{figure} 

\begin{figure}
	\centering
	\subfloat[]
	{\includegraphics[width=.38\textwidth,height=0.2\textheight]{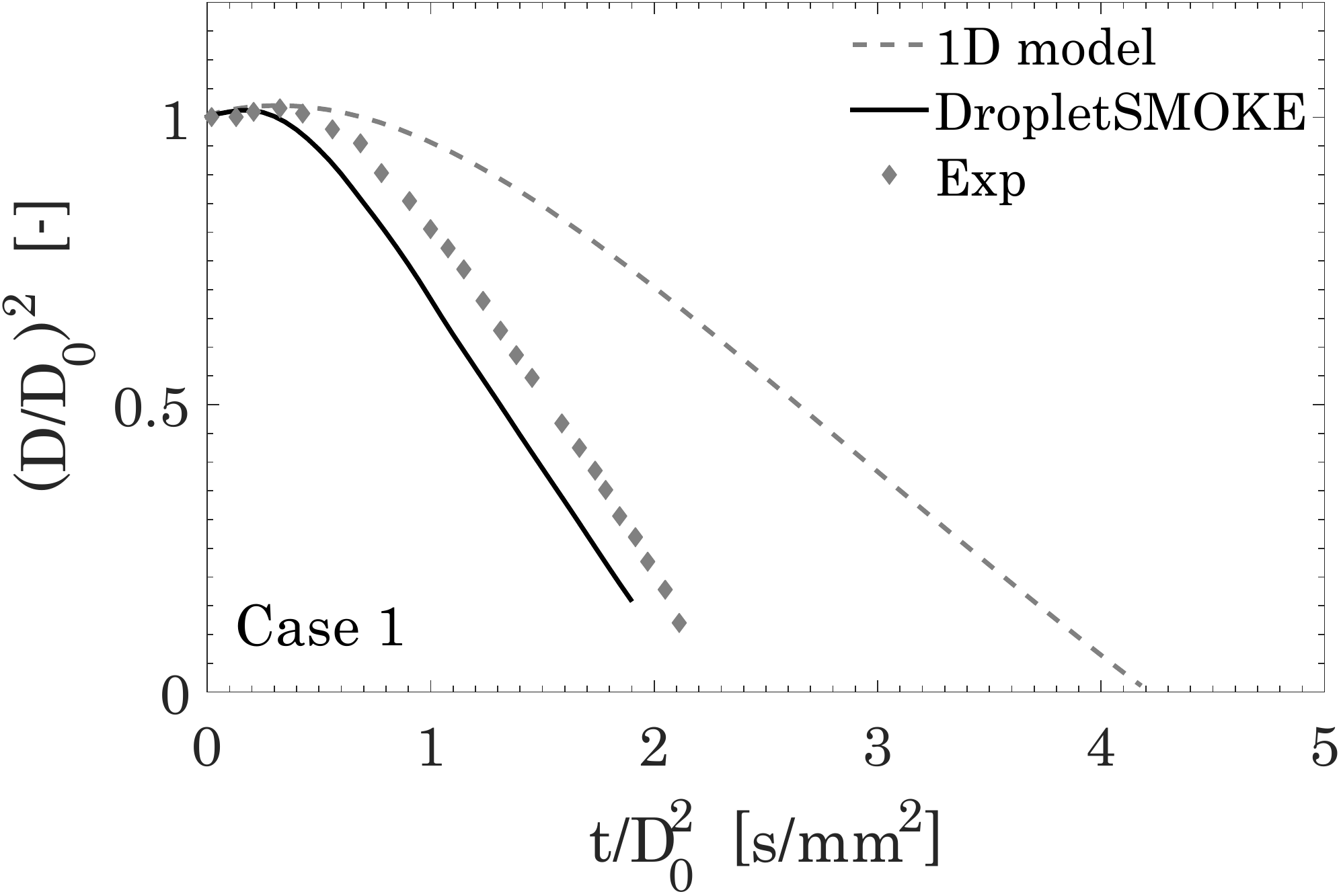}}\quad
	\subfloat[]
	{\includegraphics[width=.38\textwidth,height=0.2\textheight]{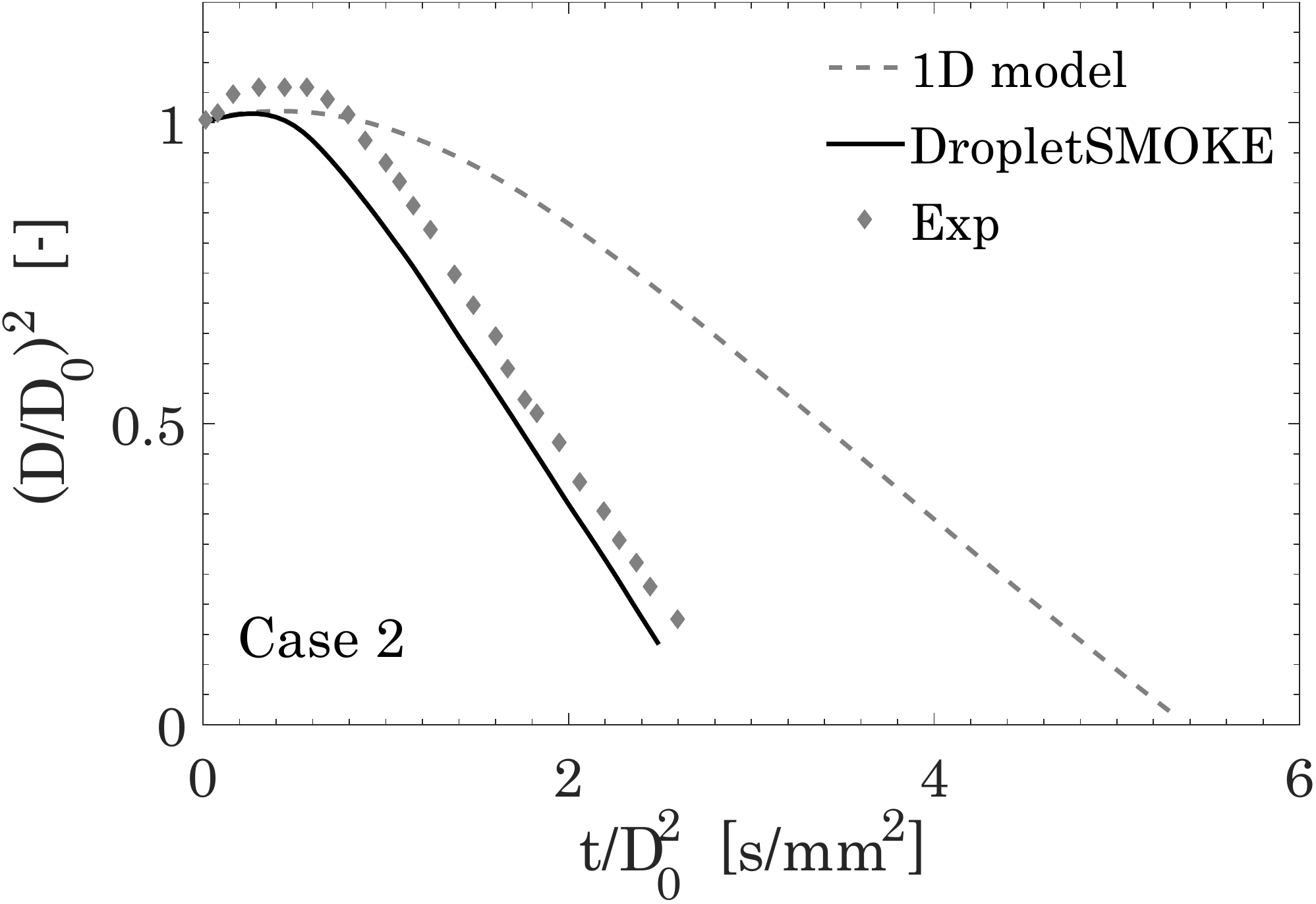}}					
	
	\caption{Model prediction of the experimental data of Ghassemi et al. \cite{ghassemi2006experimental}, cases 1, 2 of Table \ref{tableGravityCases}.}
	\label{GhassemiEtAl}
\end{figure} 

For each one of the cases presented in Table \ref{tableGravityCases}, an equivalent simulation (imposing the same initial conditions) has been performed with the microgravity solver of Cuoci et al. \cite{cuoci2005autoignition} in order to highlight the impact of gravity on the numerical simulation and better support the need of a multidimensional CFD model.\\		
In Figure \ref{Screenshots} the numerical results of the simulation of case 1 of Ghassemi et al. \cite{ghassemi2006experimental} are reported as a reference example for a qualitative analysis. The liquid volume fraction ($\alpha$), the n-heptane mass fraction field, the temperature and the vectorial velocity field are reported respectively in Figures \ref{Screenshots}.\\
The droplet is initially at an ambient temperature $T_L=300$ K and it is progressively heated by the external environment at $T_{G}=773 $ K (Figure \ref{Screenshots} c). A small amount of vapor is released from  the droplet surface (Figure \ref{Screenshots} b). The gas around the droplet is cooler than $T_{G}$ and the n-heptane vapor has a higher density. This density gradient induces a downward flow field (Figure \ref{Screenshots} d) around the droplet, generating a laminar natural convection regime. After a transient period, an equilibrium temperature is reached since the equilibrium mass fraction vapor in Figure \ref{Screenshots} (b) reaches a steady state value.\\

Figure \ref{GhassemiEtAl} reports the numerical results of the cases from Ghassemi et al. \cite{ghassemi2006experimental} (cases 1, 2), compared against the available experimental data. \texttt{DropletSMOKE++} can predict the experimental data with a reasonable accuracy for both cases. The linear behavior of the squared diameter can be easily recognized, as well as its initial increase due to the droplet heating shown in Figure \ref{Screenshots}. \\

\begin{figure}
	\centering
	\subfloat[]
	{\includegraphics[width=.35\textwidth,height=0.19\textheight]{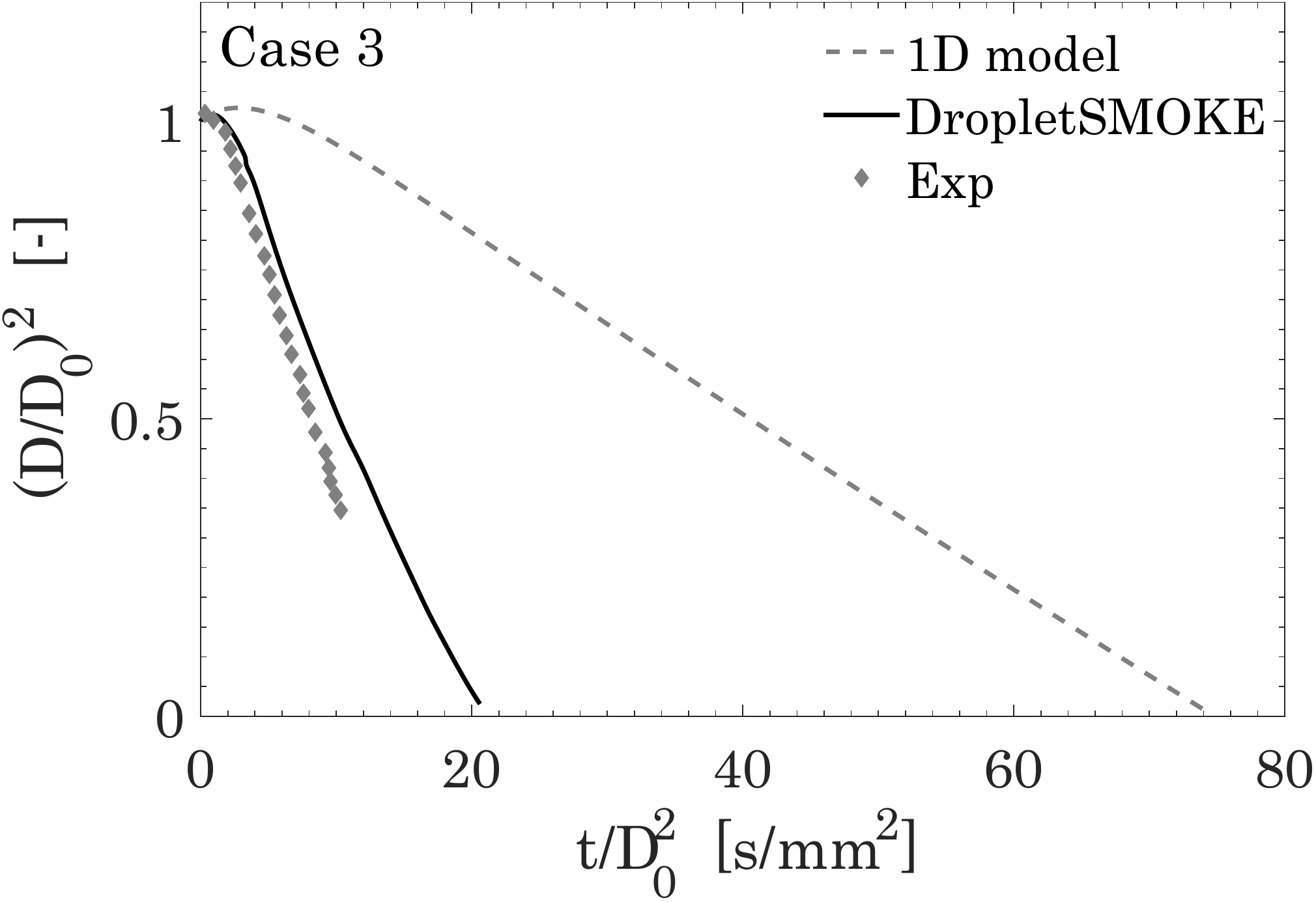}}~
	\subfloat[]
	{\includegraphics[width=.35\textwidth,height=0.19\textheight]{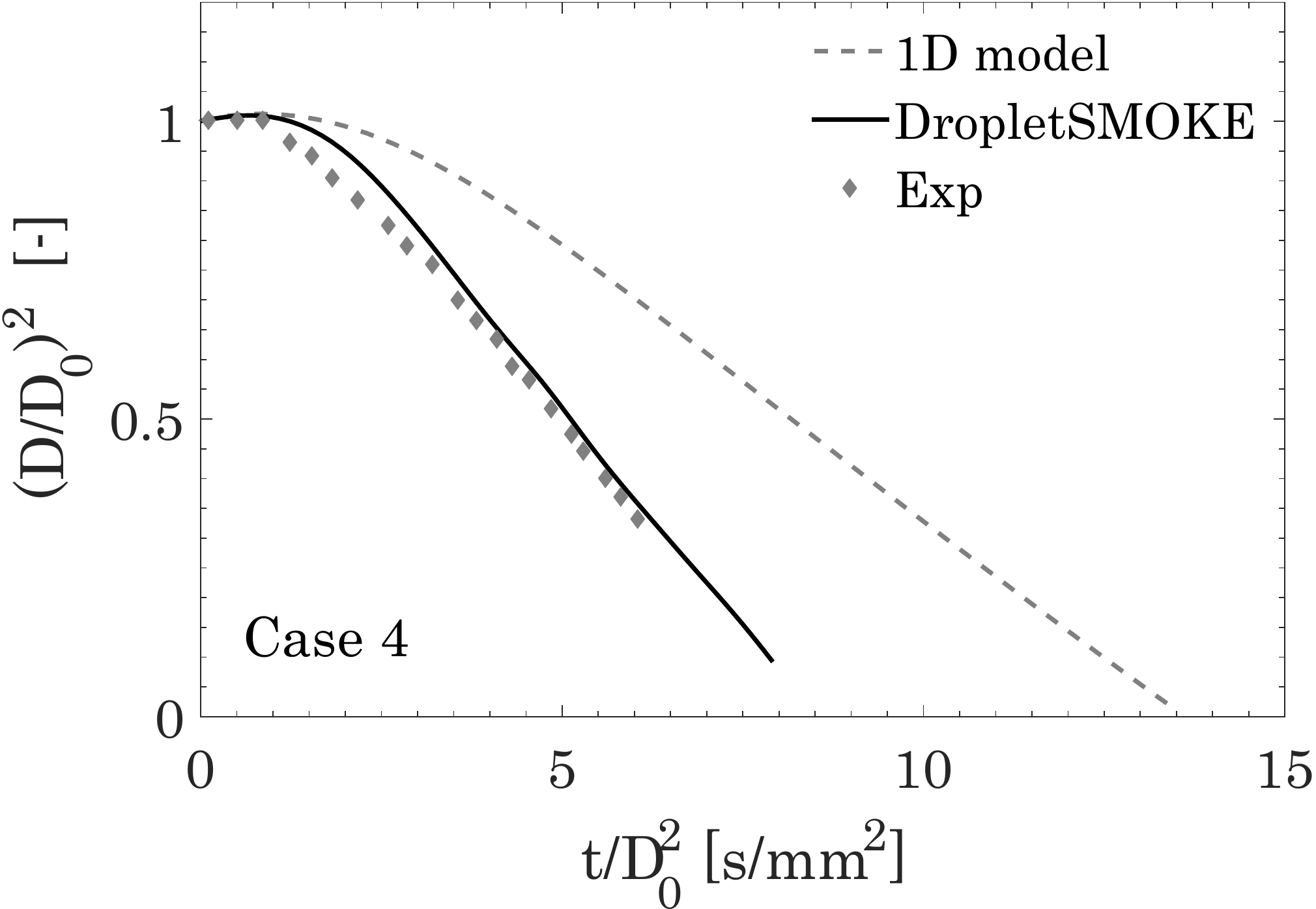}}~
	\subfloat[]
	{\includegraphics[width=.35\textwidth,height=0.19\textheight]{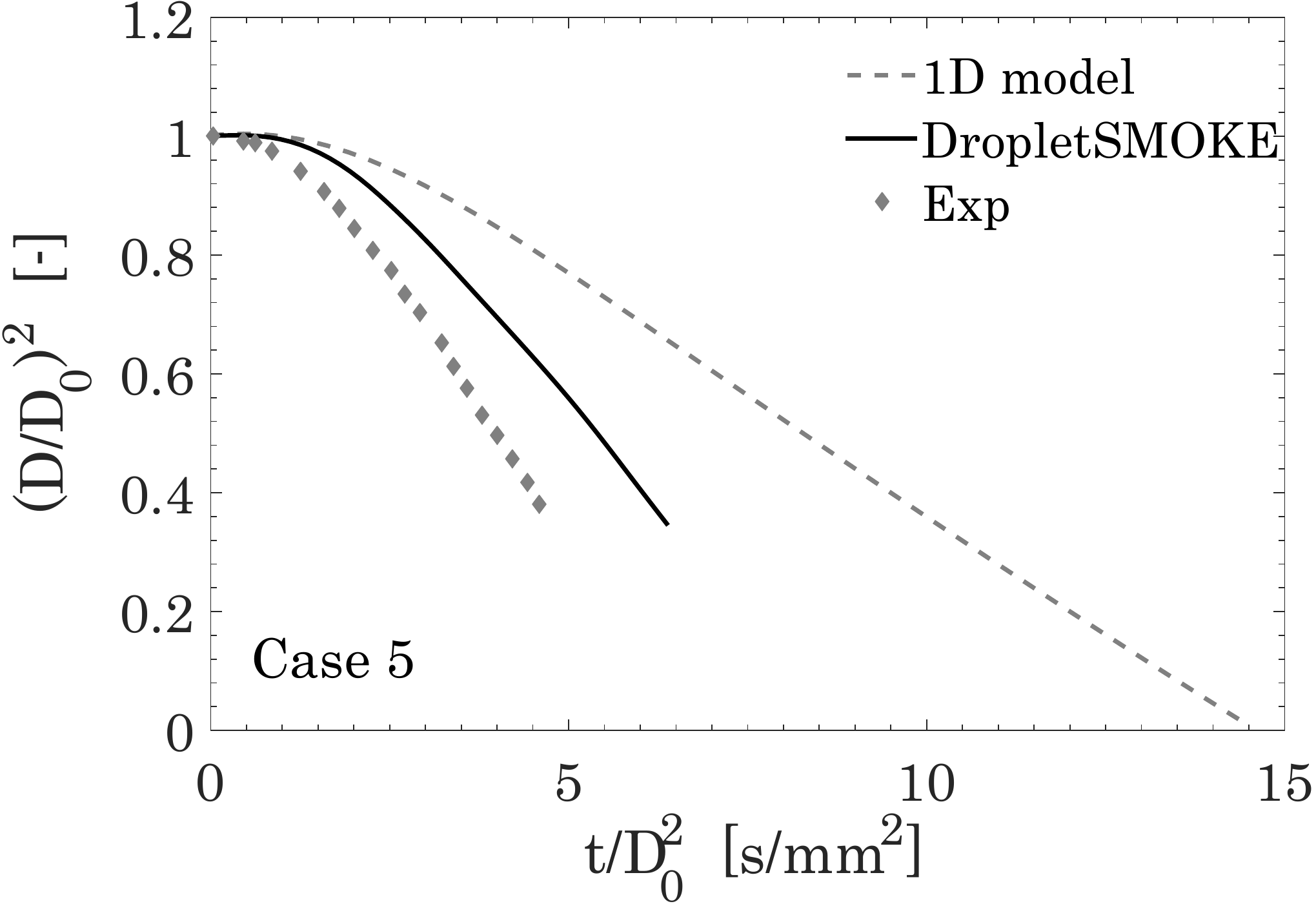}}						
	
	\caption{Model prediction of the experimental data of Nomura et al. \cite{Nomura1997}, cases 3, 4, 5 of Table \ref{tableGravityCases}.}
	\label{NomuraEtAl}
\end{figure}

The numerical simulations of the cases from Nomura et al. \cite{Nomura1997} (cases 4, 5, 6) are reported in Figure \ref{NomuraEtAl}. There is a good agreement with the experimental data, especially if compared to the model prediction of the 1D model which largely overpredicts evaporation times. The velocity field around the droplet increases both heat and mass transfer rates from the liquid surface, providing a much lower evaporation time with respect to the microgravity case. This effect is particularly enhanced at high pressure (case 3, Figure \ref{NomuraEtAl} a) where the droplet evaporates  $\sim 4$ times faster under the influence of gravity.  \\
The reason is straightforward: the Grashof number increases with  the gas density (with a power 2), which is proportional to pressure:

\begin{equation}
Gr = \frac{\rho^2 g D^3 \beta \Delta T}{\mu^2}\sim p^2
\end{equation}

Therefore, the presence of natural convection has a stronger impact in high pressures cases, if compared with the same cases in microgravity (where $Gr=0$).\\
In Case 5 (Figure \ref{NomuraEtAl} c) we slightly overpredict the droplet lifetime, especially if compared to the other cases from the same author. It is worth noticing that this experimental case has been also modeled by Gogos et al. \cite{gogos2003effects}, using a complex non-VOF axisymmetric model and assuming a perfectly spherical droplet. His model generally shows good agreement with experiments, but the results of this specific case show deviation from the experimental data very similar to ours.\\

\begin{figure}
	\centering
	\subfloat[]
	{\includegraphics[width=.38\textwidth,height=0.2\textheight]{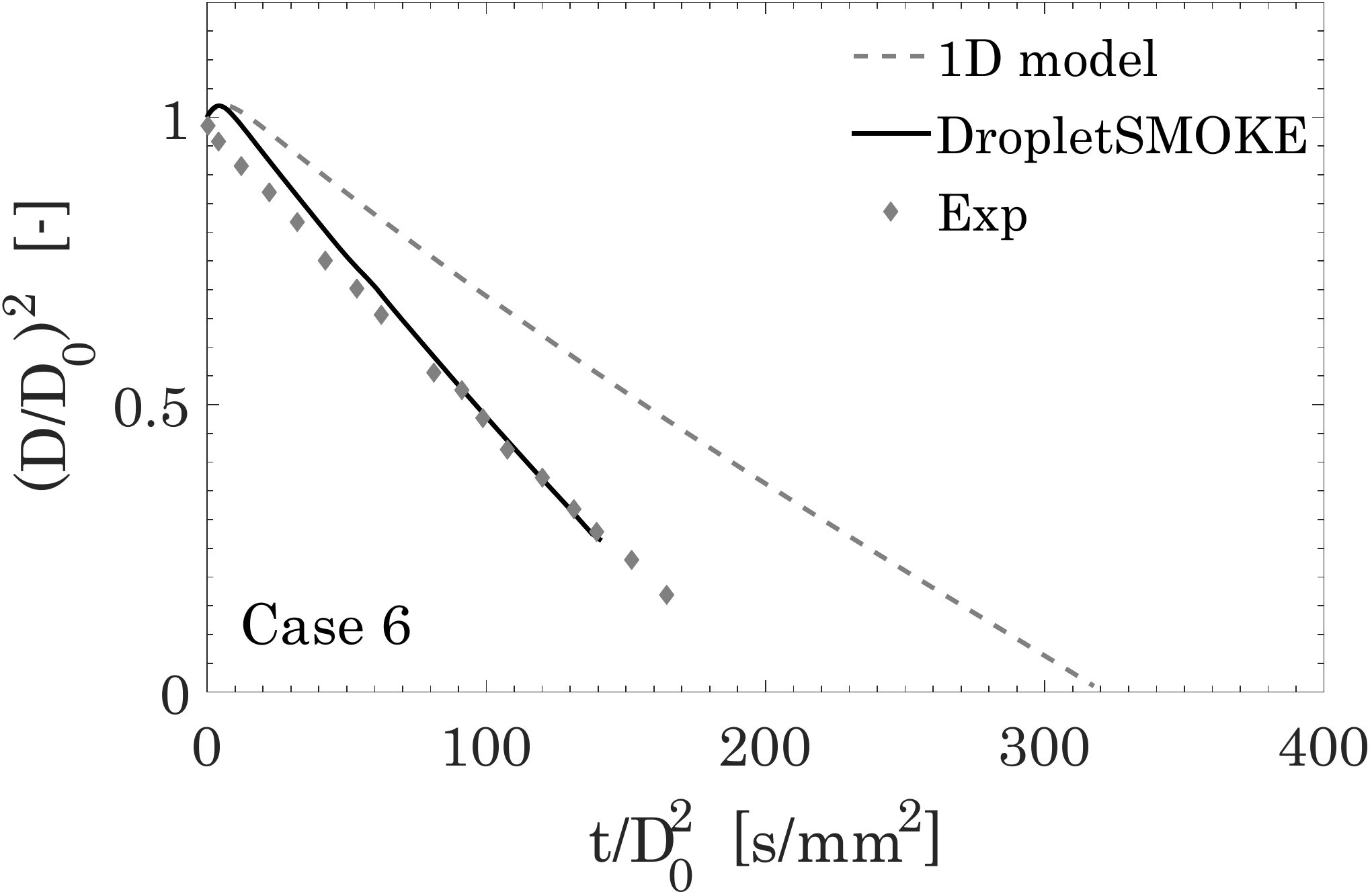}}~
	\subfloat[]
	{\includegraphics[width=.38\textwidth,height=0.2\textheight]{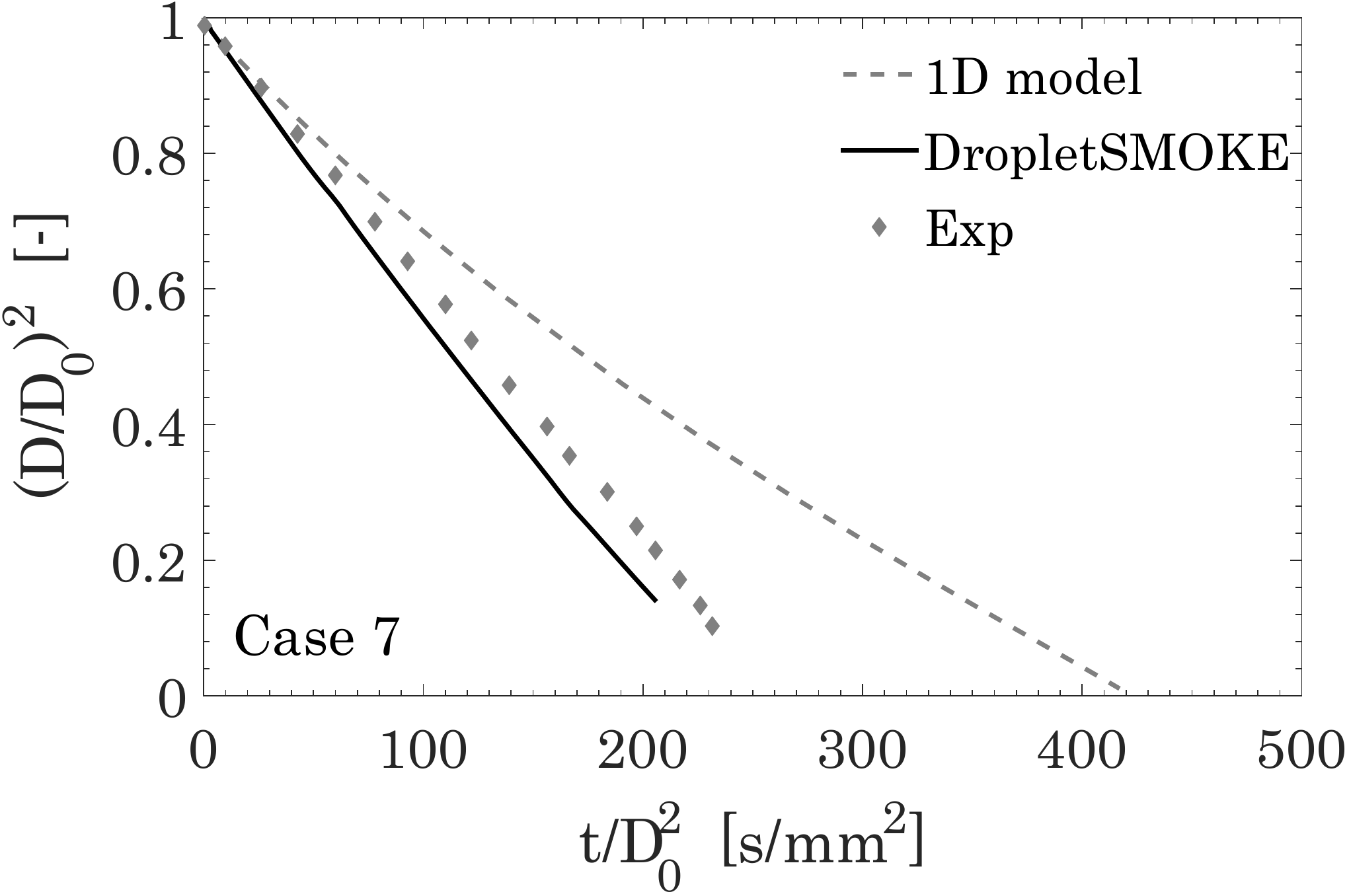}}\\
	\subfloat[]
	{\includegraphics[width=.38\textwidth,height=0.2\textheight]{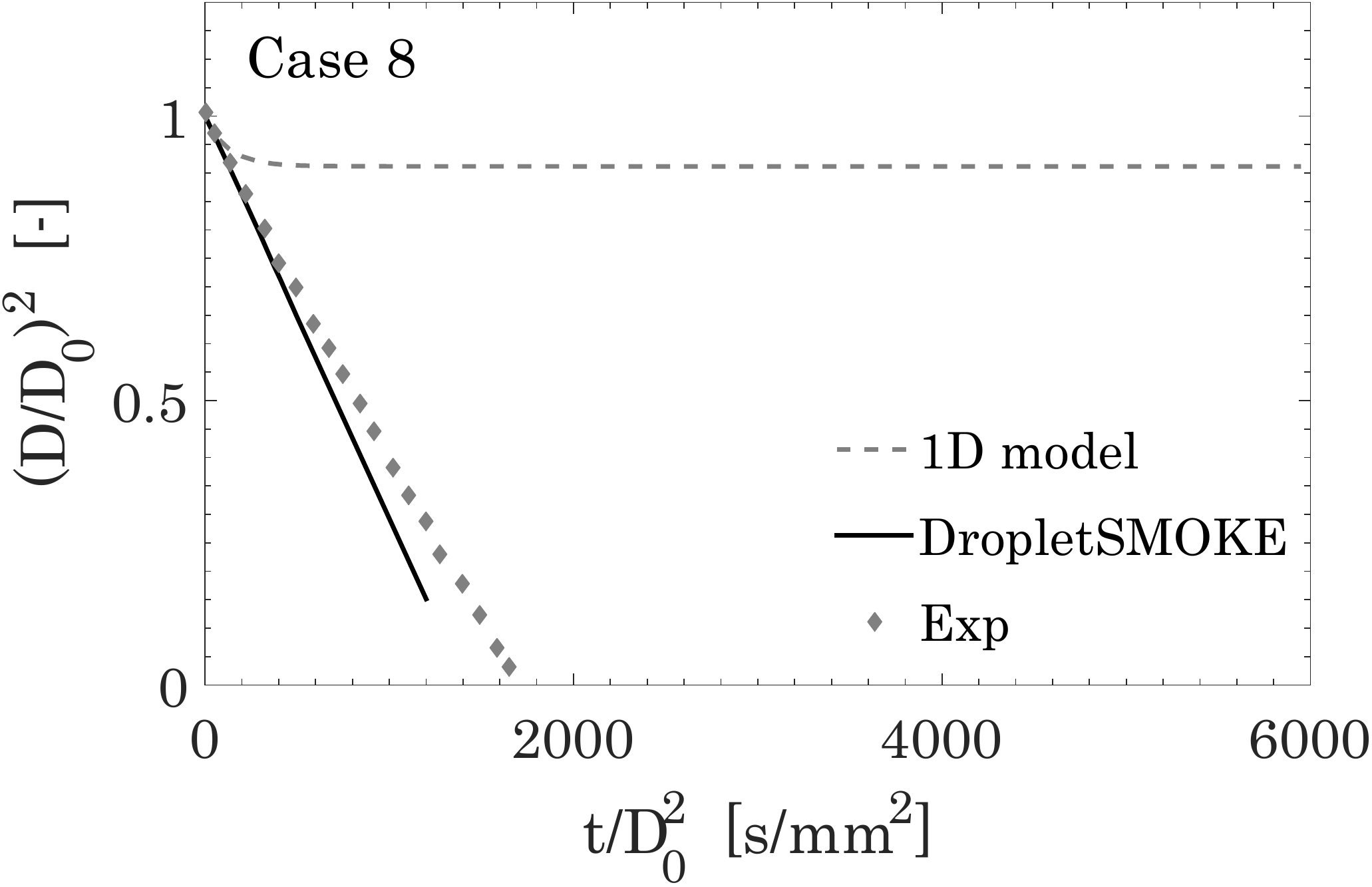}}~
	\subfloat[]
	{\includegraphics[width=.38\textwidth,height=0.2\textheight]{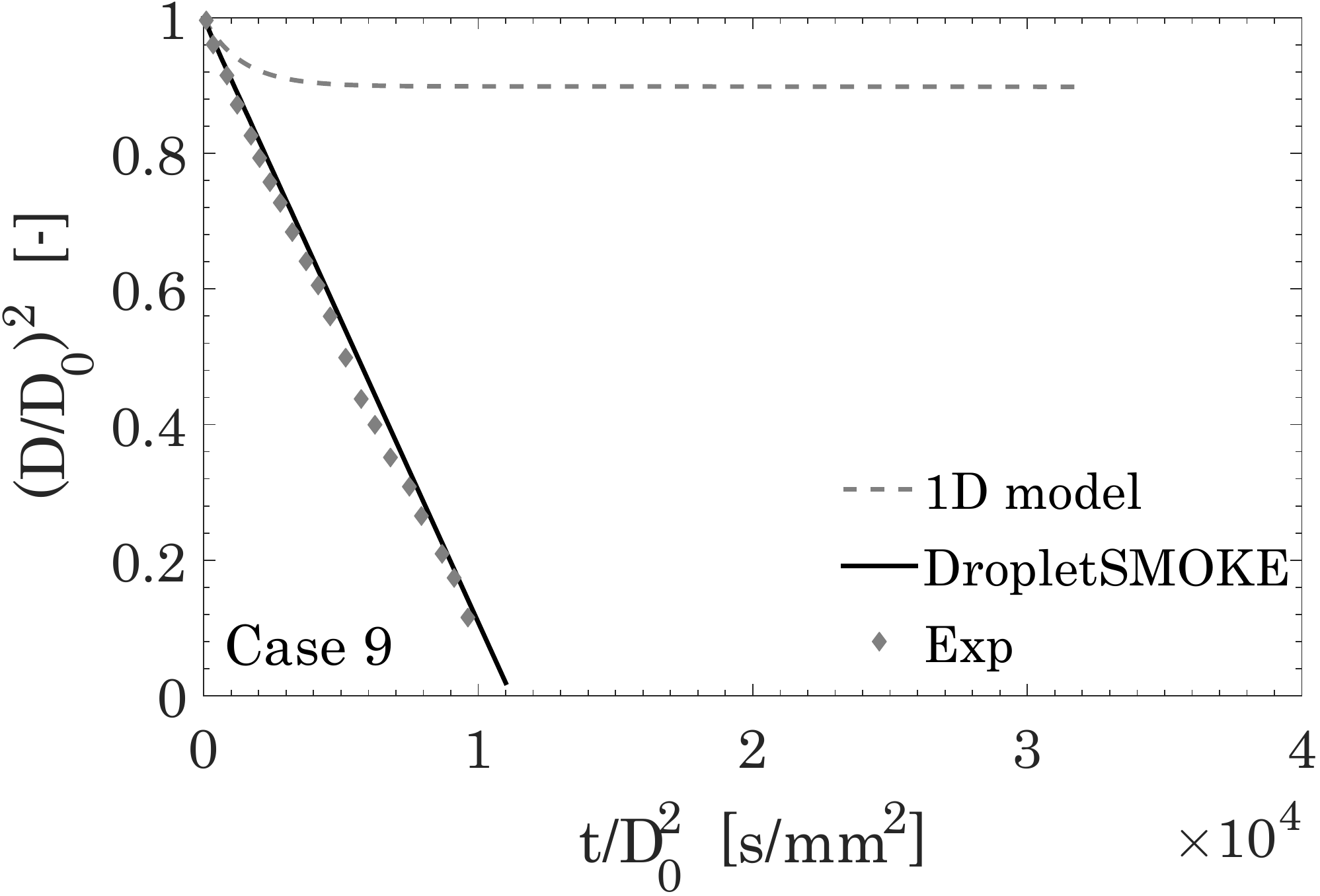}}							
	
	\caption{Model prediction of the experimental data of Verwey et al. \cite{verwey2018experimental}, cases 6, 7, 8, 9 of Table \ref{tableGravityCases}.}
	\label{VerweyEtAl}
\end{figure}

The cases from Verwey et al. \cite{verwey2018experimental} are carried out in mild conditions, characterized by a low vaporization rate due to the low temperatures and moderately high pressures involved. This is clear from Figure \ref{VerweyEtAl}, where the evaporation times are much longer with respect to the previous cases. Cases 8 and 9 are particularly interesting, since the microgravity solver predicts a quasi steady-state condition. An evaporation flux is still present, but it is extremely low due to the purely diffusive regime involved (the Stefan flow in this conditions is completely negligible). It is interesting to point out that the velocity flow around the droplet has a maximum magnitude value of only 0.5-0.6 mm/s (not shown). Despite this small convective flow, the evaporation rate is strongly enhanced.  

This can be easily analyzed by scaling arguments: in case 8 (Figure \ref{VerweyEtAl} c) the initial droplet diameter is $D_0=0.5$ mm, while the surrounding convective flow has an average magnitude of $0.5$ mm/s. This means that, taking the droplet diameter $D_0$ as a characteristic length of the system, the complete replacement of vapor cloud around the droplet lasts $\sim 1$ s, while for the pure diffusion case is practically  infinite.

\begin{figure}
	\centering
	\subfloat[]
	{\includegraphics[width=.32\textwidth,height=0.17\textheight]{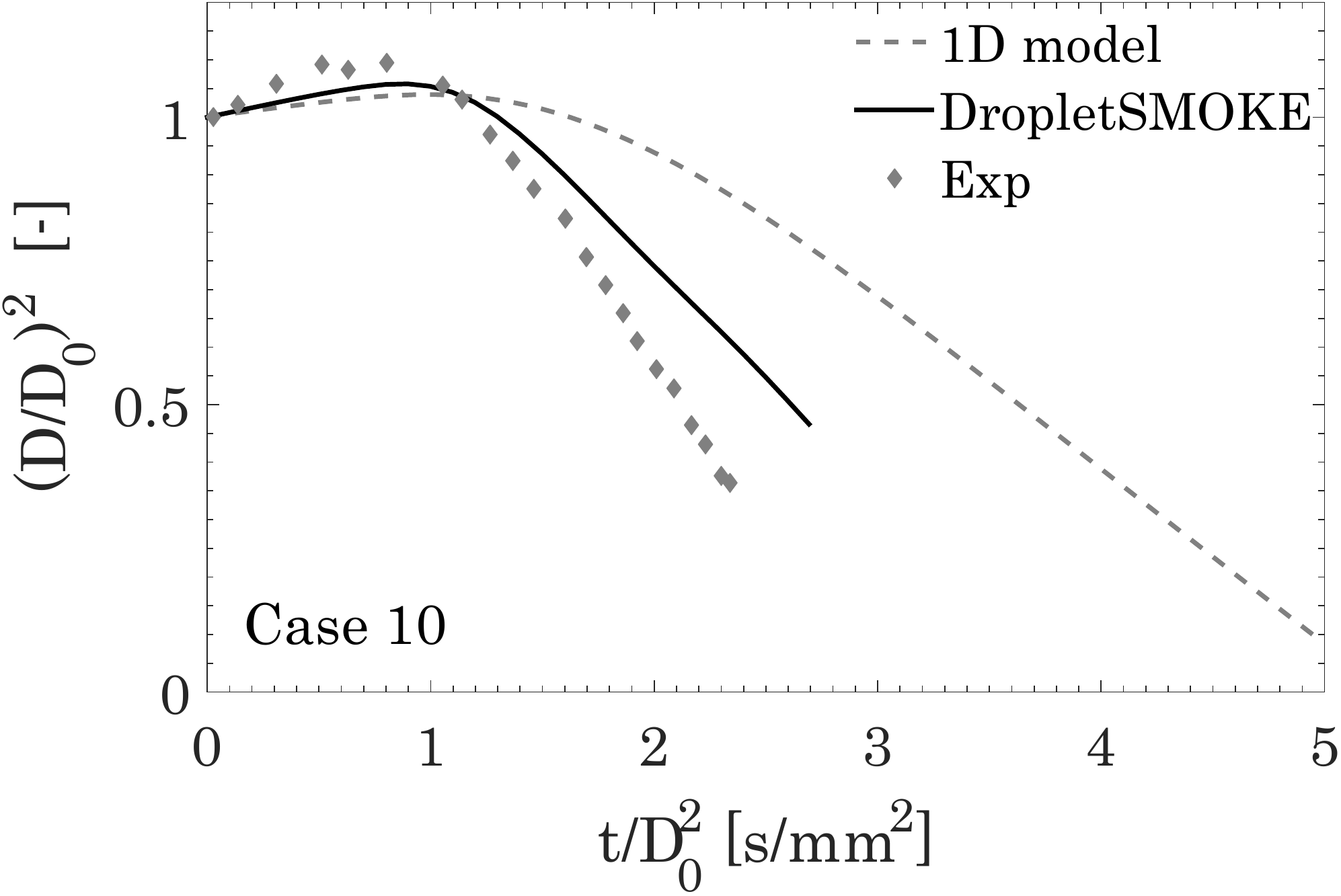}}~					
	\subfloat[]
	{\includegraphics[width=.32\textwidth,height=0.17\textheight]{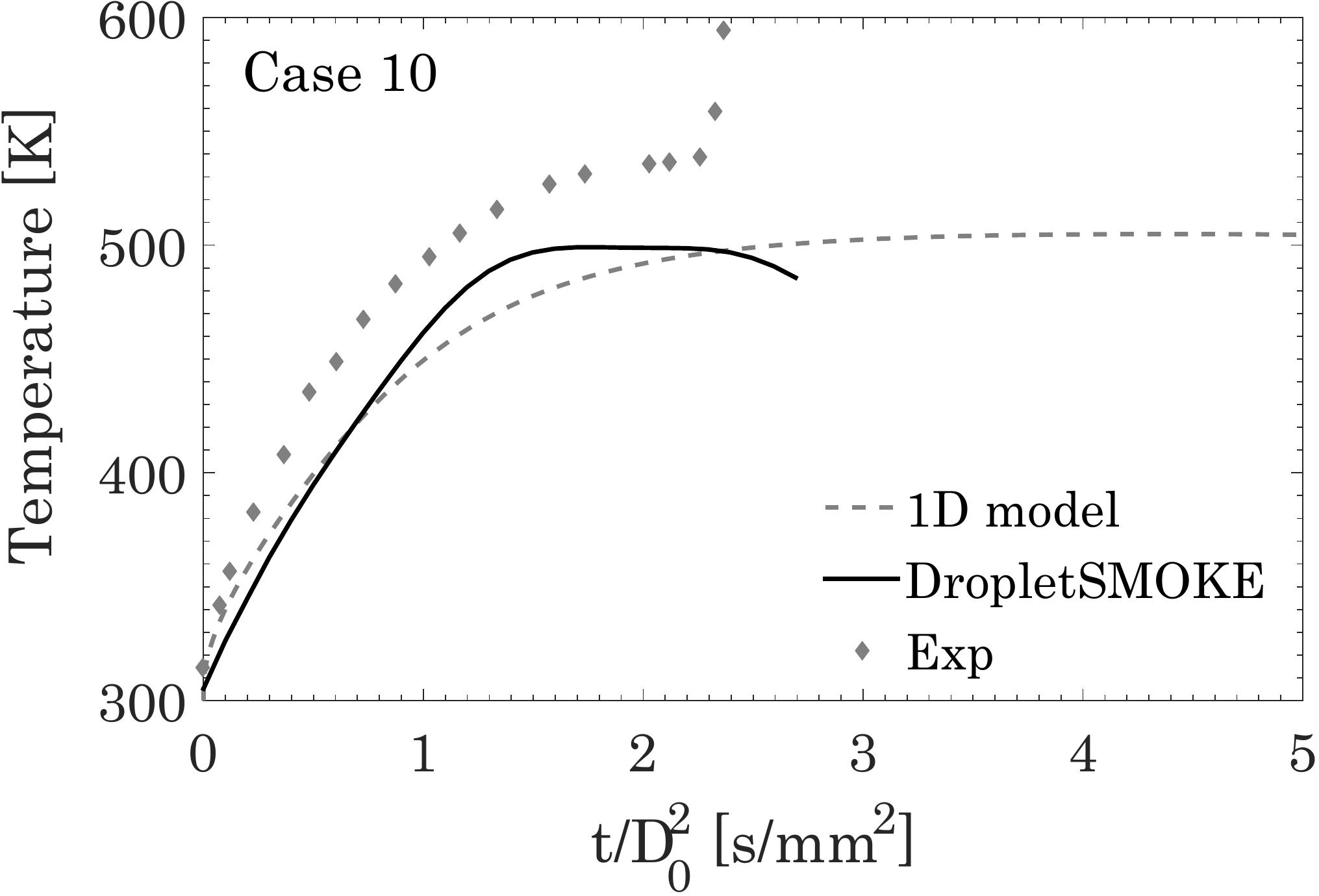}}~	
	\subfloat[]
	{\includegraphics[width=.32\textwidth,height=0.17\textheight]{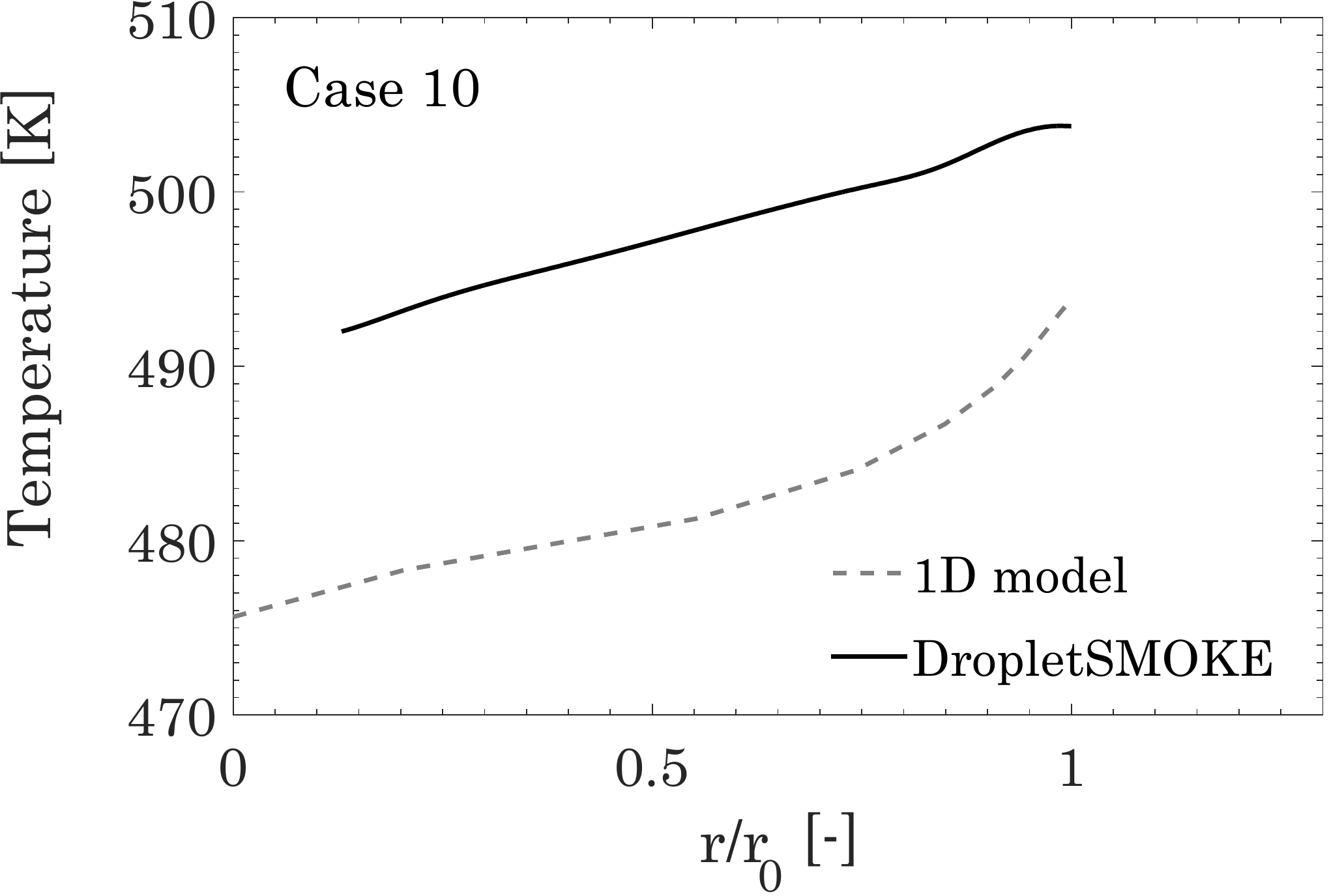}}\\	
	\subfloat[]
	{\includegraphics[width=.32\textwidth,height=0.17\textheight]{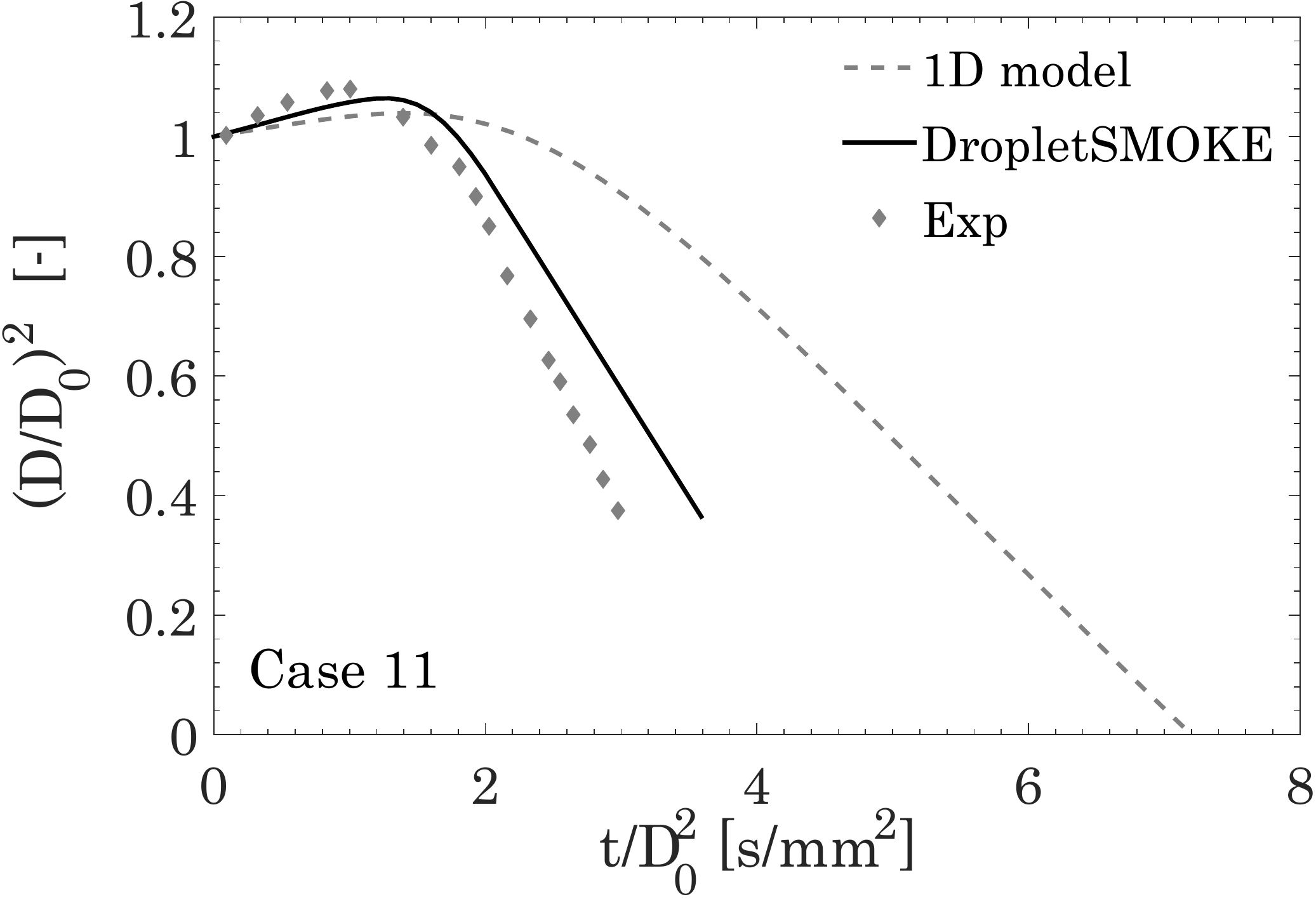}}~					
	\subfloat[]
	{\includegraphics[width=.32\textwidth,height=0.17\textheight]{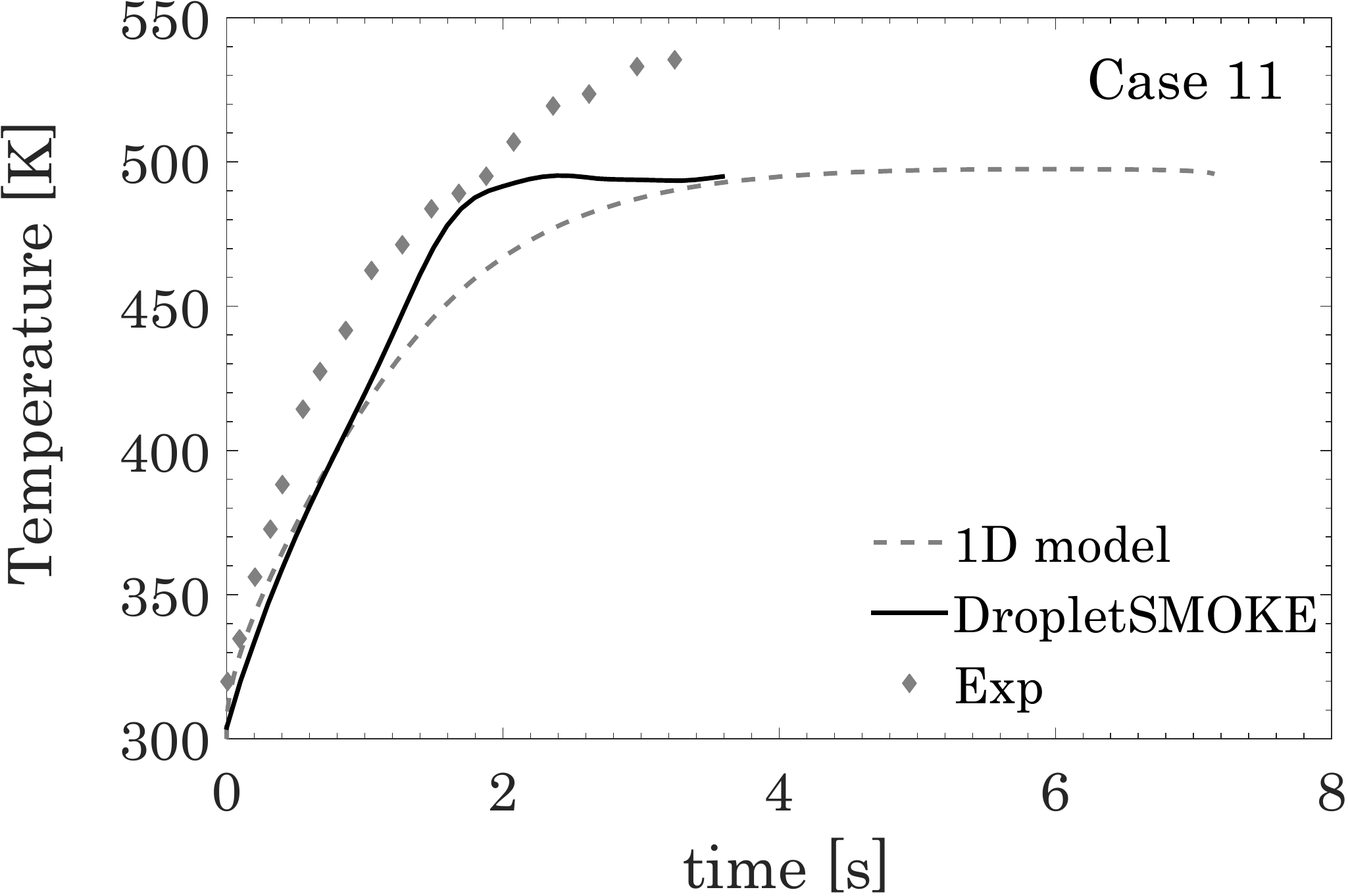}}~
	\subfloat[]
	{\includegraphics[width=.32\textwidth,height=0.17\textheight]{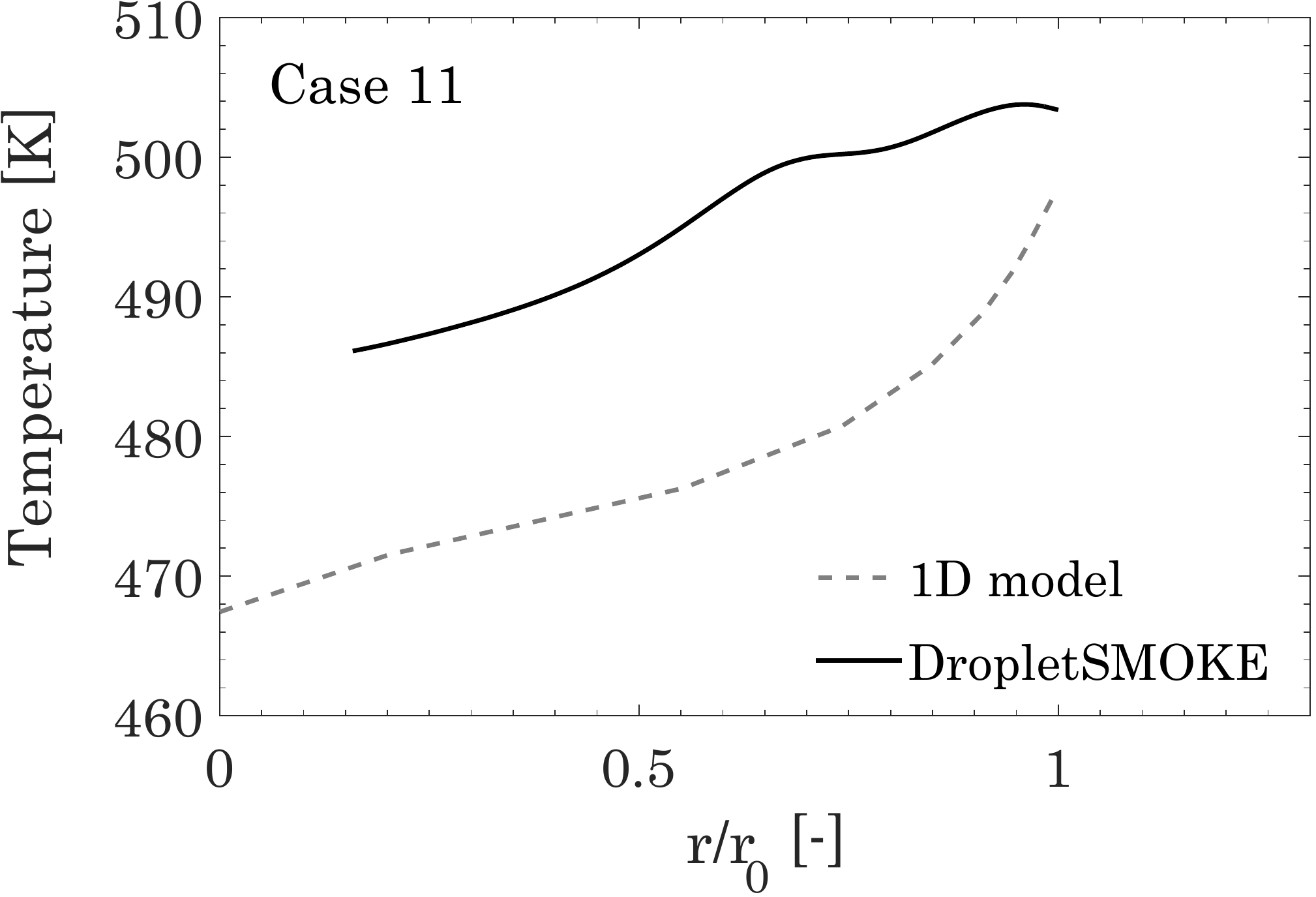}}		
	\caption{Figures (a), (b), (d), (e) show the model prediction of the experimental data of Han et al. \cite{han2015evaporation}, cases 10, 11 of Table \ref{tableGravityCases}. Figures (c), (f) report the liquid temperature distribution predited by \texttt{DropletSMOKE++} and by the 1D model for cases 10, 11 (time = 2 s).} 
	\label{HanEtAl}
\end{figure}

n-Hexadecane droplets evaporation from Han et al. \cite{han2015evaporation} have been modeled (Figure \ref{HanEtAl}). Experimental data on the mean droplet temperature are also available. 
There is a good agreement with the experimental results. The droplet temperature is slightly underestimated, especially towards the end of the simulation. Experimentally, this is due to the exposure of the thermocouple to the hot  gas when the droplet is almost entirely consumed, or to the additional heat flux on the droplet given by the heat conduction from the fiber.\\
Figures \ref{HanEtAl} (c), (f) show the liquid temperature distribution inside the droplet predicted with the 1D model and with \texttt{DropletSMOKE++}, for both Cases 10-11, at time $t=2$ s.  The heat flux on the droplet surface is rapidly redistributed inside the liquid phase because of the internal circulation, as will be further discussed later. This provides an average liquid temperature which is higher if compared to the 1D model. Moreover, the temperature distribution predicted by \texttt{DropletSMOKE++} is more homogeneous, while stronger $T$ gradients characterize the 1D model.\\

Whereas the evaporation rate and the droplet lifetime are strongly influenced by natural convection, in terms of droplet surface temperature there is no significant deviation between the 1D model and  \texttt{DropletSMOKE++}. More precisely, the initial droplet heating is faster (since evaporation is here negligible), but the steady state wet-bulb temperature is the same. The experimental data confirm this fact (Figure \ref{HanEtAl} b, e).\\
In steady state conditions, there is an equilibrium between the heat flux on the droplet surface and the evaporating flux enthalpy:

\begin{equation}
h_T\left(T_{G}-T_s\right)=h_M\left(\rho_{i,s}-\rho_{i,G}\right)\Delta h_{ev,i}
\label{equilibriumT}
\end{equation}

where $\rho_{i,s}=\frac{p_i^0\left(T_s\right) M_{w,i}}{RT_s}$ is the surface vapor concentration, while $\rho_{i,G} \sim 0$. $h_T$ is the external heat transfer coefficient, while $h_M$ is the external mass transfer coefficient. It is possible to derive one coefficient from the other through the Chilton-Colburn factor $J$ \cite{bird2002transport}:

\begin{equation}
J_M=\frac{Sh}{Re Sc^{\frac{1}{3}}}=J_T=\frac{Nu}{Re Pr^{\frac{1}{3}}}
\end{equation}

from which it is easy to derive $h_M$ in function of $h_T$:

\begin{equation}
h_M = \frac{\mathcal{D}_i}{k}\left(\frac{Sc}{Pr}\right)^{\frac{1}{3}}h_T
\end{equation}

substituting $h_M$ in Equation \ref{equilibriumT}:

\begin{equation}
T_{G}-T_s = \frac{\mathcal{D}_i}{k}\left(\frac{Sc}{Pr}\right)^{\frac{1}{3}} \frac{p_i^0\left(T_s\right) M_{w,i}}{RT_s}\Delta h_{ev,i}
\label{equilibriumTindependence}
\end{equation}

The resolution of this equation provides the equilibrium temperature $T$. Equation \ref{equilibriumTindependence} clearly indicates that the wet-bulb temperature does not depend on the external fluid flow, but only on the properties of the fluids and on the bulk temperature $T_{G}$.

\begin{table}
	\centering
	
	\begin{tabular}{lllllllll}
		\toprule
		
		Case ~~& Species~~ & $D_0$ ~~& $T_L$ ~~& $T_{G}$ ~~& p & $|\textbf{v}|$ & Refs. & Results  \\
		~~&  ~~& [mm] ~~& [K]~~ & [K] ~~& [atm] & [m/s] &   \\
		\midrule
		1 & n-Heptane     & 0.7     & 300 & 490 & 1 & 0.7 & \cite{yang2002experimental}& Figure \ref{YangEtAl} (a) \\
		2 & n-Hexadecane     & 0.7     & 300 & 490 & 1 & 0.7 & \cite{yang2002experimental}& Figure \ref{YangEtAl} (b)  \\			
		\bottomrule
		
	\end{tabular}	
	
	\caption{Experimental cases in forced convection regime examined in this work.}
	\label{tableForcedCases}
\end{table}

\section{Cases in forced convection}
\subsection{Description of the experimental cases}
To conclude this work, two cases in forced convection have been simulated and compared with experimental results taken from Yang et al. \cite{yang2002experimental}. According to our knowledge, this is the first time that these cases are numerically modeled.  n-heptane and n-hexadecane droplets have been evaporated in an upward hot gas flow generated by a heating coil. 
The experimental data are given for different supporting fibers diameters. The data chosen for comparison refer to a fiber diameter of 0.15 mm, which is the closest to the fiber dimension used in this work ($D$ = 0.1 mm).  The experimental cases are summarized in Table \ref{tableForcedCases}.

\subsection{Numerical results and discussion}
\begin{figure}
	\centering
	\subfloat[]			
	{\includegraphics[width=.38\textwidth,height=0.2\textheight]{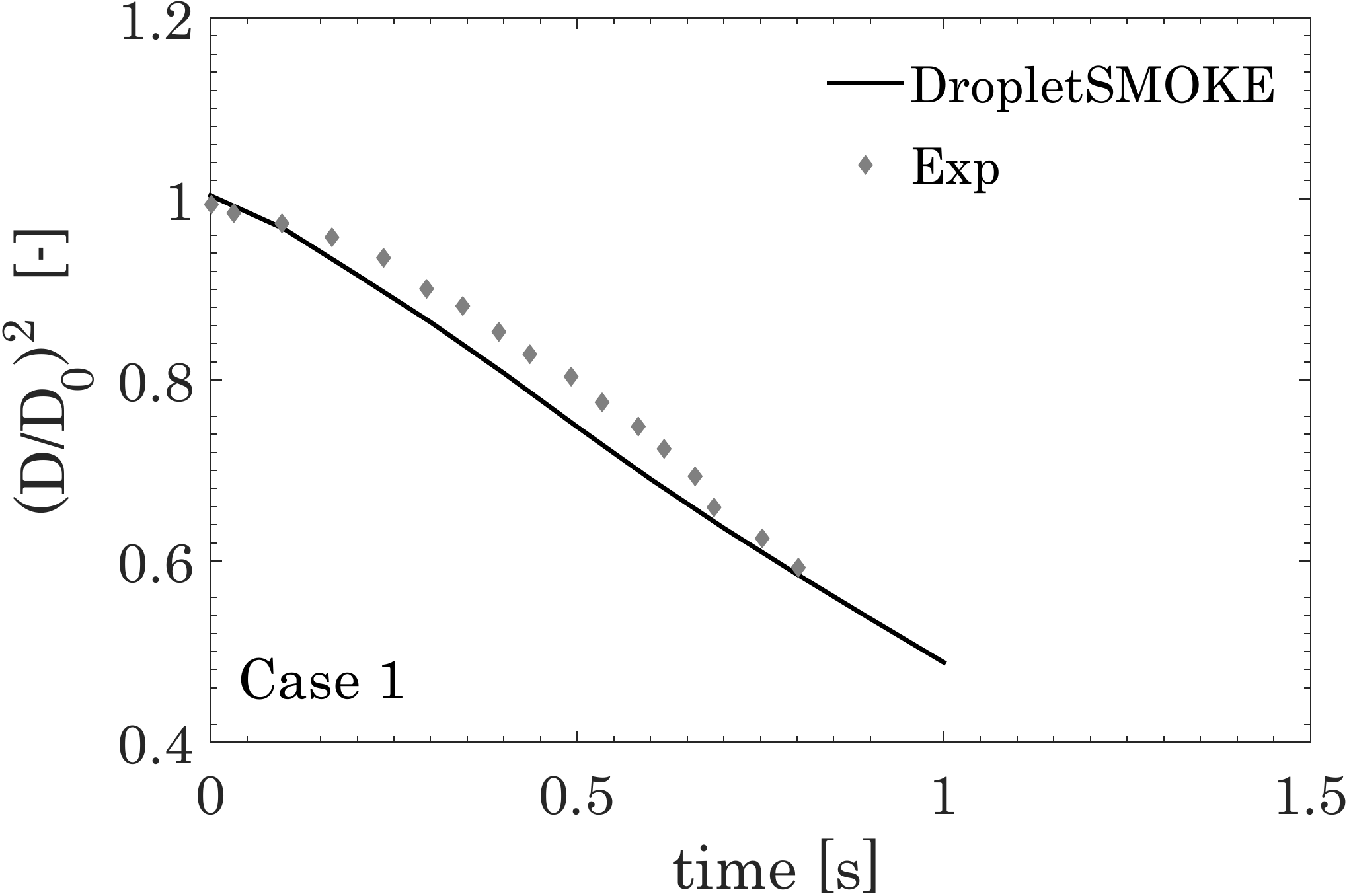}}\quad
	\subfloat[]
	{\includegraphics[width=.38\textwidth,height=0.2\textheight]{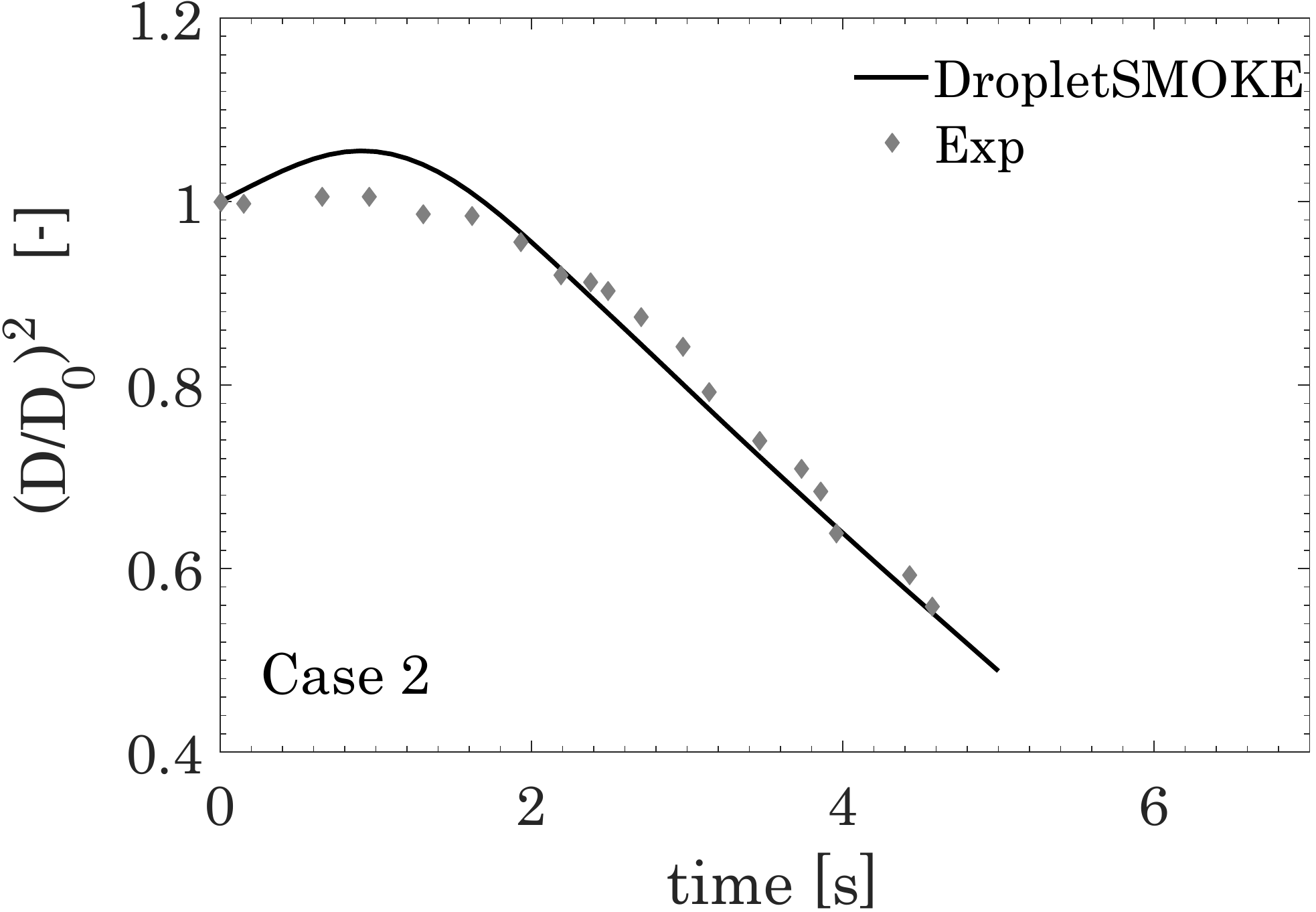}} \\
	\subfloat[]				
	{\includegraphics[width=.92\textwidth,height=0.2\textheight]{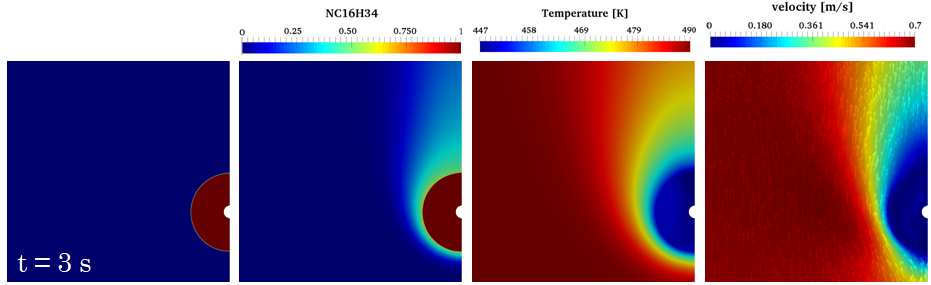}}	
	
	\caption{Figure (a): model prediction of the experimental data of Yang et al. \cite{yang2002experimental}, cases 1 (a) and 2 (b) of Table \ref{tableForcedCases}. Figure (c): $\alpha$, n-hexadecane mass fraction, temperature and velocity fields at time $t=3$ s, of the numerical simulation of case 2.}
	\label{YangEtAl}
\end{figure}

The boundary conditions presented in Table \ref{tableboundary} impose a fixed upward velocity at the $inlet$ boundary, as well as a fixed temperature value. In Figure \ref{YangEtAl} (c) we can see the upward velocity field and the relative effect on the vapor cloud around the droplet and on the temperature field.\\
The agreement with the experimental data (Figure \ref{YangEtAl} a, b) is satisfactory, with a slight overestimation of the initial heat up in case 2.

\begin{figure}
	\centering
	\subfloat[]			
	{\includegraphics[width=.37\textwidth,height=0.3\textheight]{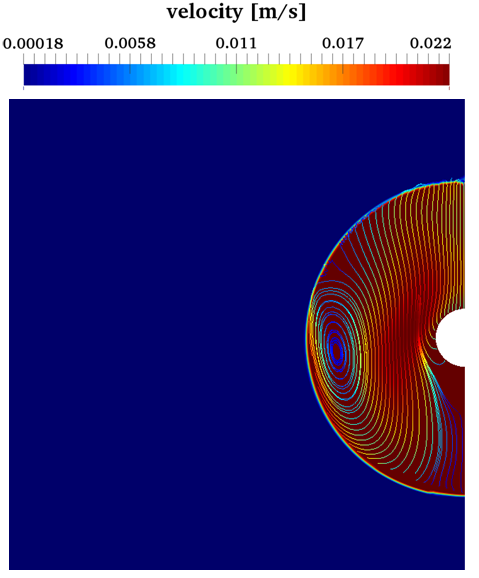}}\quad
	\subfloat[]
	{\includegraphics[width=.37\textwidth,height=0.3\textheight]{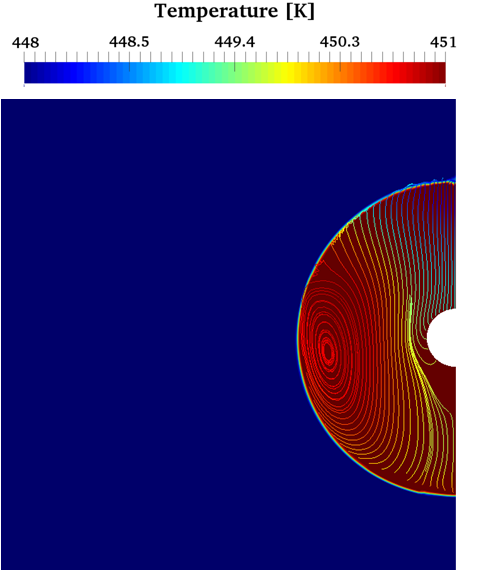}} 
	
	\caption{Flow field streamlines in the liquid phase for case 2 (Table \ref{tableForcedCases}), time $t=3$ s. Colored by velocity magnitude (a) and temperature (b).}
	\label{Streamlines}
\end{figure} 		

From Figure \ref{YangEtAl} (c), it is clear that the liquid phase can be considered almost uniform in these  conditions. This is further analyzed in Figure \ref{Streamlines}: the flow field streamlines are reported for the liquid phase, colored by the velocity magnitude (Figure \ref{Streamlines} a) and temperature  (Figure \ref{Streamlines} b) for case 2 (Table \ref{tableForcedCases}) at time $t=3$ s.  The external convective flow induces a shear stress on the surface providing a strong circulation inside the droplet, with a maximum magnitude of $\sim$ 2  cm/s. The resulting toroidal structure is well known in literature, and it is usually approximated as an inviscid Hill's spherical vortex \cite{sirignano1999fluid, lamb1945hydrodynamics} to derive sub-models of droplet heating. It has been demonstrated that internal circulation needs to be included in modeling droplet vaporization \cite{kotake1969evaporation}, and this is usually accomplished with effective conductivity models \cite{abramzon1989droplet}.  Since the entire flow field is directly solved in this work, the internal temperature field can be  predicted with no need of  such additional sub-models. The maximum temperature difference inside the liquid phase is only $\sim$ 3 $K$ (Figure \ref{Streamlines} b). As expected, a more intense heat transfer occurs at the bottom of the droplet, where the flow field impacts the liquid.\\ Finally, it is important to point out that the suppression of surface tension forces adopted in this work could in principle affect the internal motion in the liquid phase, because of the different the tangential shear stress at the interface. However, our comparative analyses in this sense, not reported in this work, show that the results obtained with \texttt{DropletSMOKE++} are in a reasonable agreement both with experimental \cite{mandal2012internal, mandal2012evidence} and numerical \cite{george2017detailed, sim2015computational} works not based on the VOF methodology. These investigations are still preliminary, but indicate  that even neglecting surface tension the internal circulation phenomenon can be quantitatively and qualitatively captured with a reasonable accuracy.

\section{Conclusions}
In this work we presented a multidimensional CFD numerical framework for the evaporation of isolated fuel droplets. The solver (called \texttt{DropletSMOKE++}) accounts for the presence of external flow field and for the influence of a gravity field, allowing to numerically model the evaporation of droplets both in natural and forced convection. The code is embedded within the well-known $\texttt{OpenFOAM}^{\textregistered}$ framework, open for contributions from the scientific community.\\
The core of the code is based on the VOF (Volume Of Fluid) methodology, widely known for its simplicity and mass conservation properties. The interface is geometrically advected and subjected to source terms to account both for evaporation and liquid expansion. This main solver is coupled with an evaporation model, which directly computes the vaporization rate from the interface molar fraction gradient, and with energy and species equations.\\
The main novelty of this work is the surface tension treatment. The parasitic currents, still an open problem for the modeling of small droplets dynamics, have been completely suppressed by-passing the modeling of surface tension effects. A centripetal force directed towards the droplet center has been introduced to stabilize the droplet position against the gravity force. This expedient allowed to accurately model the evaporation process, whatever the droplet size. By tuning this centripetal force, it is possible to modify the initial droplet shape to better match the experimental conditions (where the droplets are nearly spherical).\\
The \texttt{DropletSMOKE++} simulations has been compared with a well-known 1D microgravity solver, in order to verify the correct implementation of the governing equations, obtaining a very good agreement between the models. Afterwards an extensive validation against experimental data has been carried out, both for natural and forced convection cases available in literature, over a very wide range of temperature, pressure and initial diameter conditions.\\
The agreement with experimental data was excellent and allowed to analyze the impact of an external flow field on the evaporation process, in particularly  compared to the same cases modeled with a 1D model. The droplet lifetimes are largely overpredicted by the 1D model, while the CFD simulations carried out with  \texttt{DropletSMOKE++} are able to correctly reproduce the data. This difference between the models is particularly enhanced at high pressures and in slow evaporating conditions (low temperature and high pressure). \\
Finally, the effect of the liquid internal circulation has been highlighted.  In particular a much more uniform temperature field is provided by the internal heat transfer, strongly enhanced by the vortical structure of the internal flow field.\\
Future works will concern the introduction of a gas-phase kinetic mechanism to simulate the combustion process in presence of a convective flow, as well as the investigation of the liquid phase chemistry.  

\section*{Acknowledgments} 
The authors acknowledge the support by the European Union’s Horizon 2020 research and innovation program (Residue2Heat, G.A. No. 654650). The authors also would like to thank the Modeling and Scientific Computing (MOX) group at Politecnico di Milano for the useful and inspiring discussions.

\nomenclature[G]{$\rho$ }{density $\left[\frac{kg}{m^3}\right]$}
\nomenclature[S]{$L$}{liquid}
\nomenclature[S]{$G$}{gas}
\nomenclature[R]{$T$ }{temperature $\left[K\right]$}
\nomenclature[R]{$\textbf{v}$ }{velocity $\left[\frac{m}{s}\right]$}
\nomenclature[R]{$\dot{m}$ }{evaporative flux$\left[\frac{kg}{m^3s}\right]$}
\nomenclature[R]{$p$ }{pressure $\left[Pa\right]$}
\nomenclature[R]{$p_{rgh}$ }{dynamic pressure $\left[Pa\right]$}
\nomenclature[G]{$\alpha$ }{marker function $[-]$}
\nomenclature[R]{$k$ }{thermal conductivity $\left[\frac{W}{mK}\right]$}
\nomenclature[G]{$\beta$ }{thermal expansion coefficient  $\left[\frac{1}{K}\right]$}
\nomenclature[R]{$\textbf{j}$ }{mass flux $\left[\frac{kg}{m^2s}\right]$}
\nomenclature[S]{$i$ }{species $i$}
\nomenclature[S]{$j$ }{species $j$}
\nomenclature[S]{$c$ }{convective}
\nomenclature[S]{$d$ }{diffusive}
\nomenclature[R]{$C_p$ }{constant pressure specific heat  $\left[\frac{J}{kgK}\right]$}
\nomenclature[G]{$\Delta h_{ev}$ }{evaporation enthalpy $\left[\frac{J}{kg}\right]$}
\nomenclature[G]{$\omega$}{mass fraction $\left[-\right]$}
\nomenclature[R]{$t$}{time $\left[s\right]$}
\nomenclature[R]{$M_w$}{molecular weight $\left[\frac{kg}{mol}\right]$}
\nomenclature[R]{$\mathcal{D}$}{mass diffusion coefficient $\left[\frac{m^2}{s}\right]$}
\nomenclature[R]{$y$}{gas phase mole fraction $\left[-\right]$}
\nomenclature[R]{$x$}{liquid phase mole fraction $\left[-\right]$}
\nomenclature[R]{$p^0$}{vapor pressure $\left[Pa\right]$}
\nomenclature[G]{$\phi$}{pure gas-phase fugacity coefficient $\left[-\right]$}
\nomenclature[G]{$\hat{\phi}$}{mixture gas-phase fugacity coefficient $\left[-\right]$}
\nomenclature[R]{$Z$}{compressibility factor $\left[-\right]$}
\nomenclature[R]{$R$}{gas constant $\left[\frac{J}{mol K}\right]$}
\nomenclature[R]{$v$}{molar volume $\left[\frac{m^3}{mol}\right]$}
\nomenclature[R]{$\textbf{f}$}{force $\left[\frac{N}{m^3}\right]$}
\nomenclature[G]{$\kappa$}{curvature $\left[\frac{1}{m}\right]$}
\nomenclature[G]{$\xi$}{potential $\left[\frac{m^2}{s^2}\right]$}
\nomenclature[R]{$r$}{distance $\left[m\right]$}
\nomenclature[G]{$\mu$}{dynamic viscosity $\frac{kg}{ms}$}
\nomenclature[R]{$D$}{diameter $\left[m\right]$}
\nomenclature[R]{$Gr$}{Grashof number $\left[-\right]$}
\nomenclature[R]{$h$}{transfer coefficient $\left[\frac{W}{m^2 s}\right]$ or $\left[\frac{m}{s}\right]$}
\nomenclature[R]{$Sh$}{Sherwood number $\left[-\right]$}
\nomenclature[R]{$Ns$}{Number of species $\left[-\right]$}
\nomenclature[R]{$Nu$}{Nusselt number $\left[-\right]$}
\nomenclature[R]{$Re$}{Reynolds number $\left[-\right]$}
\nomenclature[R]{$Pr$}{Prandtl number $\left[-\right]$}
\nomenclature[R]{$W$}{basis radius of the mesh $\left[m\right]$}
\nomenclature[R]{$H$}{height of the mesh $\left[m\right]$}
\nomenclature[R]{$Sc$}{Sc number $\left[-\right]$}
\nomenclature[R]{$J$}{Chilton-Colburn factor $\left[-\right]$}
\nomenclature[R]{$\textbf{r}$}{position vector $\left[m\right]$}
\nomenclature[R]{$S$}{surface $\left[m^2\right]$}
\nomenclature[S]{$s$}{surface}
\nomenclature[S]{$M$}{mass, material}
\nomenclature[S]{$T$}{heat}
\nomenclature[S]{$m$}{centripetal}
\nomenclature[S]{$0$}{initial, reference}
\nomenclature[G]{$\sigma$}{surface tension $\left[\frac{N}{m}\right]$}
\nomenclature[S]{$x$}{$x$ direction}
\nomenclature[S]{$y$}{$y$ direction}
\nomenclature[G]{$\psi$}{sphericity $\left[-\right]$}
\nomenclature[R]{$Eo$}{E\"{o}tv\"{o}s number $\left[-\right]$}
\nomenclature[S]{$int$}{cell at the interface}
\nomenclature[S]{$adj$}{adjacent}
\nomenclature[S]{$ext$}{external}
\nomenclature[A]{VOF}{Volume Of Fluid}
\nomenclature[A]{ISS}{International Space Station}
\nomenclature[A]{MULES}{Multidimensional Universal Limiter with Explicit Solution}
\nomenclature[A]{NC\textsubscript{7}H\textsubscript{16}}{n-heptane}
\nomenclature[A]{NC\textsubscript{10}H\textsubscript{22}}{n-decane}
\nomenclature[A]{NC\textsubscript{16}H\textsubscript{34}}{n-hexadecane}

	\printnomenclature	 	 
		 	 
\section*{References}
\bibliographystyle{elsarticle-num} 
\bibliography{bibliografia}

\begin{thebibliography}{10}
\expandafter\ifx\csname url\endcsname\relax
  \def\url#1{\texttt{#1}}\fi
\expandafter\ifx\csname urlprefix\endcsname\relax\def\urlprefix{URL }\fi
\expandafter\ifx\csname href\endcsname\relax
  \def\href#1#2{#2} \def\path#1{#1}\fi

\bibitem{spalding1953combustion}
D.~B. Spalding, The combustion of liquid fuels, in: Symposium (International)
  on combustion, Vol.~4, Elsevier, 1953, pp. 847--864.

\bibitem{godsave1953studies}
G.~Godsave, Studies of the combustion of drops in a fuel spray—the burning of
  single drops of fuel, in: Symposium (International) on Combustion, Vol.~4,
  Elsevier, 1953, pp. 818--830.

\bibitem{abramzon1989droplet}
B.~Abramzon, W.~Sirignano, Droplet vaporization model for spray combustion
  calculations, International Journal of Heat and Mass Transfer 32~(9) (1989)
  1605--1618.

\bibitem{sirignano1999fluid}
W.~A. Sirignano, Fluid dynamics and transport of droplets and sprays, Cambridge
  university press, 1999.

\bibitem{dwyer1989calculations}
H.~A. Dwyer, Calculations of droplet dynamics in high temperature environments,
  Progress in Energy and Combustion Science 15~(2) (1989) 131--158.

\bibitem{dwyer1985detailed}
H.~Dwyer, B.~Sanders, Detailed computation of unsteady droplet dynamics, in:
  Symposium (International) on Combustion, Vol.~20, Elsevier, 1985, pp.
  1743--1749.

\bibitem{law1976unsteady}
C.~Law, Unsteady droplet combustion with droplet heating, Combustion and Flame
  26 (1976) 17--22.

\bibitem{kotake1969evaporation}
S.~Kotake, T.~Okazaki, Evaporation and combustion of a fuel droplet,
  International Journal of Heat and Mass Transfer 12~(5) (1969) 595--609.

\bibitem{sazhin2014modelling}
S.~Sazhin, M.~Al~Qubeissi, R.~Kolodnytska, A.~Elwardany, R.~Nasiri, M.~Heikal,
  Modelling of biodiesel fuel droplet heating and evaporation, Fuel 115 (2014)
  559--572.

\bibitem{sazhin2006advanced}
S.~S. Sazhin, Advanced models of fuel droplet heating and evaporation, Progress
  in energy and combustion science 32~(2) (2006) 162--214.

\bibitem{abramzon2005droplet}
B.~Abramzon, S.~Sazhin, Droplet vaporization model in the presence of thermal
  radiation, International Journal of Heat and Mass Transfer 48~(9) (2005)
  1868--1873.

\bibitem{marchese1999numerical}
A.~J. Marchese, F.~L. Dryer, V.~Nayagam, Numerical modeling of isolated
  n-alkane droplet flames: initial comparisons with ground and space-based
  microgravity experiments, Combustion and Flame 116~(3) (1999) 432--459.

\bibitem{cuoci2005autoignition}
A.~Cuoci, M.~Mehl, G.~Buzzi-Ferraris, T.~Faravelli, D.~Manca, E.~Ranzi,
  Autoignition and burning rates of fuel droplets under microgravity,
  Combustion and Flame 143~(3) (2005) 211--226.

\bibitem{farouk2014isolated}
T.~I. Farouk, F.~L. Dryer, Isolated n-heptane droplet combustion in
  microgravity: cool flames--two-stage combustion, Combustion and Flame 161~(2)
  (2014) 565--581.

\bibitem{okajima1975further}
S.~Okajima, S.~Kumagai, Further investigations of combustion of free droplets
  in a freely falling chamber including moving droplets, in: Symposium
  (International) on Combustion, Vol.~15, Elsevier, 1975, pp. 401--407.

\bibitem{kumagai1971combustion}
S.~Kumagai, T.~Sakai, S.~Okajima, Combustion of free fuel droplets in a freely
  falling chamber, in: Symposium (International) on Combustion, Vol.~13,
  Elsevier, 1971, pp. 779--785.

\bibitem{chauveau2000effects}
C.~Chauveau, I.~G{\"o}kalp, D.~Segawa, T.~Kadota, H.~Enomoto, Effects of
  reduced gravity on methanol droplet combustion at high pressures, Proceedings
  of the Combustion Institute 28~(1) (2000) 1071--1077.

\bibitem{farouk2017combustion}
T.~Farouk, Y.~Xu, C.~Avedisian, F.~Dryer, Combustion characteristics of primary
  reference fuel (prf) droplets: Single stage high temperature combustion to
  multistage “cool flame” behavior, Proceedings of the Combustion Institute
  36~(2) (2017) 2585--2594.

\bibitem{farouk2017isolated}
T.~Farouk, D.~Dietrich, F.~Alam, F.~Dryer, Isolated n-decane droplet
  combustion--dual stage and single stage transition to “cool flame”
  droplet burning, Proceedings of the Combustion Institute 36~(2) (2017)
  2523--2530.

\bibitem{dietrich2014droplet}
D.~L. Dietrich, V.~Nayagam, M.~C. Hicks, P.~V. Ferkul, F.~L. Dryer, T.~Farouk,
  B.~D. Shaw, H.~K. Suh, M.~Y. Choi, Y.~C. Liu, et~al., Droplet combustion
  experiments aboard the international space station, Microgravity Science and
  Technology 26~(2) (2014) 65--76.

\bibitem{cuoci2017flame}
A.~Cuoci, A.~E. Saufi, A.~Frassoldati, D.~L. Dietrich, F.~A. Williams,
  T.~Faravelli, Flame extinction and low-temperature combustion of isolated
  fuel droplets of n-alkanes, Proceedings of the Combustion Institute 36~(2)
  (2017) 2531--2539.

\bibitem{stagni2018numerical}
A.~Stagni, A.~Cuoci, A.~Frassoldati, E.~Ranzi, T.~Faravelli, Numerical
  investigation of soot formation from microgravity droplet combustion using
  heterogeneous chemistry, Combustion and Flame 189 (2018) 393--406.

\bibitem{hirt1981volume}
C.~W. Hirt, B.~D. Nichols, Volume of fluid ({VOF}) method for the dynamics of
  free boundaries, Journal of computational physics 39~(1) (1981) 201--225.

\bibitem{nabil2016interthermalphasechangefoam}
M.~Nabil, A.~S. Rattner, interthermalphasechangefoam-- {A} framework for
  two-phase flow simulations with thermally driven phase change, SoftwareX 5
  (2016) 216--226.

\bibitem{georgoulas2017enhanced}
A.~Georgoulas, M.~Andredaki, M.~Marengo, An enhanced {VOF} method coupled with
  heat transfer and phase change to characterise bubble detachment in saturated
  pool boiling, Energies 10~(3) (2017) 272.

\bibitem{nieves2015openfoam}
M.~J. Nieves-Remacha, L.~Yang, K.~F. Jensen, Open{FOAM} computational fluid
  dynamic simulations of two-phase flow and mass transfer in an advanced-flow
  reactor, Industrial \& Engineering Chemistry Research 54~(26) (2015)
  6649--6659.

\bibitem{guo2014review}
Z.~Guo, D.~Fletcher, B.~Haynes, A review of computational modelling of flow
  boiling in microchannels, The Journal of Computational Multiphase Flows 6~(2)
  (2014) 79--110.

\bibitem{schlottke2008direct}
J.~Schlottke, B.~Weigand, Direct numerical simulation of evaporating droplets,
  Journal of Computational Physics 227~(10) (2008) 5215--5237.

\bibitem{ghata2014computational}
N.~Ghata, B.~D. Shaw, Computational modeling of the effects of support fibers
  on evaporation of fiber-supported droplets in reduced gravity, International
  Journal of Heat and Mass Transfer 77 (2014) 22--36.

\bibitem{ghata2015computational}
N.~Ghata, B.~D. Shaw, Computational modeling of unsupported and fiber-supported
  n-heptane droplet combustion in reduced gravity: a study of fiber effects,
  Combustion Science and Technology 187~(1-2) (2015) 83--102.

\bibitem{brackbill1992continuum}
J.~Brackbill, D.~B. Kothe, C.~Zemach, A continuum method for modeling surface
  tension, Journal of computational physics 100~(2) (1992) 335--354.

\bibitem{cummins2005estimating}
S.~J. Cummins, M.~M. Francois, D.~B. Kothe, Estimating curvature from volume
  fractions, Computers \& structures 83~(6-7) (2005) 425--434.

\bibitem{popinet2018numerical}
S.~Popinet, Numerical models of surface tension, Annual Review of Fluid
  Mechanics 50 (2018) 49--75.

\bibitem{strotos2016predicting}
G.~Strotos, I.~Malgarinos, N.~Nikolopoulos, M.~Gavaises, Predicting the
  evaporation rate of stationary droplets with the {VOF} methodology for a wide
  range of ambient temperature conditions, International Journal of Thermal
  Sciences 109 (2016) 253--262.

\bibitem{banerjee2013numerical}
R.~Banerjee, Numerical investigation of evaporation of a single
  ethanol/iso-octane droplet, Fuel 107 (2013) 724--739.

\bibitem{george2017detailed}
O.~A. George, J.~Xiao, C.~S. Rodrigo, R.~Mercad{\'e}-Prieto, J.~Sempere, X.~D.
  Chen, Detailed numerical analysis of evaporation of a micrometer water
  droplet suspended on a glass filament, Chemical Engineering Science 165
  (2017) 33--47.

\bibitem{eisenschmidt2016direct}
K.~Eisenschmidt, M.~Ertl, H.~Gomaa, C.~Kieffer-Roth, C.~Meister,
  P.~Rauschenberger, M.~Reitzle, K.~Schlottke, B.~Weigand, Direct numerical
  simulations for multiphase flows: An overview of the multiphase code {FS3D},
  Applied Mathematics and Computation 272 (2016) 508--517.

\bibitem{popinet2009accurate}
S.~Popinet, An accurate adaptive solver for surface-tension-driven interfacial
  flows, Journal of Computational Physics 228~(16) (2009) 5838--5866.

\bibitem{roenby2016computational}
J.~Roenby, H.~Bredmose, H.~Jasak, A computational method for sharp interface
  advection, Royal Society open science 3~(11) (2016) 160405.

\bibitem{greenshields2015openfoam}
C.~J. Greenshields, Open{FOAM} user guide, Open{FOAM} Foundation Ltd, version
  3~(1).

\bibitem{damian2013extended}
S.~M. Dami{\'a}n, An extended mixture model for the simultaneous treatment of
  short and long scale interfaces, Ph.D. thesis, Universidad Nacional Del
  Litoral. Facultad de Ingenieria y Ciencias Hidricas (2013).

\bibitem{bird2002transport}
R.~B. Bird, Transport phenomena, Applied Mechanics Reviews 55~(1) (2002)
  R1--R4.

\bibitem{peng1976new}
D.-Y. Peng, D.~B. Robinson, A new two-constant equation of state, Industrial \&
  Engineering Chemistry Fundamentals 15~(1) (1976) 59--64.

\bibitem{smithintroduction}
J.~M. Smith, Introduction to chemical engineering thermodynamics, ACS
  Publications, 1950.

\bibitem{banerjee2004algorithm}
R.~Banerjee, K.~M. Isaac, An algorithm to determine the mass transfer rate from
  a pure liquid surface using the volume of fluid multiphase model,
  International Journal of Engine Research 5~(1) (2004) 23--37.

\bibitem{francois2006balanced}
M.~M. Francois, S.~J. Cummins, E.~D. Dendy, D.~B. Kothe, J.~M. Sicilian, M.~W.
  Williams, A balanced-force algorithm for continuous and sharp interfacial
  surface tension models within a volume tracking framework, Journal of
  Computational Physics 213~(1) (2006) 141--173.

\bibitem{denner2017artificial}
F.~Denner, F.~Evrard, R.~Serfaty, B.~G. van Wachem, Artificial viscosity model
  to mitigate numerical artefacts at fluid interfaces with surface tension,
  Computers \& Fluids 143 (2017) 59--72.

\bibitem{albadawi2013influence}
A.~Albadawi, D.~Donoghue, A.~Robinson, D.~Murray, Y.~Delaur{\'e}, Influence of
  surface tension implementation in volume of fluid and coupled volume of fluid
  with level set methods for bubble growth and detachment, International
  Journal of Multiphase Flow 53 (2013) 11--28.

\bibitem{raeini2012modelling}
A.~Q. Raeini, M.~J. Blunt, B.~Bijeljic, Modelling two-phase flow in porous
  media at the pore scale using the volume-of-fluid method, Journal of
  Computational Physics 231~(17) (2012) 5653--5668.

\bibitem{cuoci2015opensmoke++}
A.~Cuoci, A.~Frassoldati, T.~Faravelli, E.~Ranzi, Open{SMOKE}++: An
  object-oriented framework for the numerical modeling of reactive systems with
  detailed kinetic mechanisms, Computer Physics Communications 192 (2015)
  237--264.

\bibitem{yaws2015yaws}
C.~L. Yaws, The Yaws Handbook of Physical Properties for Hydrocarbons and
  Chemicals: Physical Properties for More Than 54,000 Organic and Inorganic
  Chemical Compounds, Coverage for C1 to C100 Organics and Ac to Zr Inorganics,
  Gulf Professional Publishing, 2015.

\bibitem{ghassemi2006experimental}
H.~Ghassemi, S.~W. Baek, Q.~S. Khan, Experimental study on binary droplet
  evaporation at elevated pressures and temperatures, Combustion science and
  technology 178~(6) (2006) 1031--1053.

\bibitem{matlosz1972investigation}
R.~Matlosz, S.~Leipziger, T.~Torda, Investigation of liquid drop evaporation in
  a high temperature and high pressure environment, International Journal of
  Heat and Mass Transfer 15~(4) (1972) 831--852.

\bibitem{morin2004vaporization}
C.~Morin, C.~Chauveau, P.~Dagaut, I.~Goekalp, M.~Cathonnet, Vaporization and
  oxidation of liquid fuel droplets at high temperature and high pressure:
  application to n-alkanes and vegetable oil methyl esters, Combustion science
  and technology 176~(4) (2004) 499--529.

\bibitem{han2015evaporation}
K.~Han, C.~Zhao, G.~Fu, F.~Zhang, S.~Pang, Y.~Li, Evaporation characteristics
  of dual component droplet of benzyl azides-hexadecane mixtures at elevated
  temperatures, Fuel 157 (2015) 270--278.

\bibitem{walton2004evaporation}
D.~Walton, The evaporation of water droplets. a single droplet drying
  experiment, Drying technology 22~(3) (2004) 431--456.

\bibitem{chauveau2008experimental}
C.~Chauveau, F.~Halter, A.~Lalonde, I.~Gokalp, An experimental study on the
  droplet vaporization: effects of heat conduction through the support fiber,
  in: Proc. of 22 nd Annual Conference on Liquid Atomization and Spray Systems
  (ILASS Europe 2008), Vol.~59, 2008, p.~61.

\bibitem{suzuki2009development}
M.~Suzuki, H.~Nomura, N.~Hashimoto, Development of apparatus for microgravity
  experiments on evaporation and combustion of palm methyl ester droplet in
  high-pressure environments, Transactions of the Japan Society for
  aeronautical and space sciences, Space technology Japan 7~(ists26) (2009)
  Ph\_43--Ph\_48.

\bibitem{Nomura1997}
H.~Nomura, H.~Rath, J.~Sato, M.~Kono, Effects of ambient pressure and natural
  convection on fuel droplet evaporation, 4th Asian-Pacific International
  Symposium on Combustion and energy Utilization (1997) 266--271.

\bibitem{gogos2003effects}
G.~Gogos, S.~Soh, D.~N. Pope, Effects of gravity and ambient pressure on liquid
  fuel droplet evaporation, International journal of heat and mass transfer
  46~(2) (2003) 283--296.

\bibitem{verwey2018experimental}
C.~Verwey, M.~Birouk, Experimental investigation of the effect of natural
  convection on the evaporation characteristics of small fuel droplets at
  moderately elevated temperature and pressure, International Journal of Heat
  and Mass Transfer 118 (2018) 1046--1055.

\bibitem{azimi2017effect}
A.~Azimi, A.~Arabkhalaj, H.~Ghassemi, R.~S. Markadeh, Effect of unsteadiness on
  droplet evaporation, International Journal of Thermal Sciences 120 (2017)
  354--365.

\bibitem{fang2017new}
W.~Fang, Y.~Jie, Y.~Shaofeng, L.~Rui, J.~Jie, A new stationary droplet
  evaporation model and its validation, Chinese Journal of Aeronautics 30~(4)
  (2017) 1407--1416.

\bibitem{yang2002experimental}
J.-R. Yang, S.-C. Wong, An experimental and theoretical study of the effects of
  heat conduction through the support fiber on the evaporation of a droplet in
  a weakly convective flow, International journal of heat and mass transfer
  45~(23) (2002) 4589--4598.

\bibitem{lamb1945hydrodynamics}
H.~Lamb, Hydrodynamics, Dover Publications, New York, 1945.

\bibitem{mandal2012internal}
D.~K. Mandal, S.~Bakshi, Internal circulation in a single droplet evaporating
  in a closed chamber, International Journal of Multiphase Flow 42 (2012)
  42--51.

\bibitem{mandal2012evidence}
D.~K. Mandal, S.~Bakshi, Evidence of oscillatory convection inside an
  evaporating multicomponent droplet in a closed chamber, Journal of colloid
  and interface science 378~(1) (2012) 260--262.

\bibitem{sim2015computational}
J.~Sim, H.~G. Im, S.~H. Chung, A computational study of droplet evaporation
  with fuel vapor jet ejection induced by localized heat sources, Physics of
  Fluids 27~(5) (2015) 053302.

\end{thebibliography}


\begin{thebibliography}{1}
\expandafter\ifx\csname url\endcsname\relax
  \def\url#1{\texttt{#1}}\fi
\expandafter\ifx\csname urlprefix\endcsname\relax\def\urlprefix{URL }\fi
\expandafter\ifx\csname href\endcsname\relax
  \def\href#1#2{#2} \def\path#1{#1}\fi

\bibitem{ref_simpo_ben:15}
A.~Attili, F.~Bisetti, M.~E. Mueller, H.~Pitsch, Formation, growth and
  transport of soot in a three-dimensional turbulent non-premixed jet flame
  \textbf{161}  1841--1865.

\end{thebibliography}

%% else use the following coding to input the bibitems directly in the
%% TeX file.

%%\begin{thebibliography}{00}

%% \bibitem[Author(year)]{label}
%% Text of bibliographic item

%%\bibitem[ ()]{}

%%\end{thebibliography}
\end{document}